\documentclass[preprint,aps]{revtex4}

\usepackage{epsfig}

\begin{document}

\title{Fitting galactic rotation curves with conformal gravity and a global quadratic potential}

\author{Philip~D.~Mannheim${}^1$ and James~G.~O'Brien${}^2$}
\affiliation{${}^1$Department of Physics\\ University of Connecticut\\ Storrs, CT 06269, USA
\\{\tt electronic address: philip.mannheim@uconn.edu}\\ 
${}^2$Department of Applied Mathematics and Sciences\\ Wentworth Institute of Technology\\ Boston, MA 02115, USA
\\{\tt electronic address: obrienj10@wit.edu}\\}

\date{January 18, 2012}

\begin{abstract}
We apply the conformal gravity theory to a sample of 111 spiral galaxies whose rotation curve data points extend well beyond the optical disk. With no free parameters other than galactic mass to light ratios, the theory is able to account for the systematics that is observed in this entire set of rotation curves without the need for any dark matter at all. In previous applications of the theory a central role was played by a universal linear potential term $V(r)=\gamma_0 c^2r/2$ that is generated through the effect of cosmology on individual galaxies, with the coefficient $\gamma_0=3.06\times 10^{-30}~{\rm cm}^{-1}$ being of cosmological magnitude. Because the current sample is so big and encompasses some specific galaxies whose data points go out to  quite substantial distances from galactic centers, we are able to identify an additional globally induced universal term in the data, a quadratic $V(r)=-\kappa c^2r^2/2$ term that is induced by inhomogeneities in the cosmic background. With $\kappa$ being found to be of magnitude $\kappa=9.54\times 10^{-54}~{\rm cm}^{-2}$, through study of the motions of particles contained within galaxies we are thus able to both detect the presence of a global de Sitter-like component and provide a specific value for its strength. Our study suggests that invoking dark matter may be nothing more than an attempt to describe global physics effects such as these in purely local galactic terms.
\end{abstract}

\maketitle

\section{Introduction}
\label{s1}

Observational studies of spiral galaxies have repeatedly established that galactic rotational velocities look nothing like the velocities that would be produced by the Newtonian gravitational potentials associated with the luminous matter in the galaxies. In consequence, it is quite widely thought that such velocity discrepancies are to be explained by the presence of copious amounts of non-luminous or dark matter in galaxies. Since the case for the presence of such dark matter rests solely on the assumption that wisdom acquired from studies on solar system distance scales can be extrapolated without modification to the much larger galactic distance scales, a few authors have ventured to suggest (see e.g. \cite{Mannheim2006} for a recent review) that dark matter may not actually exist and that instead it is the standard Newtonian description that needs modifying. In this work we apply one particular candidate alternative theory, namely conformal gravity, to a large and comprehensive sample of 111 galactic rotation curves. With only one free parameter per galaxy, the galactic mass to light ratio, we find that the conformal theory provides for a good accounting of the data without the need for any dark matter at all. Moreover, because our sample is so large, through our fitting we are able to find evidence in the data for the presence of a universal quadratic potential term that the conformal theory possesses.

As a theory, conformal gravity (see e.g. \cite{Mannheim2006}) is a completely covariant metric  theory of gravity that possesses all the general coordinate invariance and equivalence principle structure of standard Einstein gravity, but which in addition possesses a local conformal invariance in which the action is left invariant under local metric transformations of the form $g_{\mu\nu}(x)\rightarrow e^{2\alpha(x)}g_{\mu\nu}(x)$ with any arbitrary local phase $\alpha(x)$.  As a symmetry,  conformal invariance forbids the presence of any fundamental cosmological constant term in the gravitational action, with the action being uniquely prescribed by the Weyl action  
\begin{equation}
I_{\rm W}=-\alpha_g\int d^4x\, (-g)^{1/2}C_{\lambda\mu\nu\kappa}
C^{\lambda\mu\nu\kappa}
\equiv -2\alpha_g\int d^4x\, (-g)^{1/2}\left[R_{\mu\kappa}R^{\mu\kappa}-(1/3) (R^{\alpha}_{\phantom{\alpha}\alpha})^2\right],
\label{182}
\end{equation}
where 
\begin{equation}
C_{\lambda\mu\nu\kappa}= R_{\lambda\mu\nu\kappa}
-\frac{1}{2}\left(g_{\lambda\nu}R_{\mu\kappa}-
g_{\lambda\kappa}R_{\mu\nu}-
g_{\mu\nu}R_{\lambda\kappa}+
g_{\mu\kappa}R_{\lambda\nu}\right)
+\frac{1}{6}R^{\alpha}_{\phantom{\alpha}\alpha}\left(
g_{\lambda\nu}g_{\mu\kappa}-
g_{\lambda\kappa}g_{\mu\nu}\right)
\label{180}
\end{equation}
is the conformal Weyl tensor and the gravitational coupling constant $\alpha_g$ is dimensionless. Thus, unlike the standard Einstein-Hilbert action $I_{\rm EH}=-(1/16 \pi G)\int d^4x\, (-g)^{1/2}R^{\alpha}_{\phantom{\alpha}\alpha}$, which can be augmented to include a $\int d^4x\, (-g)^{1/2}\Lambda$ term, the conformal theory has a control over the cosmological constant that the standard Einstein theory does not, and because of this one is able to provide a potential solution to the cosmological constant problem \cite{Mannheim2011a,Mannheim2011b}.

\section{Local Considerations}
\label{s2}

For the Weyl action the equations of motion take the form \cite{Mannheim2006} 
\begin{equation}
4\alpha_g W^{\mu\nu}=4\alpha_g\left[
2C^{\mu\lambda\nu\kappa}_
{\phantom{\mu\lambda\nu\kappa};\lambda;\kappa}-
C^{\mu\lambda\nu\kappa}R_{\lambda\kappa}\right]=4\alpha_g\left[W^{\mu
\nu}_{(2)}-\frac{1}{3}W^{\mu\nu}_{(1)}\right]=T^{\mu\nu},
\label{188}
\end{equation}
where
\begin{eqnarray}
W^{\mu \nu}_{(1)}&=&
2g^{\mu\nu}(R^{\alpha}_{\phantom{\alpha}\alpha})          
^{;\beta}_{\phantom{;\beta};\beta}                                              
-2(R^{\alpha}_{\phantom{\alpha}\alpha})^{;\mu;\nu}                           
-2 R^{\alpha}_{\phantom{\alpha}\alpha}
R^{\mu\nu}                              
+\frac{1}{2}g^{\mu\nu}(R^{\alpha}_{\phantom{\alpha}\alpha})^2,
\nonumber\\
W^{\mu \nu}_{(2)}&=&
\frac{1}{2}g^{\mu\nu}(R^{\alpha}_{\phantom{\alpha}\alpha})   
^{;\beta}_{\phantom{;\beta};\beta}+
R^{\mu\nu;\beta}_{\phantom{\mu\nu;\beta};\beta}                     
 -R^{\mu\beta;\nu}_{\phantom{\mu\beta;\nu};\beta}                        
-R^{\nu \beta;\mu}_{\phantom{\nu \beta;\mu};\beta}                          
 - 2R^{\mu\beta}R^{\nu}_{\phantom{\nu}\beta}                                    
+\frac{1}{2}g^{\mu\nu}R_{\alpha\beta}R^{\alpha\beta}.
\label{108}
\end{eqnarray}                                 
Thus, since $W^{\mu\nu}$ vanishes when $R^{\mu\nu}$ vanishes, we see that, as well as being a vacuum solution to Einstein gravity, the Schwarzschild solution is also a vacuum solution to conformal gravity. The conformal theory thus recovers all the standard solar system Schwarzschild metric phenomenology, just as is needed for any metric theory of gravity.

However since  the vanishing of $W^{\mu\nu}$ could potentially be achieved without $R^{\mu\nu}$ needing to vanish, the conformal theory could also have some non-Schwarzschild solutions as well. To determine what such solutions might look like, Mannheim and Kazanas solved for the metric outside of a localized static, spherically symmetric source of radius $r_0$ embedded in a region with $T_{\mu\nu}(r>r_0)=0$. They found \cite{Mannheim1989} that in the conformal theory the exact, all-order classical line element is given by $ds^2=-B(r)c^2dt^2+dr^2/B(r)+r^2d\Omega_2$ where the exterior metric coefficient $B(r>r_0)$ is given by 
\begin{equation}
B(r>r_0)=w-\frac{2\beta}{r}+\gamma r -kr^2
\label{E1}
\end{equation}
with $w=(1-6\beta\gamma)^{1/2}$.
In equation (\ref{E1}) the presence of the three integration constants $\beta$, $\gamma$ and $k$ is  due to the fact that unlike the standard second-order derivative Einstein theory, the conformal theory is instead based on fourth-order derivative equations, to thus contain two additional terms. With the $\beta\gamma$ product numerically being found to be small for standard astrophysical sources (see below), we can set $w=1$. Then with the emergence of  a $1-2\beta/r$ term we see that the conformal gravity metric contains the familiar general-relativistic Schwarzschild metric solution (and thus its non-relativistic Newtonian gravitational limit as well), while departing from it only at large $r$, i.e.  departing from it in precisely the kinematic region where the dark matter problem is first encountered.

In seeking to relate the various integration constants in equation (\ref{E1}) to properties of the energy-momentum tensor $T_{\mu\nu}$ of the source, Mannheim and Kazanas found \cite{Mannheim1994} that in terms of the general source function $f(r)=(3/4\alpha_gB(r))(T^{0}_{\phantom{0}0}-T^{r}_{\phantom{r}r})$,  the exact fourth-order equation of motion given in equation (\ref{188}) reduced to the remarkably simple form
\begin{equation}
\frac{3}{B(r)}(W^{0}_{\phantom{0}0}-W^{r}_{\phantom{r}r})=\nabla^4B=B^{\prime\prime\prime\prime}+\frac{4B^{\prime\prime\prime}}{r}=
\frac{(rB)^{\prime\prime\prime}}{r}=f(r),
\label{E2}
\end{equation}
without any approximation whatsoever. (The primes here denote derivatives with respect to $r$.) Since $\nabla^4(r^2)$ vanishes identically everywhere while $\nabla^4(1/r)$ and $\nabla^4(r)$ evaluate to delta functions and their derivatives, we see that of the integration constants given in equation (\ref{E1}), only $\beta$ and $\gamma$ can be associated with properties of a local source of radius $r_0$; with the matching of the interior and exterior metrics yielding \cite{Mannheim1994} 
\begin{equation}
\gamma= -\frac{1}{2}\int_0^{r_0}dr^{\prime}\,r^{\prime 2}f(r^{\prime}),\qquad 
2\beta=\frac{1}{6}\int_0^{r_0}\,dr^{\prime}\,r^{\prime 4}f(r^{\prime}).
\label{E3}
\end{equation}
Since the $-kr^2$ term in equation (\ref{E1}) is a trivial vacuum solution ($T_{\mu\nu}(r)=0$ everywhere) to equation (\ref{E2}), as such it is not coupled to the local matter source, with the discussion here providing no basis for considering it further. (As we will show below, once we allow for matter sources in the $r>r_0$ region (i.e. $T_{\mu\nu}(r>r_0)\neq 0$), we will be able to generate a quadratic term that will be coupled to the matter in the $r>r_0$ region.) 

With the $\beta$ and $\gamma$ terms coupling to the local source, we see that in conformal gravity a given local gravitational source generates a gravitational potential
\begin{equation}
V^{*}(r)=-\frac{\beta^{*}c^2}{r}+\frac{\gamma^{*} c^2 r}{2}
\label{E4}
\end{equation}
per unit solar mass, with $\beta^{*}$ being given by the familiar $M_{\odot}G/c^2=1.48\times 10^{5}~{\rm cm}$, and with the numerical value of the solar $\gamma^{*}$ needing to be determined by data fitting. For $V^*(r)$ to generate non-relativistic motions, it is necessary that $\beta^*/r$ and $\gamma^*r$ both be very much less than one, and that the $\beta^*\gamma^*$ product thus be very much less than one too.  The domain in which we can use $V^*(r)$ is thus an intermediate one where $r$ is neither too small nor too large, and in which $\beta^*$ and $\gamma^*$ are such that $\beta^*\gamma^*$ (and thus the quantity $w-1$) are very much less than one. In our applications of $V^*(r)$ to galaxies we shall find that all of these conditions hold. However, since the form $B(r>r_0)=w-2\beta/r+\gamma r$ found above is exact without approximation, in the event that  we were not in the domain needed to use $V^*(r)$, we would have to use the exact geodesics associated with the exact $B(r)$ instead.

In conformal gravity the visible local material in a given galaxy would generate a net local gravitational potential $V_{\rm  LOC}(r)$ given by integrating $V^{*}(r)$ over the visible galactic mass distribution. Typically, the luminous material in a disk galaxy is distributed with a surface brightness $\Sigma (R)=\Sigma_0e^{-R/R_0}$ with scale length $R_0$ and total luminosity $L=2\pi \Sigma_0R_0^2$, with most of the surface brightness being concentrated in the $R \leq 4R_0$ or so optical disk region. For a galactic mass to light ratio $M/L$, one can define the total number of solar mass units $N^{*}$ in the galaxy via $(M/L)L=M=N^{*}M_{\odot}$. Then, on integrating $V^{*}(r)$ over this visible matter distribution, one obtains  \cite{Mannheim2006} the net local luminous contribution 
\begin{eqnarray}
\frac{v_{{\rm LOC}}^2}{R}&=&
\frac{N^*\beta^*c^2 R}{2R_0^3}\left[I_0\left(\frac{R}{2R_0}
\right)K_0\left(\frac{R}{2R_0}\right)-
I_1\left(\frac{R}{2R_0}\right)
K_1\left(\frac{R}{2R_0}\right)\right]
\nonumber \\
&&+\frac{N^*\gamma^* c^2R}{2R_0}I_1\left(\frac{R}{2R_0}\right)
K_1\left(\frac{R}{2R_0}\right)
\label{E5}
\end{eqnarray} 
for the centripetal accelerations  of particles in circular orbits in the plane of the galactic disk.
In the $R\gg R_0$ limit this expression simplifies to 
\begin{equation}
\frac{v_{{\rm LOC}}^2}{R} \rightarrow \frac{N^*\beta^*c^2}{R^2}\left(1+\frac{9R_0^2}{2R^2}\right)
+\frac{N^*\gamma^*c^2}{2}\left(1-\frac{3R_0^2}{2R^2}-\frac{45R_0^4}{8R^4}\right)
 \rightarrow \frac{N^*\beta^*c^2}{R^2}+
\frac{N^*\gamma^*c^2}{2},
\label{E6}
\end{equation} 
with the entire galaxy acting as if it were a point source located at the galactic center. With the surface brightness of the optical disk region essentially becoming negligible by $R=4R_0$ or so, equation (\ref{E6}) can be expected to be a good approximation to equation (\ref{E5}) at points with $R > 4R_0$. 

\section{Global Considerations}
\label{s3}

Unlike the situation that obtains in standard second-order gravity, one cannot simply use equation (\ref{E5}) as is to fit galactic rotation curve data, as one must take into consideration the effect of the rest of the material in the Universe as well. To see why this is the case, we recall that for standard gravity, the solution to the second-order Poisson equation $\nabla^2\phi(r)=g(r)$ for a general static, spherically symmetric source $g(r)$ is given by 
\begin{equation}
\phi(r)= -\frac{1}{r}\int_0^r
dr^{\prime}\,r^{\prime 2}g(r^{\prime})-\int_r^{\infty}
dr^{\prime}\,r^{\prime }g(r^{\prime}),
\label{E7}
\end{equation}                                 
with derivative 
\begin{equation}
\frac{d\phi(r)}{dr}= \frac{1}{r^2}\int_0^r
dr^{\prime}\,r^{\prime 2}g(r^{\prime}).
\label{E8}
\end{equation}                                 
As such, the import of equation (\ref{E8}) is that even though $g(r)$ could continue globally all the way to infinity, the force at any radial point $r$ is determined only by the material in the local $0< r^{\prime}< r$ region. In this sense Newtonian gravity is local, since to explain a gravitational effect in some local region one only needs to consider the material in that region. Thus in Newtonian gravity, if one wishes to explain the behavior of galactic rotation curves through the use of dark matter, one must locate the dark matter where the problem is and not elsewhere. Since the discrepancy problem in galaxies occurs primarily in the region beyond the optical disk, one must thus locate galactic dark matter in precisely the region in galaxies where there is little or no visible matter.

Despite the fact that the force in equation (\ref{E8}) is not sensitive to any material beyond the radial point of interest, this local character to Newtonian gravity is not a generic property of any gravitational potential. In particular for the fourth-order Poisson equation $\nabla^4\phi(r)=h(r)=f(r)c^2/2$ of interest to conformal gravity, the general solution is of the form 
\begin{eqnarray}
\phi(r)= -\frac{r}{2}\int_0^r
dr^{\prime}\,r^{\prime 2}h(r^{\prime})
-\frac{1}{6r}\int_0^r
dr^{\prime}\,r^{\prime 4}h(r^{\prime})
-\frac{1}{2}\int_r^{\infty}
dr^{\prime}\,r^{\prime 3}h(r^{\prime})
-\frac{r^2}{6}\int_r^{\infty}
dr^{\prime}\,r^{\prime }h(r^{\prime}).
\label{E9}
\end{eqnarray}                                 
With the derivative of the potential evaluating to 
\begin{eqnarray}
\frac{d\phi(r)}{dr}= -\frac{1}{2}\int_0^r
dr^{\prime}\,r^{\prime 2}h(r^{\prime})
+\frac{1}{6r^2}\int_0^r
dr^{\prime}\,r^{\prime 4}h(r^{\prime})
-\frac{r}{3}\int_r^{\infty}
dr^{\prime}\,r^{\prime }h(r^{\prime}),
\label{E10}
\end{eqnarray}                                 
this time we do find a contribution to the force coming from material that is beyond the radial point of interest. Thus in the third integral in equation (\ref{E10}) we recognize a potential global contribution to local motions, with a test particle in orbit in a galaxy being able to sample both the local field due to the matter in the galaxy and the global field due to the material in the rest of the Universe as well. In conformal gravity then, to determine motions of particles inside of galaxies one cannot ignore the effect of the material outside of them.

In order to determine the effect that material exterior to galaxies might have on galaxies, we note that there are actually two global effects that we need to take into consideration. Specifically, we need to consider the effects of both the homogeneous background cosmology and the inhomogeneities that are present in it. Moreover, in the conformal theory, these effects have very different geometric structures. The global background cosmology is described by a comoving Robertson-Walker (RW) geometry. Since an RW geometry is homogeneous and isotropic, its metric is conformal to flat. Thus in an RW geometry both the Weyl tensor and $W^{\mu\nu}$ vanish identically. However, since by their very nature inhomogeneities can localize in space, they are associated with geometries in which neither the Weyl tensor nor $W^{\mu\nu}$ can vanish. Indeed, in the derivation of equation (\ref{E2}) given in \cite{Mannheim1994}, it was found that in a static, spherically symmetric geometry the quantity $(3/B(r))\left(W^0_{{\phantom 0} 0} - W^r_{{\phantom r} r}\right)$ evaluates exactly to $\nabla^4B(r)$, with some components of $W^{\mu\nu}$ necessarily being non-zero in any configuration in which $\nabla^4B(r)$ is non-zero. Thus it is only inhomogeneities that contribute to the third integral in equation (\ref{E10}). 

As regards the cosmological background, we note that since the background is associated with $W^{\mu\nu}=0$, it too will contribute to the solution to $\nabla^4 B(r)=f(r)$. However, it will do so not as part of the particular integral solution given in equation (\ref{E10}) but as part of the complementary solution to $\nabla^4B=0$ instead. The background cosmology can thus have physical consequences for galactic motions provided it causes $W^{\mu\nu}$ to vanish non-trivially, i.e. provided it causes $T^{\mu\nu}$  to vanish non-trivially in equation (\ref{188}). We thus recall \cite{Mannheim1992,Mannheim2006} that in conformal gravity one can indeed construct cosmologies in which $T^{\mu\nu}$  does vanish non-trivially, and in them the scale factor $R(t)$ and the 3-curvature $K$ of the RW metric are related to the cosmological matter content, with $K$ being found \cite{Mannheim2006} to be negative.

With the Hubble flow being described in comoving coordinates and galactic rotational velocities being measured in a coordinate system in which a galaxy is at rest, to determine the effect of the Hubble flow on galactic motions we need to transform the RW metric to a static coordinate system. To this end, we recall  \cite{Mannheim1989} that the general coordinate transformation 
\begin{equation}
\rho=\frac{4r}{2(1+\gamma_0r-kr^2)^{1/2}+2 +\gamma_0 r},\qquad \tau=\int dt\,R(t)
\label{E11}
\end{equation}                                 
effects the metric transformation 
\begin{eqnarray}
&&-(1+\gamma_0r-kr^2)c^2dt^2+\frac{dr^2}{(1+\gamma_0r-kr^2)}+r^2d\Omega_2=
\nonumber \\
&&\frac{1}{R^2(\tau)}\frac{[1-\gamma_0^2\rho^2/16-k\rho^2/4]^2}
{[(1-\gamma_0\rho/4)^2+k\rho^2/4]^2}
\left[-c^2d\tau^2+\frac{R^2(\tau)}{[1-(\gamma_0^2/16+k/4)\rho^2]^2}
\left(d\rho^2+\rho^2d\Omega_2\right)\right].
\label{E12}
\end{eqnarray} 
With the transformed metric being written compactly as 
\begin{equation}
ds^2=e^{2\alpha(\tau,\rho)}\left[-c^2d\tau^2+\frac{R^2(\tau)}{[1+K\rho^2/4]^2}
\left(d\rho^2+\rho^2d\Omega_2\right)\right],
\label{E13}
\end{equation}                                 
we see that the transformed metric is conformally equivalent  to a comoving RW metric as written in spatially isotropic coordinates with spatial 3-curvature $K=-\gamma_0^2/4-k$. Since an RW geometry is conformal to flat and since it remains so under a conformal transformation, we see that when written in a static coordinate system a comoving conformal cosmology  looks just like a static metric with universal linear and quadratic terms. 

With the dynamics that leads to the $K<0$ RW metric in the first place only involving one physical cosmological scale and not two \cite{Mannheim2006}, the decomposition of just one RW scale (viz. $K$) into two static scales ($\gamma_0$ and $k$) is artificial, and it is not meaningful to keep both $\gamma_0$ and $k$. With a $K<0$ RW metric only being defined for $\rho < 2/(-K)^{1/2}$, we see that if  we keep the $k$ term alone, which would require $k=-K$ to be positive, the coordinate $r$ would become complex at $r=1/k^{1/2}$, with a transformation of the form $\rho=2r/[(1-kr^2)^{1/2}+1]$, $r=\rho/(1+k\rho^2/4)$  only being able to cover an $r \leq k^{-1/2}$ patch of a static spacetime geometry. However, if we keep the $\gamma_0$ term alone in equation (\ref{E11}), then provided $\gamma_0$ is taken to be positive, the coordinate $r$ would then be able to run all the way to infinity (just as one would want of a static, spherically symmetric geometry), with nothing being able to become complex. Thus, to be able to recover the standard static geometry with its infinite domain for the coordinate $r$, we shall retain the $\gamma_0$ term and leave out the $k$ term in the coordinate transformation. (For a discussion of a possible role for the $k$-dependent transformation in a cosmological context see \cite{Vareschi2010}.) And while we shall uncover yet another quadratic term below, viz. the $-\kappa r^2$  term that is associated with cosmological inhomogeneities, we will find that its scale is sub-cosmological and thus not to be associated with the cosmological $-kr^2$ term given in equation (\ref{E12}). (For purely phenomenological purposes, if one were to keep both of the cosmological $\gamma_0r$ and $-kr^2$ terms in equation (\ref{E12}) anyway, one could consider the quadratic term to be used in the fitting described below to be a composite of homogeneously and inhomogeneously induced quadratic terms.)

On dropping $k$ we replace equations (\ref{E11}) and (\ref{E12}) by 
\begin{equation}
\rho=\frac{4r}{2(1+\gamma_0r)^{1/2}+2 +\gamma_0 r},\qquad r= \frac{\rho}{(1-\gamma_0\rho/4)^2},\qquad \tau=\int dt\,R(t)
\label{E14}
\end{equation}                                 
and
\begin{eqnarray}
&&-(1+\gamma_0r)c^2dt^2+\frac{dr^2}{(1+\gamma_0r)}+r^2d\Omega_2=
\nonumber \\
&&\frac{1}{R^2(\tau)}\left(\frac{1+\gamma_0\rho/4}
{1-\gamma_0\rho/4}\right)^2
\left[-c^2d\tau^2+\frac{R^2(\tau)}{[1-\gamma_0^2\rho^2/16]^2}
\left(d\rho^2+\rho^2d\Omega_2\right)\right].
\label{E15}
\end{eqnarray} 
Without the $k$ term the RW 3-curvature is given by $K=-\gamma_0^2/4$, a necessarily negative quantity. Since the only way to make $K$ be positive would be to have complex $\gamma_0$, and the only way to make $K$ be zero would be to have  $\gamma_0=0$, we see that in the rest frame of a comoving galaxy (i.e. one with no peculiar velocity with respect to the Hubble flow), a topologically open comoving cosmology (viz. just the one found in \cite{Mannheim2006}), and only a topologically open one, looks just like a universal linear potential, with a strength  given by $\gamma_0/2=(-K)^{1/2}$. 

In the conformal theory then we recognize not one but two linear potential terms, a local $N^*\gamma^*$-dependent one associated with the matter within a galaxy and a global cosmological one $\gamma_0c^2r/2$ associated with the cosmological background. Thus in \cite{Mannheim1997} it was noted that in the weak gravity limit one could add the two potentials, with the total circular velocity $v_{\rm TOT}$ then being given by
\begin{equation}
v^2_{\rm TOT}=v^2_{\rm LOC}+\frac{\gamma_0 c^2R}{2},
\label{E16}
\end{equation}                                 
with asymptotic limit 
\begin{equation}
v_{{\rm TOT}}^2 \rightarrow \frac{N^*\beta^*c^2}{R}+
\frac{N^*\gamma^*c^2R}{2}+\frac{\gamma_0c^2R}{2}.
\label{E17}
\end{equation} 
In \cite{Mannheim1997}  equation (\ref{E16}) was used to fit the galactic rotation curve data of a sample of 11 galaxies, and good fits were found, with the two universal linear potential parameters being found to be given by
\begin{equation}
\gamma^*=5.42\times 10^{-41}~{\rm cm}^{-1},\qquad \gamma_0=3.06\times
10^{-30}~{\rm cm}^{-1}.
\label{E18}
\end{equation} 
The value obtained for $\gamma^*$ entails that the linear potential of the Sun is so small that there are no modifications to standard solar system phenomenology, with the values obtained for $N^*\gamma^*$ and $\gamma_0$ being so small that one has to go all the way to galactic systems before their effects can become as big as the Newtonian contribution. Moreover, the value obtained for $\gamma_0$ shows that it is indeed of cosmological magnitude, just as desired.

While the analysis described above provides no unequivocal reason for including any possible quadratic  potential term in equation (\ref{E16}), valid justification for considering it is obtained by considering not the homogeneous cosmological background, but rather the inhomogeneities in it. On large scales these inhomogeneities would typically be in the form of clusters and superclusters and would be associated with distance scales between 1 Mpc and 100 Mpc or so. Without knowing anything other than that about them, we see from equation (\ref{E9}) that  for calculating potentials at galactic distance scales (viz. scales much less than cluster scales) the inhomogeneities would contribute constant and quadratic terms multiplied by integrals that are evaluated between fixed end points, to thus be constants. (I.e. all that we require of the  $-(r^2/6)\int_r^{\infty} dr^{\prime}\,r^{\prime }h(r^{\prime})$ integral in equation (\ref{E9}) is that it begin at some minimum cluster-sized radius $r_{\rm clus}$ that is outside the galaxy and independent of it.) Thus given the quadratic term in equation (\ref{E9}), then again up to peculiar velocity effects, for weak gravity, and on scales $r < r_{\rm clus}$,  we can augment equation (\ref{E16}) to 
\begin{equation}
v^2_{\rm TOT}=v^2_{\rm LOC}+\frac{\gamma_0 c^2R}{2}-\kappa c^2R^2
\label{E20}
\end{equation}                                 
where $\kappa c^2=(1/3)\int_{r_{\rm clus}}^{\infty} dr^{\prime}\,r^{\prime }h(r^{\prime})$, with associated asymptotic limit 
\begin{equation}
v_{{\rm TOT}}^2 \rightarrow \frac{N^*\beta^*c^2}{R}+
\frac{N^*\gamma^*c^2R}{2}+\frac{\gamma_0c^2R}{2}-\kappa c^2R^2.
\label{E21}
\end{equation} 
As such, equation (\ref{E20}) can be derived from a metric with a term $B(r)=-\kappa r^2$, and thus has a de Sitter-like form. However, it is not associated with an explicit de Sitter geometry per se since the inhomogeneities that give rise to it are not distributed in a maximally 4-symmetric way. Nonetheless, a particle in orbit in a galaxy would be affected by the quadratic term, and thus behave in exactly the same way as if it had been embedded in a de Sitter background. 

Now in a conformal theory particles can only acquire mass through some scalar field symmetry breaking procedure, and thus when particles propagate they can exchange energy and momentum with such fields. However, as we show in the Appendix, even in the presence of such an exchange, circular orbits in galaxies are still of the geodesic form $rd\phi/dt=[rc^2B^{\prime}(r)/2]^{1/2}$ that leads to equation (\ref{E20}). Equation (\ref{E20}) with its universal $\kappa$ is thus our main theoretical result, and so we proceed now to apply it to galactic rotation curve data.

\section{Conformal Gravity Data Fitting}
\label{s4}

Since successful rotation curve fitting to an 11 galaxy sample was obtained in  \cite{Mannheim1997} via the use of  equation (\ref{E16}), one would initially anticipate that even if the $-\kappa c^2R$ term in equation (\ref{E20}) were to be present in principle, in practice it would be too small to have any effect. However, the sample we study here is much larger (111 galaxies) and it contains some galaxies whose data points extend to far larger distances from galactic centers than had been the case for the 11 galaxy sample originally studied in \cite{Mannheim1997}. As reported in \cite{Mannheim2011e}, it is through fitting 21 such highly extended galaxies that we were able to uncover a role for the $-\kappa c^2R$ term and extract a value for $\kappa $ given by $\kappa =9.54\times 10^{-54}~{\rm cm}^{-2}\approx (100~{\rm Mpc})^{-2}$. In the fitting to the full  111 galaxy sample we shall use this value for $\kappa$ and the values for $\gamma^*$ and $\gamma_0$ as given above in equation (\ref{E18}). For the fitting then there is just one free parameter per galaxy, namely the galactic mass to light ratio, and thus our fitting is highly constrained.

With the stars in galaxies lying within the optical disk region, to fully explore the rotation curves of galaxies one needs to study the HI gas spectra as it is only the gas in galaxies that extends well beyond the optical disk region. To get velocity measurements that are free of projection concerns one wants galaxies to be close to edge on along our line of sight, and to be able to model the gravitational contribution of the luminous disk one needs good disk photometry. Given these criteria there is a now quite substantial number of galaxies for which one can do modeling, with the  111 galaxy set that we use being a large, very varied and representative sample that contains both high surface brightness (HSB) galaxies where both $N^*$ and $\Sigma_0$ are large, and low surface brightness galaxies with small $\Sigma_0$ and dwarf galaxies with small $N^*$ (collectively referred to here as LSB galaxies since many small $N^*$ galaxies have small $\Sigma_0$ and vice versa).

Having this broad a variety of  galaxies turns out to be very instructive since one of the most interesting aspects of equations (\ref{E16}) and (\ref{E20}) is that there are situations in which departures from the luminous Newtonian prediction can be very pronounced. One situation is when $N^*$ is small, since then the net Newtonian contribution cannot compete with the fixed magnitude $\gamma_0$ and $\kappa$ terms. Another situation is when the quantity $N^*/R_0^2\sim \Sigma_0$ is small.  Specifically, since the Newtonian contribution in equation (\ref{E5}) (the $\beta^*$ dependent $I_0K_0 - I_1K_1$ term) numerically peaks at around $R=2.2R_0$, the strength of the Newtonian term at the peak will be set by the magnitude of  $N^*/R_0^2$, and when small will not be able to compete with the fixed magnitude $\gamma_0$ and $\kappa $ terms. Since the linear term dominates over the quadratic one until the largest distances, in both small $N^*$ and/or small $\Sigma_0$ galaxies one should expect the rotation curves to start rising immediately, just as is systematically seen in the data sample. The case where the luminous Newtonian contribution is not suppressed is in HSB galaxies, and here the falling Newtonian contribution can compete with the rising linear term to give a region of approximate flatness before any rise could set in, again just as is systematically seen in the data. Thus we see that the simple formula given in equation (\ref{E20}) directly captures the essence of the data, and as the fits show, the formula captures not just the qualitative trend but the actual quantitative numerical values of the velocities as well. Finally, we note that for all galaxies the quadratic term will eventually take over, to then arrest the rising linear potential terms and cause all rotation velocities to ultimately fall. Moreover, since $v^2$ cannot go negative, beyond $R\sim (N^*\gamma^*+\gamma_0)/2\kappa$ ($\sim 100~{\rm kpc}$ for $N^*=\gamma_0/\gamma^*=5.65*10^{10}$) there could no longer be any bound circular orbits, with galaxies thus having a natural way of terminating, and with the allowable sizes of galaxies being determined by an interplay between galaxies and the global structure of the Universe.

For the actual fitting we have predominantly used galaxies that were studied in large surveys. In particular for the rotation curves we have used 
18 galaxies from THINGS: The HI Nearby Galaxy Survey (as detailed in Table 1), 
30 galaxies from a study of the Ursa Major  Cluster of Galaxies (Table 2), 
20 galaxies from a study of LSB galaxies, as augmented by an extended distance study of UGC 128  (Table 3), 21 galaxies from a second study of LSB galaxies (Table 4), 
and also included some 22 miscellaneous galaxies (Table 5), with this last set containing many of the galaxies that played a significant historical role in establishing that there actually was a galactic missing mass problem in the first place.  The sample we use contains all the 11 galaxies that were studied in \cite{Mannheim1997} (DDO 154, DDO 170, NGC 1560, NGC 3109, UGC 2259, NGC 6503, NGC 2403, NGC 3198, NGC 2903, NGC 7331, and NGC 2841), with a few of them having undergone significant updates since then. Of the  111 galaxies in our sample, the 21 that extend the furthest in radial distance were reported in \cite{Mannheim2011e}, and for completeness we also include them here. In order of increasing largest radial distance the 21 galaxies are NGC 3726, NGC 3769, NGC 4013, NGC 3521, NGC 2683, UGC 1230, NGC 3198, NGC 5371, NGC 2998, NGC 5055, NGC 5033, NGC 801, NGC 5907, NGC 3992, NGC 2841, UGC 128,  NGC 5533, NGC 6674, UGC 6614, UGC 2885 and Malin 1.

For the fits we have taken photometric luminosities, optical disk scale lengths and HI gas masses  from Refs. \cite{deBlok2008} through \cite{Pickering1997}. The values we use are listed in  Tables 1 -- 5. In the last column in each of these Tables each set of four references gives the data sources for rotation velocities ($v$), luminosities ($L$), disk scale lengths ($R_0$) and HI gas mass (${\rm HI}$).

As described in the Appendix, for 11 of the galaxies (NGC 801, NGC 2998, NGC 5033, NGC 5055, NGC 5371, NGC 5533, NGC 5907, NGC 6674, UGC 2885, ESO 1440040 and Malin 1), we have also included the contribution of a central spherical bulge. Since  HI gas distributions extend well  beyond optical disk distributions, for simplicity we have modeled the gas profile in each galaxy as a single exponential disk with a scale length larger than that of its optical disk. And on finding little sensitivity to the actual gas scale lengths used (since the gas makes only a small contribution to rotation velocities), for definiteness we took the ratio of gas scale length to optical disk scale length in each galaxy to be four. Also we have multiplied the overall HI gas contribution by 1.4 to account for primordial Helium. (When an HI gas mass was not available, the HI gas mass is listed as NA in the Tables.) In the fits the gas contribution is never that significant. Specifically, in the HSB galaxies the mass in stars is much greater than the mass in gas, while in the LSB galaxies, neither the gas nor the stars are able to compete with the universal $\gamma_0$ and $\kappa$ terms. In the fits we followed the discussion in \cite{Sanders1996} and required that $M/L$ not be less than $0.2M_{\odot}/L_{\odot}$. In the Tables we have listed the fitted stellar mass to light ratios  ($M/L$) that we have obtained from our fitting,  with the $(M/L)_{\rm stars}$ values quoted in the Tables representing the total stellar disk plus bulge mass combined as divided by the total blue galactic luminosity in those cases where we have included a galactic bulge.  In almost all cases the mass to light ratios that we obtain are reasonably close to the mass to light ratio found in the local solar neighborhood, just as one would want.

In those cases where optical scale lengths have been measured in many wavelengths, by and large we have used the scale lengths as measured in the longest available wavelength band (usually the $K$ band) and have systematically  done so for the entire 30 galaxy Ursa Major sample. For seven of the galaxies  (NGC 7137, UGC 477, ESO 840411, ESO 1200211, ESO 3020120, ESO 3050090 and ESO 4880490) little or no surface photometry is available at all. As described in the Appendix, for these particular galaxies we have had to estimate scale lengths, with the sources for the scale lengths for these galaxies accordingly being listed as ES in the Tables.

The place in our theory where there is the most sensitivity to parameters is in the adopted distances to the individual galaxies, since the parameters $\gamma^*$, $\gamma_0$ and $\kappa$ that appear in equation (\ref{E20}) are given as absolute quantities. To establish a common baseline for determining adopted distances, for all the galaxies in our sample we have used the distances listed in the NASA/IPAC Extragalactic Database (NED). In this database distances are obtained either via direct visual measurements (typically Cepheids or the Tully-Fisher relation) or indirectly via redshift measurements. For the directly determined distances a world average mean value and its one standard deviation uncertainty are listed. The redshift-based determinations depend on how one models both the peculiar velocity with respect to the Hubble flow of the Milky Way Galaxy and the peculiar velocity of the galaxy of interest. Five different such models are provided in the NED, and with each one giving a mean value and uncertainty, taken together the five determinations  and their uncertainties provide a spread in values.  For definitiveness, for redshift-based distance determinations we have opted to use the mean value associated with the galactocentric  distance determination. For our entire set of  111 galaxies there was only a handful of 10 galaxies for which using the visually-determined mean or the redshift-determined galactocentric mean did not immediately give a reasonable fit. For IC 2574, NGC 2403, NGC 3621, NGC 7793 and NGC 3109 we found it advantageous to use adopted distances up to one standard deviation above the NED mean, while for NGC 2841, DDO 170, NGC 5033 and NGC 5533 we allowed up to one standard deviation below the NED mean. (For NGC 2841 the adopted distance we used coincides with the one given by Cepheid data alone, with Tully-Fisher based determinations yielding a somewhat higher value.) For NGC 6674 we used the smallest allowed distance value within the redshift determined spread in values. Thus for no less than 101 of the galaxies in our sample our theory captured the essence of the rotation curve data using the NED preferred distances as is. And moreover, despite the fact that the sensitivity to adopted distance is the most pronounced in the 21 large galaxy sample, for only four of them (NGC 5033, NGC 2841, NGC 5533  and NGC 6674) did we even need to consider not using the NED mean values as is. The fact that our fits work so well at the NED distances is thus a noteworthy achievement for our theory. In the Tables we have listed the specific adopted distances that we have used. 

Of the galaxies we fit, the data for Malin 1, a giant LSB galaxy, go out further in radial distance than any other of the galaxies in our sample, and as such this galaxy actually provides the sternest test of our ideas. Malin 1 is unusual in that its adopted distance is far larger than that of any other galaxy in our sample. In fact it is at such a high redshift  ($z=0.0824$) that its luminosity and angular diameter distances differ quite significantly. For Malin 1 the NED gives a mean galactocentric luminosity distance $D_{\rm L}=338.5$ Mpc, with the angular diameter distance thus being given by $D_{\rm A}=D_{\rm L}/(1+z)^2=288.9$ Mpc. The luminosity and HI gas masses quoted in Table 5 were evaluated using this $D_{\rm L}$, and the radial distances and scale lengths were determined using $D_{\rm A}$, with the $70''$ last data point then being at a mammoth $98.0$ kpc. The rotation curve data for Malin 1 were originally observed in \cite{Pickering1997}, and have recently been reanalyzed in \cite{Lelli2010}.  In part because of beam smearing correction considerations, the authors of \cite{Lelli2010} have revised the inner region rotation curve of \cite{Pickering1997} quite substantially, but are in reasonable agreement with \cite{Pickering1997} in the outer region, the region that is of most interest to us here. For the fit to the galaxy we thus use the first four rotation curve data points given in  \cite{Lelli2010}, and the fifth point given in  \cite{Pickering1997} (as then adjusted to the $38^{\circ}$ inclination determined by  \cite{Lelli2010}).

In regard to some specific galaxies within our  111 galaxy sample, we should note that we have some difficulty fitting NGC 7793, with the shape of its rotation curve not readily lending itself to fitting. Since the HI data for this galaxy only go out to six optical disk scale lengths or so, the fits are very sensitive to any inner region structure that would not be modeled by a single exponential disk. 

For the galaxy NGC 3109 we should note that we followed \cite{Begeman1991} and scaled up the HI gas mass by a factor of 1.67 to allow for loss of flux in the original radio observations of the galaxy given in  \cite{Jobin1990}. Even with this rescaling, at the one standard deviation NED distance of 1.5 Mpc our fit still falls a little below the observed velocities at the largest radial distances. However, as noted in our earlier fit to this galaxy \cite{Mannheim1997}, the fit falls right on the data at the slightly larger adopted distance of 1.7 Mpc. 

For the galaxy NGC 4736 the surface brightness profile was found \cite{deBlok2008} to decompose into a two-disk structure, a small disk with scale length $0.3$ kpc  that is operative in the inner $80''=1.9$ kpc region where the first 13 of the 82 rotation curve data points reported in \cite{deBlok2008} are located, together with a large disk with scale length $2.1$ kpc scale length that is operative in the region greater than $80''$. For simplicity we opted not to truncate either of the two disks so that we could use equation (\ref{E5}) as is for each of them. In the fitting the inner region disk was found to have a fitted mass $0.708\times 10^{10}~M_{\odot}$, while the dominant primary disk was found to have a mass $1.630 \times 10^{10}~M_{\odot}$. In Table 1 the reported value for $(M/L)_{\rm stars}$ for this galaxy is the total stellar mass of the two disks combined divided by the total blue luminosity of the galaxy. 

The galaxy NGC 2976 is also reported to have a two-disk structure \cite{Simon2003}, with an effective $R_0=79''=1.4$ kpc in the less than $100''$ radial region and an effective $R_0=34''=0.6$ kpc in the greater than $100''$ radial region. With the rotation curve data of \cite{deBlok2008} ending at $147''$, and with  28 of the reported 42 rotation curve data points lying in the less than $100''$ region, for the fitting we have approximated the two disks by a single disk with a blended scale length $R_0=1.2$ kpc. 

The galaxy NGC 4826 is a highly unusual galaxy in which the inner 10 of the 89 rotation curve data points reported in \cite{deBlok2008} are counter-rotating with respect to the outer 79. With the two regions being well segregated (the inner points lie within $50''$ of the center of the galaxy while the outer region points lie beyond $130''$), we provide a fit to the 79 outer region points alone. 

For two of the galaxies in our sample (UGC 5999 and F571-8) we note that even though the fits themselves are reasonable, we find fitted mass to light ratios much larger than the upper bound of $10M_{\odot}/L_{\odot}$ suggested by population synthesis models \cite{Sanders1996}. 
The galaxy UGC 5999 has a reported inclination of $14^{\circ}$. However, as noted in  \cite{deBlok1998}, since photometric data are not too sensitive to the inclination angle for close to face-on LSB galaxies, the data can permit a modest increase in the inclination to $22^{\circ}$ or so. With this latter value the rotation velocities are reduced by a factor of $\sin(14^{\circ})/\sin(22^{\circ})=0.65$, with the fitted value for  the mass to light ratio coming down to $5.4M_{\odot}/L_{\odot}$, and with the fit (not shown) even being improved. 

The galaxy F571-8 is close to edge on with an inclination close to $90^{\circ}$, and while there is then little sensitivity to inclination, because the galaxy is edge on, there instead are uncertainties in the photometry due to optical depth and projection effects \cite{deBlok2001}. For F571-8 the rotation curve data only go out to 2.7 disk scale lengths and rise to a quite high value of $143.9~{\rm km}~{\rm sec}^{-1}$. With the data being entirely in the inner optical disk region, given the large reported values for the velocities,  a fitted disk mass as high as $4.48 \times 10^{10}~M_{\odot}$  is to be expected. However for this galaxy the reported luminosity is only $0.19\times 10^{10}~L_{\odot}$, to thus lead to a large mass to light ratio. There is thus a mismatch between the large reported inner region rotation velocities and the small reported luminosity. Given the photometry uncertainties it is possible that for this galaxy the luminosity may have been underestimated.

Of the galaxies in the 21 large galaxy sample (the region were we are maximally sensitive to distance determinations) there were only three galaxies whose fitting we found challenging, viz. NGC 5533, NGC 6674 and UGC 2885, each a galaxy with a bulge. However, the fitting difficulties were mainly in the inner region where one has to make a bulge/disk decomposition of the luminosity, and not in the asymptotic region where the quadratic term contribution was still readily able to universally cancel the linear potential term contribution. Since the NED determination of the adopted distance for NGC 5533 is given as $47.7 \pm 5.7$ to one standard deviation, we found that using $42.0$ Mpc as the adopted distance gave the tightest fit. For NGC 6674 only a redshift-based adopted distance is available, and it lies in the range $42.0$ to $57.0$ Mpc. For this galaxy the fitting again preferred the smallest adopted distance value. For UGC 2885 we found that the fitting could be improved if, as described in the appendix, we used a bulge scale length somewhat larger than the one reported in the literature.  For NGC 5533 and NGC 6674 we recall \cite{Sanders1996} that NGC 5533 has significant side-to-side asymmetries and kinematic evidence for a warp, while NGC 6674 has a large scale non-axisymmetric structure and a substantial inner region bar \cite{Broeils1992}. Consequently we should not anticipate being able to do more than fit the general trend for these  two galaxies in the inner region. Nonetheless, none of these inner region luminosity structure issues affect the outer region where all the various luminous components consolidate to produce one effective $N^*$ in the asymptotic equation (\ref{E21}) that then readily controls the outer region. 

In Figs. 1 - 5 we present  the rotational velocities with their quoted errors (in ${\rm km}~{\rm sec}^{-1}$) for all of the galaxies in the  111 galaxy sample as plotted as functions of radial distances from galactic centers (in ${\rm kpc}$). For each galaxy we have exhibited the contribution due to the luminous Newtonian term alone (dashed curve), the contribution from the two linear terms alone (dot dashed curve), the contribution from the two linear terms and the quadratic terms combined (dotted curve), with the full curve showing the total contribution. As we see, the tightly constrained equation (\ref{E20}) captures the essence of the data, and does so without needing any dark matter whatsoever. 

One of the most interesting aspects  of the fits is that in the galaxies that go out to the largest radial distances the contribution of the linear potential (dot dashed curve) would actually lead to an overshoot of the data, but as the Figures show this overshoot is completely arrested by the quadratic potential term (dotted curve). Since the quadratic term would eventually cause rotation velocities to fall, to illustrate the effect, for the very small DDO 154 and for the very large UGC 128  and Malin 1 galaxies in Fig. 6 we plot the expectation of our model over an extended distance range. The anticipated ultimate fall in rotation velocities is thus a significant falsifiable diagnostic of the theory presented here, and intriguingly for the galaxy Malin 1 the  fall is expected to set in shortly beyond the current last data point. For DDO 154 we note that there actually have been some suggestions of a possible fall in the literature. However, the small fall at the end of the rotation curve that had originally been reported in \cite{Carignan1989} is not apparent in the more recent THINGS survey of the galaxy. Additionally, in \cite{Hoffman1993} it was suggested that there might  be a fall in the rotation curve at distances beyond those currently available. (The authors are indebted to Dr. M.~Milgrom for alerting them to this reference.)  However, the fall discussed in \cite{Hoffman1993} is thought to set in well before the one predicted here.

Of particular interest in the sample are the HSB galaxies NGC 3992, NGC 3198, NGC 2841 and UGC 2885, all four of which were also in the 21 large galaxy sample. While NGC 3992 is part of the Ursa Major cluster study, its NED distance of 25.6 Mpc puts it well beyond the 15.5 - 18.6 Mpc distance range that the Ursa Major cluster is thought to lie within, and yet even at this much larger distance our theory is still able to accommodate it. Both  the NGC 3198 and NGC 2841 galaxies were in the 11 galaxy sample considered in \cite{Mannheim1997}, and the rotation curves shown here are of precisely the same shape as they had been previously. However, in the interim the adopted distances to both of these galaxies have been revised upwards by as much as 50 per cent.  With the linear term contribution to $v^2$ being of the form $\gamma_0 c^2R/2$, it is extremely sensitive to distance determinations since $\gamma_0$ is given in equation (\ref{E18}) as an absolute quantity. Consequently, as the Figures show, the linear potential terms would now be requiring the NGC 3198 and NGC 2841 rotation curves to rise. That no rise is seen is due entirely to the quadratic term, with the currently observed flatness of these rotation curves being due to a natural interplay of all the various terms involved. 

While the rotation curves of all of the galaxies in the sample are obtained from HI radio studies that extend beyond the optical disk region, the rotation curve of UGC 2885 had originally been obtained from HII optical studies \cite{Rubin1980} that were thus restricted to the optical disk region where hot stars can ionize hydrogen gas. Now even though the UGC 2885 HII rotation curve data were found to quickly rise to flat (to thereby immediately suggest a missing mass problem), because the optical disk is highly extended, within the optical disk it is actually possible to fit the UGC 2885 HII rotation curve data using only the Newtonian contributions of the luminous disk and bulge and visible HI gas, without the need to invoke dark matter or alternate gravity at all \cite{Kent1986}.  Such a fit would have to be a maximum disk fit in which the luminous disk $N^*$ is taken to be as large as it possibly can be in equation (\ref{E5}), with optical disk region flatness thus not necessarily being an indicator of any failure of the luminous Newtonian expectation. However, because the UGC 2885 optical disk region does go out so far (not relatively in disk scale lengths but absolutely in kpc), the inner region rotation curve is sensitive to the linear and quadratic terms in equation (\ref{E20}), and as is seen in our fit to UGC 2885, they force the normalization of the Newtonian disk term to be less than maximal. Our work here thus supports the notion that the UGC 2885 optical disk region HII data do in fact serve as an indicator of the failure of the luminous Newtonian expectation.

In total, our fits here and in \cite{Mannheim2011e} are noteworthy in that the universal $\gamma_0$ and $\kappa$ terms have no dependence on individual galactic properties whatsoever and yet have to work in every single case. Our fits are also noteworthy in that we have captured the essence of the rotation curve data even though we have imposed some rather strong constraints on the input parameters. For adopted distances in most cases we have used NED mean values. We have not used actual surface brightness distributions or actual gas profiles but have treated these distributions simply as exponentials. Moreover, for the optical disk scale lengths we have mainly used those associated with the longest  wavelength bands available, and have taken gas scale lengths to be four times disk scale lengths. Additionally, we have not included the effects of a disk thickness  or taken any galactic inclination angle uncertainties into consideration.  On the theoretical side our fits are noteworthy in that equation (\ref{E20}) is not simply a phenomenological or empirical formula that was extracted solely from consideration of the systematics of galactic rotation curves. Rather, equation (\ref{E20}) was explicitly derived from first principles in a fundamental, uniquely prescribed metric-based theory of gravity, namely conformal gravity. Moreover, conformal gravity  itself was not even advanced for the purposes of addressing the dark matter problem. Rather, before it was known what its static, spherically symmetric solutions might even look like, it was advanced by one of us \cite{Mannheim1990} simply because it had a symmetry that could control the cosmological constant. Our fitting is thus quite non-trivial.

\section{General Comments}
\label{s5}

While beyond the scope of the present paper, we note that since the scale we find for $\kappa$ is of order $1/(100~{\rm Mpc})^2$, our work potentially has some interesting implications for clusters of galaxies. For clusters one can make measurements using either interior or exterior probes. The interior probe involves measuring galaxy kinematics and X-ray kinematics, while the exterior probe involves measuring lensing by clusters. For the interior case we need to use equation (\ref{E10}) for points within the cluster, and for lensing we need to use equation (\ref{E9}) for points exterior to the cluster, and in both cases we need to include the global effect due to all of the other clusters in the Universe. Since previous applications of conformal gravity to clusters (velocity dispersions \cite{Mannheim1995}, X-rays in clusters \cite{Horne2006,Diaferio2009}, and lensing  \cite{Walker1994,Edery1998,Pireaux2004a,Pireaux2004b,Sultana2010}) did not include this global effect,  studies of its possible impact on clusters and also on gamma ray bursters \cite{Schaefer2003,Schaefer2007,Speirits2007,Diaferio2011} could be instructive. 

A second area of interest is the growth of inhomogeneities in conformal cosmology, to see if one can generate a theoretical expectation for the matter distributions $f(r)$ and $h(r)$ that appear in equations (\ref{E2}) and (\ref{E9}). While a theory for inhomogeneity growth in conformal gravity is only in the initial stages of development with only tensor gravitational fluctuations having so far been studied \cite{Mannheim2011c}, by providing a measurement of $\kappa c^2=(1/3)\int_{r_{\rm clus}}^{\infty} dr^{\prime}~r^{\prime} h(r^{\prime})$ in this paper we have determined one of the moment integrals of the matter source $h(r)$. Our paper will thus provide an immediate test for the theory of matter fluctuations once it is developed. An essential first step toward developing such a theory has recently been taken in \cite{Mannheim2011a,Mannheim2011b,Mannheim2011d}, where it was noted that conformal cosmological perturbation theory has to be developed as a power series in Planck's constant rather than as a power series in the gravitational coupling constant. One is able to make such a quantum-mechanical development since a realization of fourth-order derivative theories such as conformal gravity has recently been found in which the quantum theory is unitary and ghost free \cite{Bender2008a,Bender2008b,Mannheim2011a,Mannheim2011d}.

Another area where our theory could potentially be tested is in the behavior of satellite galaxies around primary galaxies. Specifically, in the above we had noted that the interplay of the linear and quadratic potentials in equation (\ref{E21}) would lead to a cut-off in bound circular orbits at the point at which $R^2(d\phi/dt)^2=N^*\beta^*c^2/R+(N^*\gamma+\gamma_0)c^2R/2 -\kappa c^2R^2$ would vanish, with galaxies not being able to support bound circular orbits beyond this cut-off. Nonetheless, as we show in the Appendix, it is still possible for galaxies to support some trajectories beyond the cut-off, they just would not be circular. In a non-circular trajectory with energy $U$ and angular momentum $J$ the quantity $(dr/dt)^2/2+J^2/2r^2=U+N^*\beta^*c^2/R-(N^*\gamma+\gamma_0)c^2R/2 +\kappa c^2R^2/2$ would need to be positive. Such trajectories  are thus of relevance when the $\kappa c^2R^2$ term is large, while the circular orbits occur when $\kappa c^2R^2$ is small. A possible way to explore any such large $R$ switch over would be through the use of satellite galaxies as they are located well outside the primary galaxies. In making any such application one would have to allow for the fact that in the conformal theory the satellites themselves also put out linearly growing potentials, which would cause the satellite galaxies to interact with each other far more than they do in Newtonian gravity where potentials fall with distance.  The requirement of a switch over in the conformal theory conforms with the existence of a gap between the luminous material within each primary galaxy and the luminous material in its satellites. Actually determining the specific way in which the switch over occurs could be quite instructive, and especially since there is no such switch over in the dark matter case, where, as noted for instance in \cite{Yegorova2011}, the extrapolation for Newtonian dark matter halos is quite straightforward and smooth. (We are indebted to a referee to our paper for alerting us to the issue of satellite galaxies and to this reference.)

It is also of interest to compare our work with some other  alternative theories that have been proposed. Of them, two other non-dark-matter theories have also had success when applied to large samples of galaxies. One is the Modified Newtonian Dynamics (MOND) theory of Milgrom \cite{Milgrom1983a,Milgrom1983b,Milgrom1983c}, and the other is the Metric Skew Tensor Gravity (MSTG) theory of Moffat \cite{Moffat2005,Moffat2006}. In MOND one modifies the connection between acceleration and force by setting
\begin{equation}
 \mu\left(\frac{a}{a_0}\right)\frac{v^2}{r}=\frac{dV}{dr}
\label{E22}
\end{equation}                                 
where $a=v^2/r$ is the ordinary centripetal acceleration, $dV/dr$ is the standard Newtonian gravitational force, and $\mu(a/a_0)$ is the modification as defined in terms of some new universal parameter $a_0$ with the dimensions of acceleration. Milgrom introduced this modification because of his empirical discovery that in all those cases where the standard Newtonian theory needed dark matter, the measured centripetal accelerations were found to fall below a common value $a_0=1.2\times 10^{-8}~{\rm cm~sec}^{-2}$. Through use of the simple expression
\begin{equation}
\mu(x)= \frac{x}{(1+x^2)^{1/2}}
\label{E23}
\end{equation}                                 
for the function $\mu(x)$, Milgrom was able to construct a function that interpolated between standard Newton-Kepler behavior at $a \gg a_0$ and departures from it in the $a \ll a_0$ MOND regime where it led to asymptotically flat rotation velocities. In the years since Milgrom first introduced MOND many rotation curves of many different varieties of galaxy have been measured, and to a remarkable degree (see e.g. \cite{Begeman1991,Sanders1996,Sanders1998,deBlok1998,Sanders2002} and references therein) they have been successfully fitted by equation (\ref{E23}) without the need to include any dark matter at all. 

In Moffat's MSTG theory a skew-symmetric tensor field is coupled to Einstein gravity, with the centripetal accelerations that result being given by the simple formula
\begin{equation}
\frac{v^2}{r}=\frac{GM}{r^2}\left\{1+\left(\frac{M_0}{M}\right)^{1/2}\left[1-\left(1+\frac{r}{r_0}\right)
{\rm exp}\left(-\frac{r}{r_0}\right)\right]\right\}
\label{E24}
\end{equation} 
for a galaxy of mass $M$. In applications of equation (\ref{E24})  the parameters $M_0$ and $r_0$ are found to be given by $M_0=9.60\times 10^{11}M_{\odot}$ and $r_0=4.30\times 10^{22}~{\rm cm}$; and together they combine with Newton's constant $G$ to give a universal acceleration parameter $GM_0/r_0^2=6.90\times 10^{-8}~{\rm cm~sec}^{-2}$. In equation (\ref{E24}) the velocity obeys $v^2=GM/r$ for $r \ll r_0$ and obeys $v^2=GM[1+(M_0/M)^{1/2}]/r$ for $r \gg r_0$, to thus be Kepler in both limits, albeit  with different effective Newton constants. Via equation (\ref{E24}) successful fitting to a wide variety of galaxies has been obtained without dark matter \cite{Brownstein2006}.

That conformal gravity, MOND and MSTG can all succeed in fitting the data is because not only does each one of them possess  a universal (i.e. galaxy independent) parameter with the dimensions of an inverse length (viz. $a_0/c^2=1.33\times 10^{-29}~{\rm cm}^{-1}$ for MOND, $G_0M_0/r_0^2c^2=7.67\times 10^{-29}~{\rm cm}^{-1}$ for MSTG, and $\gamma_0=3.06\times 10^{-30}~{\rm cm}^{-1}$ for conformal gravity), the data do too. Specifically, in the Tables we have listed the value of the quantity $(v^2/c^2R)_{\rm last}$ at the last data point for each of the galaxies in the sample. As we see, despite the huge variation in luminosity and central surface brightness across the sample, within one order of magnitude all the $(v^2/c^2R)_{\rm last}$ values cluster around a value of $3\times 10^{-30}~{\rm cm}^{-1}$ or so. (In all the galaxies where $(v^2/c^2R)_{\rm last} $ is greater than $10\times 10^{-30}~{\rm cm}^{-1}$, the luminous Newtonian contribution is dominating $v^2_{\rm last}$, with those galaxies not being asymptotic enough to be in the region where the universal linear potential $\gamma_0$ term would dominate.) Now different theories cannot agree for ever, and since equations (\ref{E20}), (\ref{E22}) and (\ref{E24}) predict differing behaviors at large $R$, study of rotation curves at large enough $R$ could enable us to distinguish between them. 

As regards the near universality of $(v^2/c^2R)_{\rm last}$, we should note that this is an empirical property of the raw data themselves. Moreover, while there may be some uncertainties in the adopted distances to the galaxies, such uncertainties are never more than a factor of two or so. With the velocities being uncertain to no more than 10 to 20 per cent or so, the near universality of $(v^2/c^2R)_{\rm last}$ is thus a genuine property of the data. It should thus be regarded as an important empirical clue for galactic dynamics. 

It is important to recognize that the fits provided by conformal gravity (and likewise by MOND and MSTG) are predictions. Specifically, for all these theories the only input one needs is the optical and gas spectra, and the only free parameter is the $M/L$ ratio for each given galaxy, with rotation velocities then being determined. Moreover, the $M/L$  ratios are highly constrained by the data in the inner rotation curve region where departures from the Newtonian expectation are at their minimum, and as the Tables show, they are all by and large found to be of order the mass to light ratio found in the local solar neighborhood, just as one would want. It is important to stress this point since dark matter fitting to galactic data works very differently. There one first needs to know the velocities so that one can then ascertain the needed amount of dark matter, i.e. in its current formulation dark matter is only a parametrization of the velocity discrepancies that are observed and is not a prediction of them. Dark matter theory has yet to develop to the point where it is able to predict rotation velocities given a knowledge of the luminous distribution alone (or explain the near universality found for $(v^2/c^2R)_{\rm last}$). Thus dark matter theories, and in particular those theories that produce dark matter halos in the early Universe, are currently unable to make an a priori determination as to which halo is to go with which particular luminous matter distribution, and need to fine-tune halo parameters to luminous parameters galaxy by galaxy. (In the NFW CDM simulations \cite{Navarro1996,Navarro1997} for instance, one finds generic spherical halo profiles close in form to $\sigma(r)=\sigma_0/[r(r+r_0)^2]$ (as then cut off at some $r_{\rm max}$), but with the halo parameters needing to be fixed galaxy by galaxy.) No such fine-tuning shortcomings appear in conformal gravity, and if standard gravity is to be the correct  description of gravity, then a universal formula akin to the one given in equation (\ref{E20}) would need to be derived by dark matter theory. However, since our study establishes that global physics has an influence on local galactic motions, the invoking of dark matter in galaxies could potentially be nothing more than an attempt to describe global physics effects in purely local galactic terms.

The authors wish to thank Dr.~J.~R.~Brownstein, Dr.~W.~J.~G.~ de Blok, Dr.~J.~W.~Moffat, and Dr.~ S.~S.~McGaugh for helpful communications, and especially for providing their galactic data bases. This research has made use of the NASA/IPAC Extragalactic Database (NED) which is operated by the Jet Propulsion Laboratory, California Institute of Technology, under contract with the National Aeronautics and Space Administration. 

\appendix
\setcounter{equation}{0}
\def\theequation{A\arabic{equation}}

\section{Galaxies with  Bulges or without Photometry}

\subsection{Spherical Bulge Formalism}

For a spherically symmetric matter distribution with radial matter number density $\sigma(r)$
and $N=4\pi \int  dr^{\prime}\,r^{\prime 2}\sigma(r^{\prime})$ stars, as follows directly from equation (\ref{E8}) and equation (\ref{E10}) the rotational velocities associated with  the Newtonian and linear potentials are given by
\begin{eqnarray}
v^2_{\beta}(r)&=&{4\pi\beta^* c^2\over r}\int_0^r
dr^{\prime}\,\sigma(r^{\prime}) r^{\prime 2},
\nonumber\\
v^2_{\gamma}(r)&=&{2\pi\gamma^* c^2\over 3r}\int_0^r
dr^{\prime}\,\sigma(r^{\prime}) (3r^2r^{\prime 2}-r^{\prime 4})
+{4\pi\gamma^* c^2r^2\over 3}\int_r^{\infty} dr^{\prime}\,\sigma(r^{\prime})
r^{\prime }.
\label{A30}
\end{eqnarray} 
Ordinarily it is not the 3-dimensional $\sigma(r)$ which is directly
measured in spherical astronomical systems. Rather, it is only the
two-dimensional surface matter distribution $I(R)$ which is measured,
with $\sigma(r)$ having to be extracted from it via an Abel transform
\begin{equation}
\sigma(r)=-{1 \over \pi} \int _r^{\infty} dR\,{ I^{\prime}(R) 
\over (R^2-r^2)^{1/2}},\qquad
I(R)=2\int _R^{\infty} dr\, {\sigma(r) r \over (r^2-R^2)^{1/2}}.
\label{A31}
\end{equation} 
In terms of $I(R)$ the Newtonian integral in equation (\ref{A30}) can be rewritten as \cite{Kent1986}
\begin{equation}
v^2_{\beta}(r)={2 \pi \beta^* c^2\over r} \int_0^rdR\,RI(R) 
+{4 \beta^* c^2\over r}\int_r^{\infty}dR\,RI(R)
\left[ {\rm arcsin} \left({r \over R} \right) -{r \over
(R^2-r^2)^{1/2}}\right],
\label{A36}
\end{equation} 
while the linear potential integral reduces to  \cite{Mannheim2006}
\begin{eqnarray}
v^2_{\gamma}(r)&=&
{\gamma^* c^2 \pi \over 2r} \int_0^rdR\,RI(R)(2r^2-R^2)
\nonumber\\
&+&{\gamma^* c^2  \over r}\int_r^{\infty}dR\,RI(R)
\left[ (2r^2-R^2){\rm arcsin} \left({r \over R} \right)
+r(R^2-r^2)^{1/2}\right].
\label{A37}
\end{eqnarray} 

For the very convenient exponential surface density
\begin{equation}
I(R)=\frac{N}{2\pi t^2}e^{-R/t}
\label{B5}
\end{equation}
considered in \cite{Andredakis1994}, the Abel transform can be performed analytically, to yield
\begin{equation}
\sigma(r)=\frac{N}{2\pi^2 t^3}K_0(r/t).
\label{B6}
\end{equation}
For particles orbiting such a spherical bulge at radius $r$ we immediately obtain circular velocities of the form
\begin{equation}
v^2_{\beta}(r)={2 N\beta^* c^2\over \pi r} \int_0^{r/t}dz\, z^2K_0(z)
\label{B7}
\end{equation} 
\begin{eqnarray}
v^2_{\gamma}(r)&=&{N\gamma^* c^2 r\over \pi} \int_0^{r/t}dz\, z^2K_0(z)
\nonumber\\
&-&{ N\gamma^* c^2 t^2 \over 3 \pi r} \int_0^{r/t}dz\, z^4K_0(z)
+{2 N\gamma^* c^2 r^2 \over 3\pi t}\int_{r/t}^{\infty}dz\, zK_0(z)
\nonumber\\
&=&{N\gamma^* c^2 r\over \pi} \int_0^{r/t}dz\, z^2K_0(z)
\nonumber\\
&-&{ N\gamma^* c^2 t^2 \over 3 \pi r} \int_0^{r/t}dz\, z^4K_0(z)
+{2 N\gamma^* c^2 r^3 \over 3\pi t^2}K_1(r/t).
\label{B8}
\end{eqnarray} 

\subsection{Applications to Galaxies with Spherical Bulges}

We considered bulges for 11 galaxies. At our adopted distances the measured bulge scale lengths for NGC 801 and NGC 2998 are $t=0.9$ kpc and $t=0.9$ kpc \cite{Andredakis1994,Sanders1996},  for NGC 5371, NGC 5533 and NGC 6674 the measured values are  $t=0.9$ kpc, $t=1.3$ kpc, and $t=0.9$ kpc \cite{Sanders1996}, and for ESO 0140040  $t=1.36$ kpc \cite{Beijersbergen1999}.  For NGC 5033, NGC 5055, NGC 5907 and Malin 1 we determined respective best values of $t=1.73$ kpc, $t=0.35$ kpc, $t=1.84$ kpc, and $t=1.0$ kpc from fitting the rotation curve themselves. The fitting could generally accommodate fairly broad ranges around these particular fitted values, and could do so while only affecting the fitting in the inner rotation curve region.  (The expression for $v^2_{\beta}(r)$ given in equation (\ref{B7}) peaks at around $r=2.7t$ and becomes Keplerian by about $r=5t$. Thus for all but the innermost of the points on the rotation curve, $v^2_{\beta}(r)$ acts just like a point Newtonian source at the center of the galaxy. In addition, just like a linear potential point source, in the innermost region the contribution of the $v^2_{\gamma}(r)$ term given in equation (\ref{B8}) is negligible.) For UGC 2885, we could readily fit the outer 16 of the 19 rotation curve data points using the $t=0.6$ kpc scale length given in \cite{Andredakis1994,Sanders1996}. Given the uncertainties inherent in bulge/disk decompositions, we can vary the bulge scale length somewhat, to find that we can improve the fit for the innermost three points while still being able to account for the other 16 points. In the Figures we report the fit with $t=1.0$ kpc. In the fitting we obtained  fitted bulge masses  for the 11 galaxies that are respectively given by   4.29, 1.93, 2.38, 11.12, 10.44, 3.52, 9.75, 0.73, 7.71, 9.46, and 8.72 (in units of $10^{10}M_{\odot}$). While there may be some uncertainties in bulge/disk decompositions, these uncertainties only affect the inner rotation curve region and do not impact on the behavior of rotation curves at the largest radial distances where the missing mass problem is the most pronounced and where the linear and quadratic potential terms are dominant.

\subsection{Treatment of Galaxies with no Photometry}

For two of the galaxies in our sample (NGC 7137 and UGC 477) there appears to be no surface photometry reported in the literature, while for five of them (ESO 840411, ESO 1200211, ESO 3020120, ESO 3050090 and ESO 4880490) there is only a minimal amount. In the absence of any surface brightness photometry our strategy is to assume that the surface brightness can be described as a disk with exponential $\Sigma(R)=\Sigma_0e^{-R/R_0}$, and simply do a fit to the rotation curve data using $R_0$ and $N^*$ as two free parameters. To constrain such fits we follow \cite{Sanders1996} and require that $M/L$ not be less than $0.2M_{\odot}/L_{\odot}$. Additionally, we require that $R_0$ be less than the measured value of $R_{\rm last}$ at the last data point. On imposing these constraints,  we will regard a fit as acceptable, though of course only indicative, if we can find a range of such constrained values of $R_0$ and $N^*$ for which the fitting is reasonable.  Interestingly, this prescription is found to work for all seven of the galaxies, with there being a range of allowed values in each case. In the Tables we present some typical fitted values within the allowed ranges for each of the seven galaxies,  and then use these values to generate the associated Figures. Just as with the spherical bulges, we should note that none of these photometry concerns affect the outer region rotation curve fitting.

To support the $R_0$ values that we obtained this way, we note that for the five ESO galaxies listed above, some limited surface brightness data actually are available. Specifically, in the ESO Lauberts-Valentijn Archive (as accessed at  http://archive.eso.org/wdb/wdb/eso/esolv/form) both a red band total apparent magnitude $m_{\rm T}$ and a red band mean  central surface brightness $\bar{m}_0$ (in magnitudes per square arc second) are listed for the ESO 840411, ESO 1200211, ESO 3020120 and ESO 4880490 galaxies, while a red band mean central surface brightness is listed for ESO 3050090. The quantity $\bar{m}_0$ is not precisely the apparent central surface brightness $m_0$ itself, but rather the average apparent surface brightness in a 10 arc second circular aperture. If we nonetheless now approximate $m_0$ by $\bar{m}_0$, then from $2.5{\rm log}_{10}(2\pi R_0^2)=m_0-m_{\rm T}$ (a quantity that conveniently is not affected by extinction corrections) we can extract an approximate value for $R_0$ in arc seconds. Doing this is found to yield red band scale lengths $R_0({\rm ESO}~840411)=9.4''$, $R_0({\rm ESO}~1200211)=18.9''$, $R_0({\rm ESO}~3020120)=13.5''$, and $R_0({\rm ESO}~4880490) =11.1''$; and thus respective scale lengths of $3.8$, $1.4$, $4.6$, and $1.6$ kpc at the adopted distances listed in the Tables.  For ESO 3050090 only a blue band total apparent magnitude of $13.08$ is listed. Taking the red band total apparent magnitude to be equal  to 13.5, 13.0 and 12.5 (i.e. to be within 0.5 magnitudes of the blue band value, a reasonable enough expectation) respectively yield $R_0({\rm ESO}~3050090)=15.9''$, $R_0({\rm ESO}~3050090)=20.0''$ and $R_0({\rm ESO}~3050090)=25''$, with the $20''$ value corresponding to $R_0({\rm ESO}~3050090)=1.3$ kpc  at the adopted distance listed in the Tables. For all five of the ESO galaxies then, the $R_0$ values are compatible with the allowed ranges of values for $R_0$ that we found from fitting the rotation curves.

\subsection{Double-Counting in the Bulge-Disk Overlap Region}

Since bulges and disks of spiral galaxies overlap in the galactic center region, there could be some double counting. A possible way to allow for this would be to truncate the bulge contribution so that it is only non-zero in the galactic center region, and another possibility would be to truncate the disk contribution so that it is only non-zero outside the galactic center region. We describe the formalism for doing this in the case where there are both Newtonian and linear potentials. However, for the bulge galaxies of interest to us in this paper, we found that neither of the two truncation procedures had that much of an impact on the fits (mainly because only the innermost rotation curve points could be affected by the bulge/disk decomposition in the first place), and only present the formalism here for reference purposes.

For bulges the most straightforward truncation is of the form 
\begin{equation}
\sigma(r,t_0)=\frac{N}{2\pi^2 t^3}K_0\left(\frac{r}{t}\right)\theta(t_0-r),
\label{B9}
\end{equation}
with the volume density being truncated at $r=t_0$. (We truncate $\sigma(r)$ rather than $I(R)$ with a step function since as noted in \cite{Mannheim1995}, truncating $I(R)$ with a step function would generate singularities in the Abel transform.) Given the truncated equation (\ref{B9}) the orbital velocities are then  given by
\begin{equation}
v^2_{\beta}(r<t_0)={2 N\beta^* c^2\over \pi r} \int_0^{r/t}dz\, z^2K_0(z),\qquad
v^2_{\beta}(r>t_0)={2 N\beta^* c^2\over \pi r} \int_0^{t_0/t}dz\, z^2K_0(z)
\label{B10}
\end{equation} 
\begin{eqnarray}
v^2_{\gamma}(r<t_0)&=&{N\gamma^* c^2 r\over \pi} \int_0^{r/t}dz\, z^2K_0(z)
\nonumber\\
&-&{ N\gamma^* c^2 t^2 \over 3 \pi r} \int_0^{r/t}dz\, z^4K_0(z)
+{2 N\gamma^* c^2 r^2 \over 3\pi t^2}\left[rK_1\left(\frac{r}{t}\right)-t_0K_1\left(\frac{t_0}{t}\right)\right],
\nonumber\\
v^2_{\gamma}(r>t_0)&=&{N\gamma^* c^2 r\over \pi} \int_0^{t_0/t}dz\, z^2K_0(z)
-{N\gamma^* c^2 t^2 \over 3 \pi r} \int_0^{t_0/t}dz\, z^4K_0(z).
\label{B11}
\end{eqnarray} 

For disks the most straightforward truncation is to truncate the surface density according to $\Sigma(R)\theta(R-s_0)$ so that it starts at some minimum value $s_0$ near to the galactic center. Following the procedure described in \cite{Mannheim2006} for the arbitrary $\Sigma(R)$, we find that for a truncated $\Sigma(R)$ the Newtonian potential contribution is given by
\begin{eqnarray}
V_{\beta}(R)&=&-2 \pi \beta^* c^2\int_0^{\infty}dk\,\int_{s_0}^{\infty}dR^{\prime}\,R^{\prime}\Sigma(R^{\prime})J_0(kR^{\prime})J_0(kR)
\nonumber\\
&=&-2 \pi \beta^* c^2\int_0^{\infty}dk\,\int_{0}^{\infty}dR^{\prime}\,R^{\prime}\Sigma(R^{\prime})J_0(kR^{\prime})J_0(kR)
\nonumber\\
&&+2 \pi \beta^* c^2\int_0^{\infty}dk\,\int_{0}^{s_0}dR^{\prime}\,R^{\prime}\Sigma(R^{\prime})J_0(kR^{\prime})J_0(kR)
\label{B12}
\end{eqnarray} 
For an exponential disk  with $\Sigma(R)=\Sigma_0e^{-R/R_0}$ (and thus a truncated number of stars $N^*_{\rm TR}=2\pi\Sigma_0(R_0^2+s_0R_0)e^{-s_0/R_0}$ in the $R>s_0$ region), the first of the last two integrals in equation (\ref{B12}) can be done analytically (and leads to the Newtonian term given in equation (\ref{E5})). However, the $dk$ integration range in the second of the last two integrals in equation (\ref{B12}) has to be broken into two separate $R<R^{\prime}$, $R>R^{\prime}$ regions. But when $s_0$ is less than the positions $R$ of the points of interest for the rotation curves (as would typically be the case) we only need the $R>R^{\prime}$ region,  with equation (\ref{B12}) then simplifying to
\begin{eqnarray}
V_{\beta}(R>s_0)&=&-\pi \Sigma_0 \beta^*c^2 R\left[I_0\left(\frac{R}{2R_0}
\right)K_1\left(\frac{R}{2R_0}\right)-
I_1\left(\frac{R}{2R_0}\right)
K_0\left(\frac{R}{2R_0}\right)\right]
\nonumber\\
&+&\frac{4 \Sigma_0\beta^* c^2}{R}\int_0^{s_0}dR^{\prime}\,R^{\prime}e^{-R^{\prime}/R_0}{\bf K}\left(\frac{R^{\prime 2}}{R^2}\right),
\label{B13}
\end{eqnarray} 
where ${\bf K}(x^2)=\int_0^{\pi/2}dy\,(1-x^2{\rm sin}^2y)^{-1/2}$ is the complete elliptic integral of the first kind. Finally, on differentiating equation (\ref{B13}) with respect to $R$ and recalling that
\begin{equation}
\frac{d{\bf K}(x^2)}{dx}=\frac{{\bf E}(x^2)-(1-x^2){\bf K}(x^2)}{x(1-x^2)}
\label{B14}
\end{equation}
where  ${\bf E}(x^2)=\int_0^{\pi/2}dy\,(1-x^2{\rm sin}^2y)^{1/2}$ is the complete elliptic integral of the second kind, we then obtain for the Newtonian contribution to the orbital velocities
\begin{eqnarray}
v^2_{\beta}(R > s_0)&=&\frac{\pi \Sigma_0 \beta^*c^2 R^2}{R_0}\left[I_0\left(\frac{R}{2R_0}
\right)K_0\left(\frac{R}{2R_0}\right)-
I_1\left(\frac{R}{2R_0}\right)
K_1\left(\frac{R}{2R_0}\right)\right]
\nonumber\\
&-&\frac{4 \Sigma_0\beta^* c^2}{R}\int_0^{s_0}dR^{\prime}\,R^{\prime }e^{-R^{\prime}/R_0}\left[\frac{{\bf E}(x^2)}{(1-x^2)}\right]_{x=R^{\prime }/R}.
\label{B15}
\end{eqnarray} 

Similarly, again following \cite{Mannheim2006}, for the linear potential contribution we obtain the general
\begin{eqnarray}
V_{\gamma}(R)&=& \pi \gamma^* c^2\int_0^{\infty}dk\,\int_{s_0}^{\infty}dR^{\prime}\,R^{\prime}\Sigma(R^{\prime})\left[(R^2+R^{\prime 2})J_0(kR^{\prime})J_0(kR)
-2RR^{\prime}J_1(kR^{\prime})J_1(kR)\right]
\nonumber\\
&=&\pi \gamma^* c^2\int_0^{\infty}dk\,\int_{0}^{\infty}dR^{\prime}\,R^{\prime}\Sigma(R^{\prime})\left[(R^2+R^{\prime 2})J_0(kR^{\prime})J_0(kR)
-2RR^{\prime}J_1(kR^{\prime})J_1(kR)\right]
\nonumber\\
&-&\pi \gamma^* c^2\int_0^{\infty}dk\,\int_{0}^{s_0}dR^{\prime}\,R^{\prime}\Sigma(R^{\prime})\left[(R^2+R^{\prime 2})J_0(kR^{\prime})J_0(kR)
-2RR^{\prime}J_1(kR^{\prime})J_1(kR)\right].
\nonumber\\
\label{B16}
\end{eqnarray} 
Thus for a truncated exponential disk we obtain
\begin{eqnarray}
V_{\gamma}(R>s_0)&=& \pi \Sigma_0 \gamma^*c^2 RR_0^2\left[I_0\left(\frac{R}{2R_0}
\right)K_1\left(\frac{R}{2R_0}\right)-
I_1\left(\frac{R}{2R_0}\right)
K_0\left(\frac{R}{2R_0}\right)\right]
\nonumber\\
&+& \frac{\pi \Sigma_0 \gamma^*c^2 R^2R_0}{2}\left[I_0\left(\frac{R}{2R_0}
\right)K_0\left(\frac{R}{2R_0}\right)+
I_1\left(\frac{R}{2R_0}\right)
K_1\left(\frac{R}{2R_0}\right)\right]
\nonumber\\
&-&2 \Sigma_0\gamma^* c^2\int_{0}^{s_0}dR^{\prime}\,R^{\prime}e^{-R^{\prime}/R_0}\left[\frac{(R^{\prime 2}-R^2)}{R}{\bf K}\left(\frac{R^{\prime 2}}{R^2}\right)
+2R{\bf E}\left(\frac{R^{\prime 2}}{R^2}\right)\right]
\label{B17}
\end{eqnarray} 
for points with $R>s_0$. Finally, on differentiating equation (\ref{B17}) with respect to $R$ and recalling that
\begin{equation}
\frac{d{\bf E}(x^2)}{dx}=\frac{{\bf E}(x^2)-{\bf K}(x^2)}{x},
\label{B18}
\end{equation}
we then obtain for the linear potential contribution to the orbital velocities
\begin{eqnarray}
v^2_{\gamma}(R>s_0)&=& \pi \Sigma_0 \gamma^*c^2 R^2R_0I_1\left(\frac{R}{2R_0}
\right)K_1\left(\frac{R}{2R_0}\right)
\nonumber\\
&-&2 \Sigma_0\gamma^* c^2R\int_{0}^{s_0}dR^{\prime}\,R^{\prime}e^{-R^{\prime}/R_0}
\left[{\bf E}(x^2)\right]_{x=R^{\prime }/R}.
\label{B19}
\end{eqnarray} 

For most cases of interest the disk scale length $R_0$ will typically be much larger than the truncation point $s_0$, with the $e^{-R^{\prime}/R_0}$ term thus being very close to one in the integration ranges needed for both equation (\ref{B15}) and equation (\ref{B19}). Then, when we do approximate the $e^{-R^{\prime}/R_0}$ term to one, the integrals in equation (\ref{B15}) and equation (\ref{B19}) can be done analytically, and yield the simple and convenient expressions
\begin{eqnarray}
v^2_{\beta}(R > s_0)&=&\frac{\pi \Sigma_0 \beta^*c^2 R^2}{R_0}\left[I_0\left(\frac{R}{2R_0}
\right)K_0\left(\frac{R}{2R_0}\right)-
I_1\left(\frac{R}{2R_0}\right)
K_1\left(\frac{R}{2R_0}\right)\right]
\nonumber\\
&-&4 \Sigma_0\beta^* c^2R\left[{\bf K}(x^2)-{\bf E}(x^2)\right]_{x=s_0/R},
\label{B20}
\end{eqnarray} 
\begin{eqnarray}
v^2_{\gamma}(R>s_0)&=& \pi \Sigma_0 \gamma^*c^2 R^2R_0I_1\left(\frac{R}{2R_0}
\right)K_1\left(\frac{R}{2R_0}\right)
\nonumber\\
&-&\frac{2 \Sigma_0\gamma^* c^2R^3}{3}
\left[(1+x^2){\bf E}(x^2)-(1-x^2){\bf K}(x^2)\right]_{x=s_0/R}.
\label{B21}
\end{eqnarray} 

\subsection{The Effect of Mass Generation on Galactic Orbits}

In a strictly conformal invariant theory such as conformal gravity there can be no mass scales at the level of the Lagrangian, with all such scales needing to be generated dynamically. Similarly, in the $SU(3)\times SU(2)\times U(1)$ standard model of elementary particle physics there are no fundamental mass scales in the gauge boson and fermion sectors of the theory, with mass scales being generated by scalar fields that could either be fundamental q-number fields or c-number matrix elements of fermion bilinear condensates. Since the double-well scalar field potential with its tachyonic mass would violate the conformal symmetry, in the conformal case the symmetry breaking must be done by fermion condensates, but in both the conformal and the  $SU(3)\times SU(2)\times U(1)$ cases there must be some form of scale-breaking scalars. (As noted in \cite{Mannheim2006},  it is because of the contribution of the energy and momentum of the scale breaking fields that the traceless matter field energy-momentum tensor required of a conformal theory is able to support a non-zero dynamically induced matter field mass term.) Moreover, when one couples to grand-unified theories or to cosmology, there must be additional scalar fields that set the scale for the early Universe. These latter fields are not involved in particle mass generation (they are already operative at temperatures above the early Universe $SU(2)\times U(1)$  phase transition) and can be modeled by expectation values of fermion quadrilinears of the form $\sum_i\langle \bar{\psi}_i\psi_i\bar{\psi}_i\psi_i\rangle$, as summed over all the fermions in the Universe. Masses for individual fermions would be generated by bilinear condensates of the form $\langle \bar{\psi}_i\psi_i\rangle$, while masses of systems of $N$ fermions would be generated by $\sum_i\langle \bar{\psi}_i\psi_i\rangle$ as summed over the $N$ fermions (i.e. the mass of an $N$-particle system is of order $N$ times that of each individual particle in it). On stepping up from microscopic fermions to macroscopic systems such as stars and galaxies, we see that if we associate an effective scalar field with each solar mass unit of material, then the scalar field associated with an entire galaxy of mass $N^*M_{\odot}$ would be $N^*$ times larger. Since it is this latter scalar field that sets the scale for the galactic geometry (by contributing to the source integrals given in equation (\ref{E3}) as it generates galactic masses in the first place and enables particles to localize), it will completely overwhelm the contribution that a one solar mass generating scalar field could make to the galactic orbit of a one solar mass object. (As noted in \cite{Mannheim2007}, the contribution of a radial scalar field $S(r)$ to the source function $f(r)$ in equation (\ref{E3}) is given by $f(r)=(SS^{\prime\prime}-2S^{\prime 2})/4\alpha_g$, with it thus being the departure of the scalar field from a constant value that generates the potential produced by a localized source, with the $f(r)$ of an $N$-particle system being $N$ times larger than the $f(r)$ of a one-particle system.)

To explore the explicit effect that a scalar field has on a particle orbit we follow the discussion given in \cite{Mannheim2007} and introduce the test particle action
\begin{equation}                                                                              
I_T=-h\int d\tau S(x)
\label{108a}
\end{equation}                                                  
that describes the coupling of a particle to its own scalar field as it propagates in a background geometry. The utility of this action is that it not only reduces to the conventionally used massive test particle action when $S(x)$ is constant, for varying $S(x)$ it is actually fully conformal invariant, since the $e^{-\alpha(x)}$ change in the scalar field is compensated by an accompanying $e^{\alpha(x)}$ change in the proper time. Variation of the action of equation (\ref{108a}) with respect to the coordinates of the test particle thus yields trajectories that are conformal invariant, with their specific form being given as \cite{Mannheim2006}
\begin{equation}                                                                              
hS\left(
\frac{d^2x^{\lambda} }{ d\tau^2} +\Gamma^{\lambda}_{\mu \nu} 
\frac{dx^{\mu}}{d\tau}\frac{dx^{\nu } }{ d\tau} \right) 
= -hS_{;\beta} \left( g^{\lambda
\beta}+
\frac{dx^{\lambda}}{d\tau}                                                      
\frac{dx^{\beta}}{d\tau}\right),
\label{109}
\end{equation}                                                  
where the proper time $d\tau=(-g_{\mu\nu}dx^{\mu}dx^{\nu})^{1/2} $ is positive in test particle orbits.

For the general static, spherically symmetric metric of the form $d\tau^2=B(r)c^2dt^2-A(r)dr^2-r^2d\Omega_2$, and for a scalar field that only depends on the radius $r$, the four equations of motion contained in equation (\ref{109}) take the form
\begin{eqnarray}
c\frac{d^2t}{d\tau^2}+\frac{cB^{\prime}}{B}\frac{dt}{d\tau}\frac{dr}{d\tau}
&=&-\frac{cS^{\prime}}{S}\frac{dt}{d\tau}\frac{dr}{d\tau},
\nonumber \\
\frac{d^2r}{d\tau^2}+\frac{A^{\prime}}{2A}\left(\frac{dr}{d\tau}\right)^2
-\frac{r}{A}\left(\frac{d\theta}{d\tau}\right)^2
- \frac{r{\rm sin}^2\theta}{A}\left(\frac{d\phi}{d\tau}\right)^2
+\frac{c^2B^{\prime}}{2A}\left(\frac{dt}{d\tau}\right)^2
&=&-\frac{S^{\prime}}{AS} 
-\frac{S^{\prime}}{S}\left(\frac{dr}{d\tau}\right)^2,
\nonumber \\
\frac{d^2\theta}{d\tau^2}
+\frac{2}{r}\frac{d\theta}{d\tau}\frac{dr}{d\tau}
-{\rm sin}\theta{\rm cos}\theta\left(\frac{d\phi}{d\tau}\right)^2
&=&-\frac{S^{\prime}}{S}\frac{d\theta}{d\tau}\frac{dr}{d\tau},
\nonumber \\
\frac{d^2\phi}{d\tau^2}
+\frac{2}{r}\frac{d\phi}{d\tau}\frac{dr}{d\tau}
+2\frac{{\rm cos}\theta}{{\rm sin}\theta}\frac{d\phi}{d\tau}\frac{d\theta}{d\tau}
&=&-\frac{S^{\prime}}{S}\frac{d\phi}{d\tau}\frac{dr}{d\tau},
\label{110}
\end{eqnarray}                                 
with the prime denoting differentiation with respect to $r$. Equatorial plane solutions can be found in which $\theta$ is fixed to $\theta=\pi/2$, with the equations of motion for the three other coordinates given in equation (\ref{110}) then being found to admit of exact first integrals. Since
\begin{equation}
\frac{d}{d\tau}\left[S^2+AS^2\left(\frac{dr}{d\tau}\right)^2\right]
=2AS^2\frac{dr}{d\tau}\left[\frac{d^2r}{d\tau^2}+\frac{A^{\prime}}{2A}\left(\frac{dr}{d\tau}
\right)^2+\frac{S^{\prime}}{AS} 
+\frac{S^{\prime}}{S}\left(\frac{dr}{d\tau}\right)^2\right],
\label{110a}
\end{equation}                                 
the structure of these first integrals depends on whether or not $dr/d\tau$ is zero. In trajectories with non-zero $dr/d\tau$ in which $d\phi/d\tau$ is non-zero too the integrals are of the form
\begin{eqnarray}
cBS\frac{dt}{d\tau}&=&C,
\label{111a} \\
S^2+AS^2\left(\frac{dr}{d\tau}\right)^2
+\frac{K^2}{r^2}-\frac{C^2}{B}&=&D,
\label{111b} \\
r^2S\frac{d\phi}{d\tau}&=&K,
\label{111c}
\end{eqnarray}                                 
where $C$, $D$ and $K$ are integration constants.  In circular orbits in which the radius $r$ is fixed the integrals are of the form 
\begin{equation}
c\frac{dt}{d\tau}=E,~~~~
\frac{d\phi}{d\tau}=F,~~~ -rF^2
+\frac{B^{\prime}E^2}{2}
+\frac{S^{\prime}}{S}=0,
\label{111e}
\end{equation}                                 
where $E$ and $F$ are integration constants.  In radial trajectories in which the angle $\phi$ is fixed the integrals are of the form 
\begin{eqnarray}
cBS\frac{dt}{d\tau}&=&G,
\label{111e2} \\
S^2+AS^2\left(\frac{dr}{d\tau}\right)^2
-\frac{G^2}{B}&=&H,
\label{111e3} 
\end{eqnarray}                                 
where $G$ and $H$ are integration constants

On eliminating the dependence on $d\tau$ from the above sets of equations, for trajectories with both $dr/d\tau$ and $d\phi/d\tau$ non-zero we obtain 
\begin{eqnarray}
S^2+\frac{AC^2}{c^2B^2}\left(\frac{dr}{dt}\right)^2
+\frac{K^2}{r^2}-\frac{C^2}{B}&=&D,
\label{112a} \\
S^2+\frac{AK^2}{r^4}\left(\frac{dr}{d\phi}\right)^2
+\frac{K^2}{r^2}-\frac{C^2}{B}&=&D,
\label{112b} \\
\frac{r^2}{cB}\frac{d\phi}{dt}&=&\frac{K}{C},
\label{112c}
\end{eqnarray}                                 
with equation (\ref{112b}) correcting a typographical error in \cite{Mannheim2007}. Similarly, in circular orbits we obtain
\begin{equation}
\frac{1}{c}\frac{d\phi}{dt}=\frac{F}{E},
\label{112d}
\end{equation}
and in radial trajectories we obtain 
\begin{equation}
S^2+\frac{AG^2}{c^2B^2}\left(\frac{dr}{dt}\right)^2
-\frac{G^2}{B}=H.
\label{112d1} 
\end{equation}                                 

In addition, we note that substituting equations (\ref{111a}), (\ref{111b}), (\ref{111c}), (\ref{111e}), (\ref{111e2}) and (\ref{111e3}) into $-g_{\mu\nu}(dx^{\mu}/d\tau)(dx^{\nu}/d\tau)=1$ yields 
\begin{equation}
D=0
\label{111f}
\end{equation}
in trajectories with both $dr/d\tau$ and $d\phi/d\tau$ non-zero,  and yields 
\begin{equation}
BE^2-r^2F^2=1
\label{111g}
\end{equation}
in circular orbits. In radial trajectories we obtain
\begin{equation}
H=0,
\label{111g2}
\end{equation}
with the radial velocity then being given by 
\begin{equation}
\frac{dr}{dt}=\left[\frac{c^2B}{A}\left(1-\frac{BS^2}{G^2}\right)\right]^{1/2}.
\label{111g3}
\end{equation}

For the circular orbits, on using equations (\ref{111e}) and (\ref{111g}) to express $E$ and $F$ in terms of the scalar field we obtain
\begin{equation}
\frac{r}{c}\frac{d\phi}{dt}=\left(\frac{rB^{\prime}}{2}+\frac{rS^{\prime}(2B-rB^{\prime})}{2(S+rS^{\prime})}\right)^{1/2},
\label{114}
\end{equation}
an expression for circular orbits that is exact and without approximation. Finally, with the one-particle $S^{\prime}$ term being negligible in the presence of the $B^{\prime}$ that is generated by an $N$-particle background geometry, for a one-particle system in an $N$-particle background equation (\ref{114}) reduces to the standard circular orbit geodesic equation
\begin{equation}
r\frac{d\phi}{dt}=\left(\frac{rc^2B^{\prime}}{2}\right)^{1/2}.
\label{115}
\end{equation}
With the dependence on the metric coefficient $A(r)$ dropping out of equation (\ref{110}) in orbits in which $dr/d\tau=0$, and with $A(r)$ not appearing at all in equation (\ref{111e}),  equations (\ref{114}) and (\ref{115}) will hold in geometries in which $A(r)$ has any form whatsoever. In particular, equations (\ref{114}) and (\ref{115}) will hold in the geometry explicitly considered in this paper in which $A(r)=1/B(r)$. We thus justify the use of equation (\ref{115}) in the fits presented in this paper.

As regards the circular orbits, we additionally note that since equations (\ref{114}) and (\ref{115}) are generic equations that hold for any radial scalar field and any choice of $B(r)$ and $A(r)$, these equations will hold for circular orbits not just in conformal gravity but in standard Einstein gravity as well. Thus even in standard gravity in the presence of $SU(3)\times SU(2)\times U(1)$ scalar fields, to order $1/N$ the circular orbits of one-particle systems in $N$-particle background geometries can be well-approximated by equation (\ref{115}). The only situation in which one could not immediately drop the $S^{\prime}$-dependent term in (\ref{114}) would be in a few-particle system, and it would thus be of interest to see if binary pulsar systematics might show some sensitivity to the scalar fields that are responsible for the masses that the two stars in the binary possess.

To recover the standard non-relativistic limits for the above orbits and trajectories, to first order in weak gravity we set $S^2/C^2=1-2U/c^2$, $J=cK/C$, and $B(r)=1+2V(r)/c^2$ where $U$, $J$ and $V(r)$ are the first-order energy, angular momentum and potential of the test particle. On requiring only that $A(r)=1+O(V(r))$ (in both the conformal gravity solution and in the standard Schwarzschild solution we have $A(r)=1/B(r)\sim 1-2V(r)/c^2$), then from equations (\ref{112a}) and (\ref{112c}) as evaluated with $D=0$ we obtain
\begin{equation}
\frac{1}{2}\left(\frac{dr}{dt}\right)^2+\frac{J^2}{2r^2} +V(r)=U,~~~~r^2\frac{d\phi}{dt}=J,
\label{115e}
\end{equation}
to lowest order for a unit mass particle in a non-circular trajectory.  On differentiating equation (\ref{115e}) we find that in a non-circular trajectory the acceleration is given by
\begin{equation}
\frac{d^2r}{dt^2}-r\left(\frac{d\phi}{dt}\right)^2=-\frac{dV}{dr}.
\label{115f}
\end{equation}

As regards weak gravity circular orbits, we note that since the source function $f(r)$ in equation (\ref{E3}) is given by $f(r)=(SS^{\prime\prime}-2S^{\prime 2})/4\alpha_g$, the departure of $B(r)$ from its flat spacetime value of one is correlated with the departure of $S(r)$ from a constant value. With the leading term in $S(r)$ being the mass of the test particle, for weak gravity the $rS^{\prime}/S$ term in equation (\ref{114}) is negligible, with equation (\ref{114}) then reducing to 
\begin{equation}
r^2\left(\frac{d\phi}{dt}\right)^2=r\frac{dV}{dr}
\label{115g}
\end{equation}
for  circular orbits in weak gravity (to thus also satisfy equation (\ref{115f}) since $d^2r/dt^2=0$).

For the generic potential $V(r)=-\beta c^2/r+\gamma c^2 r/2-\kappa c^2 r^2/2$ of interest to us in this paper, we see that for circular orbits we require 
\begin{equation}
\frac{\beta c^2}{r}+\frac{\gamma c^2r}{2}-\kappa c^2r^2 >0,
\label{115h}
\end{equation}
while for non-circular trajectories we require
\begin{equation}
U +\frac{\beta c^2}{r}-\frac{\gamma c^2r}{2}+\frac{\kappa c^2r^2}{2} >0.
\label{115i}
\end{equation}
Thus when the $\kappa c^2r^2$ term is small the conformal theory supports circular orbits, and when the $\kappa c^2r^2$ term is large the theory supports non-circular trajectories, with the non-circular trajectories not specifically needing to be purely radial.

\bigskip

{\bf Added note.} Since completing this paper we have extended our study of rotation curve fitting based on equation (\ref{E20}) to include 27 additional galaxies, the bulk of which are dwarf spirals. As reported in \cite{Obrien2012},  on using the  same values for $\gamma^*$, $\gamma_0$ and $\kappa$ that are used here, and again with no need for any dark matter whatsoever, we find acceptable fits of the same quality as the ones we present here. This latest study brings to 138 the number of rotation curves of galaxies that have successfully been fitted by the conformal gravity theory.

\vfill\eject

{}

\vfill\eject

\hoffset=-0.5in

\begin{table}
\caption{Properties of the THINGS 18 Galaxy Sample}
\centering
\begin{tabular}{l c c c c c c c c c c} 
\hline\hline
 \phantom{00}Galaxy  & \phantom{0}Type \phantom{0}&Distance  & $L_{\rm  B}$ & $R_0$  & $R_{\rm last} $ &  $M_{\rm HI} $ & $M_{\rm disk}$ &  $ 
(M/L) _{\rm stars}$ & $(v^2 / c^2 R)_{\rm last}$ & Data~Sources \\  
& &  (Mpc)  &  $(10^{10}{\rm L}_{\odot})$&(kpc) & (kpc) & {$(10^{10} M_\odot)$} & {$(10^{10}
M_\odot)$} & ({$M_{\odot}/L_{\odot}$}) & {$(10^{-30}~\texttt{cm}^{-1})$} & $v~~~~L~~~R_0~~{\rm HI}$\\
\hline
DDO 0154 &LSB& \phantom{0}4.2 &   0.007   & 0.8 & \phantom{0}8.1 & 0.03  &
\phantom{0}0.003 & 0.45 & \phantom{0}1.12&\cite{deBlok2008} \cite{Walter2008}  \cite{Leroy2008}  \cite{Walter2008}  
\\
\phantom{NI}IC 2574 &LSB & \phantom{0}4.5  &  0.345    &4.2 & 13.1 &0.19 &
\phantom{0}0.098 & 0.28 &\phantom{0}1.69&\cite{deBlok2008} \cite{Walter2008} \cite{Pasquali2008} \cite{Walter2008}   
\\
NGC 0925 &LSB& \phantom{0}8.7 &   1.444   & 3.9 & 12.4 & 0.41 & \phantom{0}1.372 & 0.95
&\phantom{0}4.17&\cite{deBlok2008} \cite{Walter2008} \cite{Leroy2008} \cite{Walter2008}    
 \\
NGC 2403 & HSB &\phantom{0}4.3 &  1.647    & 2.7 & 23.9 & 0.46 & \phantom{0}2.370 &1.44 & \phantom{0}2.89&\cite{deBlok2008} \cite{Walter2008} \cite{Wevers1986} \cite{Walter2008}
\\
NGC 2841 &HSB& 14.1 &  4.742    & 3.5 & 51.6 & 0.86 & 19.552 & 4.12& \phantom{0}5.83&\cite{deBlok2008} \cite{Walter2008} \cite{Begeman1987} \cite{Walter2008}  
\\
NGC 2903 &HSB& \phantom{0}9.4 &  4.088    &  3.0 & 30.9 & 0.49 & \phantom{0}7.155 &
1.75 &\phantom{0}3.75&\cite{deBlok2008} \cite{Walter2008} \cite{Wevers1986} \cite{Walter2008}
\\
NGC 2976 &LSB& \phantom{0}3.6 &  0.201   & 1.2 & \phantom{0}2.6 & 0.01 &
\phantom{0}0.322 &1.60&10.43&\cite{deBlok2008} \cite{Walter2008} \cite{Simon2003} \cite{Walter2008}     
\\
NGC 3031 &HSB& \phantom{0}3.7 &   3.187  & 2.6 & 15.0 & 0.38 & \phantom{0}8.662 & 2.72
&\phantom{0}9.31&\cite{deBlok2008} \cite{Walter2008} \cite{Murphy2008} \cite{Walter2008}
\\
NGC 3198 &HSB& 14.1 &   3.241   & 4.0 & 38.6 & 1.06 & \phantom{0}3.644 &1.12& \phantom{0}2.09&\cite{deBlok2008} \cite{Walter2008} \cite{Wevers1986} \cite{Walter2008}  
\\
NGC 3521 &HSB& 12.2&   4.769  & 3.3 & 35.3 & 1.03 & \phantom{0}9.245 & 1.94
&\phantom{0}4.21&\cite{deBlok2008} \cite{Walter2008} \cite{Leroy2008} \cite{Walter2008}   
\\
NGC 3621 &HSB& \phantom{0}7.4 &   2.048   & 2.9 & 28.7 & 0.89 & \phantom{0}2.891 & 1.41
&\phantom{0}3.18&\cite{deBlok2008} \cite{Walter2008} \cite{deBlok2008} \cite{Walter2008}
\\
NGC 3627 &HSB&10.2 &  3.700    & 3.1 &\phantom{0}8.2 & 0.10 & \phantom{0}6.622 & 1.79 &
15.64&\cite{deBlok2008} \cite{Walter2008} \cite{Leroy2008} \cite{Walter2008}    
\\
NGC 4736 &HSB&\phantom{0}5.0 &    1.460  & 2.1 & 10.3 & 0.05 & \phantom{0}1.630 & 1.60
&\phantom{0}4.66&\cite{deBlok2008} \cite{Walter2008} \cite{deBlok2008} \cite{Walter2008}   
\\
NGC 4826 &HSB& \phantom{0}5.4 &   1.441   & 2.6 & 15.8 & 0.03 & \phantom{0}3.640 & 2.53
&\phantom{1}5.46&\cite{deBlok2008} \cite{Walter2008} \cite{Regan2006} \cite{Walter2008}
\\
NGC 5055 &HSB& \phantom{0}9.2&  3.622    &2.9 & 44.4 &0.76& \phantom{0}6.035 & 1.87& 
\phantom{0}2.36&\cite{deBlok2008} \cite{Walter2008} \cite{Leroy2008} \cite{Walter2008}   
\\
NGC 6946 &HSB& \phantom{0}6.9 & 3.732     & 2.9 & 22.4 & 0.57 & \phantom{0}6.272 &
1.68&\phantom{0}6.39 &\cite{deBlok2008} \cite{Walter2008} \cite{Leroy2008} \cite{Walter2008}   
\\
NGC 7331 &HSB& 14.2 &  6.773    & 3.2 & 24.4 & 0.85 & 12.086 & 1.78&\phantom{0}9.61&\cite{deBlok2008} \cite{Walter2008} \cite{Leroy2008} \cite{Walter2008}   
\\
NGC 7793 &HSB& \phantom{0}5.2 &   0.910   & 1.7 & 10.3 & 0.16 &\phantom{0}0.793 & 0.87
&\phantom{1}3.61 &\cite{deBlok2008} \cite{Walter2008} \cite{Leroy2008} \cite{Walter2008} 
\\
\hline
\end{tabular}
\label{table:things}
\end{table}

\voffset=-0.5in

\begin{table}
\caption{Properties of the Ursa Major 30 Galaxy Sample}
\centering
\begin{tabular}{l c c c c c c c c c c} 
\hline\hline
\phantom{00}Galaxy  & \phantom{0}Type \phantom{0}&Distance  & $L_{\rm  B}$ & $R_0$  & $R_{\rm last} $ &  $M_{\rm HI} $ & $M_{\rm disk}$ &  $ 
(M/L) _{\rm stars}$ & $(v^2 / c^2 R)_{\rm last}$ & Data~Sources \\  
& &  (Mpc)  &  $(10^{10}{\rm L}_{\odot})$&(kpc) & (kpc) & {$(10^{10} M_\odot)$} & {$(10^{10}
M_\odot)$} & ({$M_{\odot}/L_{\odot}$}) & {$(10^{-30}~\texttt{cm}^{-1})$} & $v~~~~L~~~R_0~~{\rm HI}$\\
\hline
NGC 3726 &HSB& 17.4 & 3.340    & 3.2 & 31.5 & 0.60 & \phantom{0}3.82 & 1.15 &
\phantom{0}3.19 &   \cite{Verheijen2001} \cite{Sanders1998} \cite{Tully1996} \cite{Sanders1998}  \\
NGC 3769 &HSB&15.5& 0.684   &1.5 & 32.2 &0.41 &\phantom{0}1.36 &1.99 &\phantom{0}1.43
&   \cite{Verheijen2001} \cite{Sanders1998} \cite{Tully1996} \cite{Sanders1998}    
\\
NGC 3877&HSB& 15.5 &   1.948   & 2.4 & \phantom{0}9.8 & 0.11& \phantom{0}3.44 & 1.76 &
10.51 &     \cite{Verheijen2001} \cite{Sanders1998} \cite{Tully1996} \cite{Sanders1998}  
\\
NGC 3893&HSB & 18.1 &   2.928   & 2.4 & 20.5 & 0.59& \phantom{0}5.00 & 1.71 &
\phantom{0}3.85 &   \cite{Verheijen2001} \cite{Sanders1998} \cite{Tully1996} \cite{Sanders1998}    \\
NGC 3917&LSB& 16.9 &  1.334   &2.8 & 13.9 & 0.17 & \phantom{0}2.23 &1.67 &
\phantom{0}4.85 &   \cite{Verheijen2001} \cite{Sanders1998} \cite{Tully1996} \cite{Sanders1998}    \\
NGC 3949&HSB& 18.4 &  2.327    & 1.7 & \phantom{0}7.2 & 0.35 & \phantom{0}2.37 &1.02
&14.23&   \cite{Verheijen2001} \cite{Sanders1998} \cite{Tully1996} \cite{Sanders1998}   
 \\
NGC 3953 &HSB& 18.7 &  4.236    & 3.9 & 16.3 & 0.31 &\phantom{0}9.79 &2.31& 10.20 &   \cite{Verheijen2001} \cite{Sanders1998} \cite{Tully1996} \cite{Sanders1998}    
\\
NGC 3972&HSB& 18.6 &    0.978  & 2.0 &\phantom{0}9.0 & 0.13 & \phantom{0}1.49&1.53 &
\phantom{0}7.18 &    \cite{Verheijen2001} \cite{Sanders1998} \cite{Tully1996} \cite{Sanders1998}   
\\
NGC 3992 &HSB& 25.6 &   8.456   & 5.7 & 49.6 & 1.94 & 13.94 &  1.65 & \phantom{0}4.08 &   \cite{Verheijen2001} \cite{Sanders1998} \cite{Tully1996} \cite{Sanders1998}    
\\
NGC 4010 &LSB& 18.4 &  0.883    & 3.4 &  10.6 & 0.29 &\phantom{0}2.03 & 2.30 &
\phantom{0}5.03 &     \cite{Verheijen2001} \cite{Sanders1998} \cite{Tully1996} \cite{Sanders1998} 
 \\
NGC 4013&HSB & 18.6 &  2.088    & 2.1 & 33.1 & 0.32 & \phantom{0}5.58 & 2.67
&\phantom{0}3.14 &   \cite{Verheijen2001} \cite{Sanders1998} \cite{Tully1996} \cite{Sanders1998}    
\\
NGC 4051 &HSB& 14.6 &   2.281   & 2.3 &  \phantom{0}9.9& 0.18 & \phantom{0}3.17 & 1.39 &
\phantom{0}8.52 &     \cite{Verheijen2001} \cite{Sanders1998} \cite{Tully1997} \cite{Sanders1998} 
 \\
NGC 4085&HSB& 19.0 &   1.212   &1.6 & \phantom{0}6.5 & 0.15 & \phantom{0}1.34 &1.11 &
10.21 &    \cite{Verheijen2001} \cite{Sanders1998} \cite{Tully1996} \cite{Sanders1998}   
\\
NGC 4088&HSB & 15.8 &   2.957  & 2.8 & 18.8 & 0.64 &\phantom{0}4.67 & 1.58 &
\phantom{0}5.79&   \cite{Verheijen2001} \cite{Sanders1998} \cite{Tully1996} \cite{Sanders1998}    \\
NGC 4100&HSB & 21.4 &  3.388   & 2.9 & 27.1 & 0.44 & \phantom{0}5.74 & 1.69 &
\phantom{0}3.35 &   \cite{Verheijen2001} \cite{Sanders1998} \cite{Tully1996} \cite{Sanders1998}    \\
NGC 4138 &LSB&15.6 &   0.827   & 1.2 & 16.1 & 0.11 & \phantom{0}2.97 & 3.59 &
\phantom{0}5.04 &   \cite{Verheijen2001} \cite{Sanders1998} \cite{Tully1996} \cite{Sanders1998}    \\
NGC 4157&HSB & 18.7 &   2.901   & 2.6 & 30.9 & 0.88 & \phantom{0}5.83 & 2.01 &
\phantom{0}3.99 &   \cite{Verheijen2001} \cite{Sanders1998} \cite{Tully1996} \cite{Sanders1998}    \\
NGC 4183 &HSB& 16.7 &    1.042 & 2.9 & 19.5 & 0.30 &\phantom{0}1.43 & 1.38 &
\phantom{0}2.36 &    \cite{Verheijen2001} \cite{Sanders1998} \cite{Tully1996} \cite{Sanders1998}   \\
NGC 4217&HSB & 19.6 &   3.031   & 3.1 & 18.2 & 0.30 & \phantom{0}5.53 &1.83 &
\phantom{0}6.28 &   \cite{Verheijen2001} \cite{Sanders1998} \cite{Tully1996} \cite{Sanders1998}    \\
NGC 4389&HSB& 15.5 &   0.610   & 1.2 & \phantom{0}4.6 &0.04 & \phantom{0}0.42 & 0.68 &
\phantom{0}9.49 &    \cite{Verheijen2001} \cite{Sanders1998} \cite{Tully1996} \cite{Sanders1998}   
\\
UGC 6399&LSB& 18.7 &   0.291  & 2.4 & \phantom{0}8.1 &0.07 & \phantom{0}0.59 &2.04 &
\phantom{0}3.42&    \cite{Verheijen2001} \cite{Sanders1998} \cite{Tully1996} \cite{Sanders1998}   \\
UGC 6446&LSB &15.9 &  0.263    & 1.9 &13.6 &0.24 &\phantom{0}0.36 & 1.36 &
\phantom{0}1.70 &    \cite{Verheijen2001} \cite{Sanders1998} \cite{Tully1997} \cite{Sanders1998}  
\\
UGC 6667&LSB& 19.8 &   0.422   &3.1& \phantom{0}8.6 &0.10 & \phantom{0}0.71 & 1.67&
\phantom{0}3.09 &   \cite{Verheijen2001} \cite{Sanders1998} \cite{Tully1996} \cite{Sanders1998}    \\
UGC 6818&LSB& 21.7 &   0.352   &2.1 & \phantom{0}8.4 & 0.16 & \phantom{0}0.11 & 0.33 &
\phantom{0}2.35 &   \cite{Verheijen2001} \cite{Sanders1998} \cite{Tully1996} \cite{Sanders1998}    \\
UGC 6917&LSB& 18.9 &  0.563    & 2.9 & 10.9 &0.22 &\phantom{0}1.24 &2.20
&\phantom{0}4.05 &    \cite{Verheijen2001} \cite{Sanders1998} \cite{Tully1996} \cite{Sanders1998}  
 \\
UGC 6923&LSB& 18.0 &  0.297    & 1.5 & \phantom{0}5.3 &0.08 & \phantom{0}0.35 & 1.18 &
\phantom{0}4.43 &    \cite{Verheijen2001} \cite{Sanders1998} \cite{Tully1997} \cite{Sanders1998}  
 \\
UGC 6930 & LSB & 17.0 &   0.601  & 2.2 & 15.7 & 0.29 & \phantom{0}1.02 &
1.69 & \phantom{0}2.68 &     \cite{Verheijen2001} \cite{Sanders1998} \cite{Tully1996} \cite{Sanders1998}  
\\
UGC 6973&HSB & 25.3 &  1.647    & 2.2 & 11.0 & 0.35 & \phantom{0}3.99 & 2.42
& 10.58&     \cite{Verheijen2001} \cite{Sanders1998} \cite{Tully1997} \cite{Sanders1998}  
\\
UGC 6983&LSB & 20.2&   0.577   & 2.9 &17.6 & 0.37 & \phantom{0}1.28 & 2.22 &
\phantom{0}2.43 &   \cite{Verheijen2001} \cite{Sanders1998} \cite{Tully1996} \cite{Sanders1998}    \\
UGC 7089&LSB& 13.9 &   0.352   & 2.3 & \phantom{0}7.1 &0.07 & \phantom{0}0.35 & 0.98 &
\phantom{0}3.18 &   \cite{Verheijen2001} \cite{Sanders1998} \cite{Tully1996} \cite{Sanders1998}    \\
\hline
\end{tabular}
\label{table:uma}
\end{table}

\begin{table}
\caption{Properties of the LSB 20 Galaxy Sample}
\centering
\begin{tabular}{l c c c c c c c c c c} 
\hline\hline
\phantom{00}Galaxy  & \phantom{0}Type \phantom{0}&Distance  & $L_{\rm  B}$ & $R_0$  & $R_{\rm last} $ &  $M_{\rm HI} $ & $M_{\rm disk}$ &  $ 
(M/L) _{\rm stars}$ & $(v^2 / c^2 R)_{\rm last}$ & Data~Sources \\  
& &  (Mpc)  &  $(10^{10}{\rm L}_{\odot})$&(kpc) & (kpc) & {$(10^{10} M_\odot)$} & {$(10^{10}
M_\odot)$} & ({$M_{\odot}/L_{\odot}$}) & {$(10^{-30}~\texttt{cm}^{-1})$} & $v~~~~L~~~R_0~~{\rm HI}$\\
\hline
DDO 0064 &LSB &\phantom{0}6.8 &  0.015    & 1.3 & \phantom{0}2.1 & 0.02 & 0.04 & \phantom{0}2.87 &\phantom{0}6.05 &    \cite{Kuzio2008} \cite{deBlok2002} \cite{deBlok2002} \cite{Stil2002}  
\\
F563-1 &LSB &46.8 &   0.140  &2.9 & 18.2 & 0.29 & 1.35 & \phantom{0}9.65 & \phantom{0}2.44 &  \cite{McGaugh2001} \cite{deBlok1997} \cite{deBlok1997} \cite{deBlok1996}     
\\
F563-V2 &LSB & 57.8 &    0.266  & 2.0 &\phantom{0}6.3 & 0.20 & 0.60 & \phantom{0}2.26 & \phantom{0}6.15 &    \cite{Kuzio2006} \cite{deBlok1997} \cite{deBlok1997} \cite{deBlok1996}    
\\
F568-3 & LSB& 80.0 &   0.351   & 4.2 & 11.6 & 0.30 & 1.20 & \phantom{0}3.43 & \phantom{0}3.16 &    \cite{McGaugh2001} \cite{deBlok1997} \cite{deBlok1997} \cite{deBlok1996}   
 \\
F583-1 & LSB & 32.4 &  0.064    & 1.6 &  14.1 &0.18 &   0.15 & \phantom{0}2.32 & \phantom{0}1.92&  \cite{McGaugh2001} \cite{deBlok1997} \cite{deBlok1997} \cite{deBlok1996}      
 \\
F583-4 & LSB &50.8 &   0.096   & 2.8 & \phantom{0}7.0 & 0.06 & 0.31 & \phantom{0}3.25 & \phantom{0}2.52 &   \cite{McGaugh2001}  \cite{deBlok1997} \cite{deBlok1997} \cite{deBlok1996}    
\\
NGC 0959 & LSB & 13.5 & 0.333    & 1.3& \phantom{0}2.9 &   0.05 & 0.37 &\phantom{0}1.11&\phantom{0}7.43 &   \cite{Kuzio2008}  \cite{Fisher1981} \cite{Esipov1991}   \cite{Fisher1981} 
\\
NGC 4395 &LSB & \phantom{0}4.1 &0.374&2.7& \phantom{0}0.9 &   0.13 &0.83 &\phantom{0}2.21 &\phantom{0}2.29&\cite{Kuzio2006} \cite{Swaters2002a} \cite{deBlok2002} \cite{Swaters2002a}     
\\
NGC 7137 &LSB & 25.0 &  0.959    & 1.7 & \phantom{0}3.6 & 0.10 & 0.27 & \phantom{0}0.28 & \phantom{0}3.91 &   \cite{Kuzio2008}  \cite{Bieging1977} ~ES \cite{Bieging1977}  
\\
UGC 0128 &LSB&  64.6 &    0.597  &  6.9 & 54.8 & 0.73 & 2.75 & \phantom{0}4.60 & \phantom{0}1.03 &  \cite{Verheijen1999} \cite{deBlok1997} \cite{vanderHulst1993} \cite{vanderHulst1993}     
\\
UGC 0191 &LSB &15.9 &  0.129    & 1.7 & \phantom{0}2.2 & 0.26 & 0.49 &\phantom{0}3.81 &15.48 &   \cite{Kuzio2008} \cite{vanZee2000} \cite{vanZee2000} \cite{vanZee1997}    
\\
UGC 0477 & LSB &35.8 &  0.871    &3.5 &10.2 & 1.02& 1.00 & \phantom{0}1.14 & \phantom{0}4.42 &     \cite{Kuzio2006} \cite{Fisher1981} ~ES   \cite{Fisher1981}
\\
UGC 1230 &LSB& 54.1&   0.366   & 4.7 & 37.1 & 0.65 & 0.67 &\phantom{0}1.82 & \phantom{0}0.97 &    \cite{deBlok1997} \cite{deBlok1997} \cite{vanderHulst1993} \cite{vanderHulst1993}
\\
UGC 1281 & LSB & \phantom{0}5.1 &0.017& 1.6& \phantom{0}1.7 &0.03&0.01 & \phantom{0}0.53 &\phantom{0}3.02 &    \cite{Kuzio2006}  \cite{vanZee2000} \cite{deBlok2002} \cite{Swaters2002a}    
\\
UGC 1551 &LSB & 35.6 &   0.780   & 4.2 & \phantom{0}6.6 & 0.44 & 0.16 &\phantom{0}0.20 & \phantom{0}3.69 &    \cite{Kuzio2008} \cite{Swaters2002b} \cite{deJong1996} \cite{Swaters2002b}    
\\
UGC 4325 &LSB& 11.9 &  0.373    & 1.9 & \phantom{0}3.4 & 0.10 &0.40 & \phantom{0}1.08 & \phantom{0}7.39&   \cite{Kuzio2008} \cite{Swaters2002a} \cite{deBlok2002} \cite{Swaters2002a}    
\\
UGC 5005 & LSB & 51.4 &   0.200   &  4.6 & 27.7 &0.28 &1.02&\phantom{0}5.11 &\phantom{0}1.30 &  \cite{deBlok1997} \cite{deBlok1997}  \cite{vanderHulst1993} \cite{vanderHulst1993}     
\\
UGC 5750 &LSB &56.1 &   0.472   &3.3 &\phantom{0}8.6 & 0.10 & 0.10 & \phantom{0}0.21 & \phantom{0}1.58 &\cite{Kuzio2006} \cite{deBlok1997}  \cite{vanderHulst1993}  \cite{vanderHulst1993}  
\\
UGC 5999 & LSB & 44.9 &   0.170   &4.4 & 15.0 & 0.18 & 3.36 &19.81 & \phantom{0}5.79 & \cite{deBlok1997} \cite{deBlok1997}  \cite{vanderHulst1993} \cite{vanderHulst1993}     
\\
UGC 11820 & LSB & 17.1 &   0.169   & 3.6 & \phantom{0}3.7 &0.40  & 1.68 & \phantom{0}9.95 &\phantom{0}8.44 &   \cite{Kuzio2008} \cite{vanZee1997} \cite{Kim2007}  \cite{vanZee1997}   
\\
\hline
\end{tabular}
\label{table:lsb20}
\end{table}

\begin{table}
\caption{Properties of the LSB 21 Galaxy Sample}
\centering
\begin{tabular}{l c c c c c c c c c c} 
\hline\hline
\phantom{00}Galaxy  & \phantom{0}Type \phantom{0}&Distance  & $L_{\rm  B}$ & $R_0$  & $R_{\rm last} $ &  $M_{\rm HI} $ & $M_{\rm disk}$ &  $ 
(M/L) _{\rm stars}$ & $(v^2 / c^2 R)_{\rm last}$ & Data~Sources \\  
& &  (Mpc)  &  $(10^{10}{\rm L}_{\odot})$&(kpc) & (kpc) & {$(10^{10} M_\odot)$} & {$(10^{10}
M_\odot)$} & ({$M_{\odot}/L_{\odot}$}) & {$(10^{-30}~\texttt{cm}^{-1})$} & $v~~~~L~~~R_0~~{\rm HI}$\\
\hline
ESO 0140040&LSB& 217.8 &  \phantom{0}7.169   &10.1 & 30.0 & ~~ &20.70&\phantom{0}3.38 &8.29 & \cite{McGaugh2001} \cite{deBlok2001} \cite{Beijersbergen1999} NA     
\\
ESO 0840411 &LSB &\phantom{0}82.4 &  \phantom{0}0.287    &3.5 & \phantom{0}9.1 & ~~
&\phantom{0}0.06 & \phantom{0}0.21 &1.49 & \cite{McGaugh2001} \cite{deBlok2001} ~ES NA    
\\
ESO 1200211 &LSB & \phantom{0}15.2 &   \phantom{0}0.028   &2.0 & \phantom{0}3.5 & ~~
&\phantom{0}0.01 & \phantom{0}0.20 &0.66 & \cite{McGaugh2001} \cite{deBlok2001} ~ES NA   
 \\
ESO 1870510 & LSB& \phantom{0}16.8 &    \phantom{0}0.054  & 2.1 & \phantom{0}2.8 &~~
&\phantom{0}0.09 & \phantom{0}1.62 &2.02 & \cite{McGaugh2001} \cite{deBlok2001} \cite{Bell2000} NA    
\\
ESO 2060140 & LSB & \phantom{0}59.6 &    \phantom{0}0.735 & 5.1 & 11.6 & ~~ &  \phantom{0}3.51 & \phantom{0}4.78 & 4.34 & \cite{McGaugh2001} \cite{deBlok2001} \cite{Beijersbergen1999} NA    
\\
ESO 3020120 & LSB & \phantom{0}70.9 &   \phantom{0}0.717   & 3.4 & 11.2 & ~~ & \phantom{0}0.77 & \phantom{0}1.07 & 2.37 &\cite{McGaugh2001} \cite{deBlok2001} ~ES NA   
\\
ESO 3050090 & LSB & \phantom{0}13.2 &  \phantom{0}0.186    &1.3& \phantom{0}5.6 & ~~ &
\phantom{0}0.06 & \phantom{0}0.32 & 1.87 & \cite{McGaugh2001} \cite{deBlok2001} ~ES NA   
\\
ESO 4250180 & LSB & \phantom{0}88.3 &  \phantom{0}2.600    & 7.3 & 14.6 & ~~ &\phantom{0}4.79 &\phantom{0}1.84 &5.17 
& \cite{McGaugh2001} \cite{deBlok2001} \cite{Beijersbergen1999} NA   
\\
ESO 4880490 &LSB & \phantom{0}28.7 &  \phantom{0}0.139    & 1.6 & 7.8 & ~~ &\phantom{0}0.43 &\phantom{0}3.07 & 4.34 & \cite{McGaugh2001} \cite{deBlok2001} ~ES NA     
\\
F571-8 & LSB & \phantom{0}50.3 &   \phantom{0}0.191   & 5.4 & 14.6 & 0.16& \phantom{0}4.48 &
 23.49& 5.10 & \cite{McGaugh2001} \cite{deBlok1996} \cite{deBlok1997} \cite{deBlok1996}     
  \\
F579-V1 & LSB & \phantom{0}86.9&   \phantom{0}0.557   & 5.2 & 14.7 & 0.21 & \phantom{0}3.33 &
\phantom{0}5.98 &  3.18 & \cite{McGaugh2001} \cite{deBlok1996} \cite{deBlok1997} \cite{deBlok1996}    
\\
F730-V1 &LSB & 148.3 &    \phantom{0}0.756 & 5.8 & 12.2 & ~~ & \phantom{0}5.95 & \phantom{0}7.87 & 6.22 & \cite{McGaugh2001} \cite{Kim2007} \cite{Kim2007} NA   
 \\
UGC 04115& LSB & \phantom{00}5.5 &  \phantom{0}0.004    & 0.3 & \phantom{0}1.7 & ~~ & \phantom{0}0.01 & \phantom{0}0.97 & 3.42&\cite{McGaugh2001} \cite{deBlok2001} \cite{Bazarra2001} NA 
\\
UGC 06614 &LSB&\phantom{0}86.2 &    \phantom{0}2.109 & 8.2 & 62.7 & 2.07 & \phantom{0}9.70 &\phantom{0}4.60& 2.39 &\cite{McGaugh2001} \cite{deBlok2001} \cite{vanderHulst1993} \cite{vanderHulst1993} 
\\
UGC 11454 & LSB & \phantom{0}93.9 &   \phantom{0}0.456   & 3.4 & 12.3 &~~ & \phantom{0}3.15 & \phantom{0}6.90& 6.79&     \cite{McGaugh2001} \cite{deBlok2001} \cite{Kim2007} NA   
\\
UGC 11557 & LSB & \phantom{0}23.7 &   \phantom{0}1.806   & 3.0 & \phantom{0}6.7 &0.25 &\phantom{0}0.37& \phantom{0}0.20& 3.49 &  \cite{McGaugh2001} \cite{deBlok2001} \cite{Swaters2002a} \cite{Swaters2002a}    
\\
UGC 11583&LSB & \phantom{00}7.1 &  \phantom{0}0.012    & 0.7 & \phantom{0}2.1 & ~~
&\phantom{0}0.01 & \phantom{0}0.96 & 2.15&    \cite{McGaugh2001} \cite{deBlok2001} \cite{Kim2007} NA     
 \\
UGC 11616 &LSB&\phantom{0}74.9 &   \phantom{0}2.159   &3.1 & \phantom{0}9.8 & ~~ & \phantom{0}2.43 &\phantom{0}1.13 & 7.49 &     \cite{McGaugh2001} \cite{deBlok2001} \cite{Kim2007} NA  
\\
UGC 11648 &LSB & \phantom{0}49.0 &  \phantom{0}4.073    &4.0 &13.0 &~~ &\phantom{0}2.57 &\phantom{0}0.63 & 5.79&    \cite{McGaugh2001} \cite{deBlok2001} \cite{Kim2007} NA    
\\
UGC 11748 &LSB &\phantom{0}75.3 &  23.930  & 2.6 & 21.6 & ~~ &\phantom{0}9.67 &\phantom{0}0.40 &1.01 &     \cite{McGaugh2001} \cite{deBlok2001} \cite{Kim2007} NA  
\\
UGC 11819 &LSB &\phantom{0}61.5 &    \phantom{0}2.155  & 4.7 & 11.9 & ~~ &\phantom{0}4.83 & \phantom{0}2.24& 7.03 &   \cite{McGaugh2001} \cite{deBlok2001} \cite{Kim2007} NA    
\\
\hline
\end{tabular}
\label{table:lsb21}
\end{table}

\begin{table}
\caption{Properties of the Miscellaneous 22 Galaxy Sample}
\centering
\begin{tabular}{l c c c c c c c c c c} 
\hline\hline
 \phantom{00}Galaxy  & \phantom{0}Type \phantom{0}&Distance  & $L_{\rm  B}$ & $R_0$  & $R_{\rm last} $ &  $M_{\rm HI} $ & $M_{\rm disk}$ &  $ 
(M/L) _{\rm stars}$ & $(v^2 / c^2 R)_{\rm last}$ & Data~Sources \\  
& &  (Mpc)  &  $(10^{10}{\rm L}_{\odot})$&(kpc) & (kpc) & {$(10^{10} M_\odot)$} & {$(10^{10}
M_\odot)$} & ({$M_{\odot}/L_{\odot}$}) & {$(10^{-30}~\texttt{cm}^{-1})$} & $v~~~~L~~~R_0~~{\rm HI}$\\
\hline
DDO 0168 & LSB&\phantom{00}4.5& \phantom{0}0.032    &\phantom{0}1.2 & \phantom{0}4.4 & 0.03 &
\phantom{0}0.06 & 2.03& 2.22 & \cite{Broeils1992} \cite{Broeils1992} \cite{Broeils1992} \cite{Broeils1992}
\\
DDO 0170 & LSB& \phantom{0}16.6 &  \phantom{0}0.023    & \phantom{0}1.9 & 13.3 & 0.09 &
\phantom{0}0.05 & 1.97 &1.18 & \cite{Lake1990} \cite{Lake1990} \cite{Lake1990} \cite{Lake1990}
\\
\phantom{CC}M 0033& HSB &\phantom{00}0.9 &  \phantom{0}0.850   &\phantom{0}2.5 & \phantom{0}8.9 & 0.11 &\phantom{0}1.13 &1.33& 4.62 &\cite{Rhee1996} \cite{Rhee1996} \cite{Kent1987b} \cite{Rhee1996}
\\
NGC 0055 & LSB& \phantom{00}1.9 &  \phantom{0}0.588  & \phantom{0}1.9 & 12.2 & 0.13 &  \phantom{0}0.30 & 0.50 & 2.22 & \cite{Puche1991}  \cite{Puche1991}  \cite{Puche1991}  \cite{Puche1991}
\\
NGC 0247 & LSB & \phantom{00}3.6 &  \phantom{0}0.512    & \phantom{0}4.2 & 14.3 & 0.16 &
\phantom{0}1.25 & 2.43 & 2.94& \cite{Carignan1990} \cite{Carignan1985a} \cite{Carignan1985a} \cite{Carignan1990}
\\
NGC 0300 &LSB &\phantom{00}2.0&   \phantom{0}0.271  &\phantom{0}2.1&11.7& 0.08 &\phantom{0}0.65
&2.41 & 2.69& \cite{Puche1990} \cite{Carignan1985a} \cite{Carignan1985a} \cite{Puche1990}
\\
NGC 0801 & HSB &\phantom{0}63.0 &   \phantom{0}4.746   & \phantom{0}9.5 & 46.7 & 1.39 & \phantom{0}6.93 &
2.37 & 3.59 & \cite{Broeils1992} \cite{Broeils1992} \cite{Kent1986} \cite{Broeils1992} 
\\
NGC 1003 &LSB& \phantom{0}11.8 &  \phantom{0}1.480    & \phantom{0}1.9 & 31.2 & 0.63 &  \phantom{0}0.66 &0.45 & 1.53& \cite{Rhee1996} \cite{Rhee1996} \cite{Broeils1991} \cite{Sanders1996} 
\\
NGC 1560 & LSB & \phantom{00}3.7 &  \phantom{0}0.053    & \phantom{0}1.6 & 10.3 & 0.12 &
\phantom{0}0.17 & 3.16 & 2.16& \cite{Broeils1992}  \cite{Broeils1992}  \cite{Broeils1992}  \cite{Broeils1992}
\\
NGC 2683 &HSB& \phantom{0}10.2 &   \phantom{0}1.882   & \phantom{0}2.4 & 36.0 &0.15 & \phantom{0}6.03 & 3.20
& 2.28 & \cite{Casertano1991} \cite{Casertano1991} \cite{Kent1985} \cite{Sanders1996}
\\
NGC 2998 &HSB& \phantom{0}59.3&   \phantom{0}5.186   &\phantom{0}4.8 & 41.1 & 1.78 & \phantom{0}7.16 & 1.75
& 3.43  & \cite{Broeils1992}  \cite{Broeils1992}  \cite{Kent1986}  \cite{Broeils1992}
\\
NGC 3109 & LSB &\phantom{00}1.5 &  \phantom{0}0.064    & \phantom{0}1.3 & \phantom{0}7.1 & 0.06 &
\phantom{0}0.02 &0.35 & 2.29& \cite{Jobin1990}  \cite{Carignan1985b}  \cite{Carignan1985b} \cite{Jobin1990}
\\
NGC 5033 & HSB & \phantom{0}15.3&   \phantom{0}3.058   & \phantom{0}7.5 & 45.6& 1.07 & \phantom{0}0.27 &
3.28 & 3.16 & \cite{Begeman1987}  \cite{Begeman1987}  \cite{Wevers1986} \cite{Begeman1987}
\\
NGC 5371 &HSB& \phantom{0}35.3 &   \phantom{0}7.593   & \phantom{0}4.4 & 41.0 & 0.89 & \phantom{0}8.52 & 1.44 & 3.98 & \cite{Begeman1987} \cite{Begeman1987} \cite{Wevers1986} \cite{Begeman1987}
\\
NGC 5533 &HSB& \phantom{0}42.0 &  \phantom{0}3.173    & \phantom{0}7.4 & 56.0 & 1.39 & \phantom{0}2.00 &4.14  &3.31 & \cite{Broeils1992} \cite{Broeils1992}  \cite{Kent1985}  \cite{Broeils1992}
\\
NGC 5585 & HSB &\phantom{00}9.0 &   \phantom{0}0.333   & \phantom{0}2.0 & 14.0 & 0.28 &
\phantom{0}0.36 & 1.09 & 2.06 & \cite{Cote1991} \cite{Cote1991} \cite{Cote1991} \cite{Cote1991}
\\
NGC 5907 & HSB &\phantom{0}16.5      &  \phantom{0}5.400&\phantom{0}5.5 & 48.0 & 1.90 &\phantom{0}2.49 &1.89 & 3.44& \cite{Barnaby1994} \cite{Sanders1996} \cite{Barnaby1994} \cite{Sanders1996}
\\
NGC 6503 & HSB&\phantom{00}5.5 &  \phantom{0}0.417   & \phantom{0}1.6 & 20.7 & 0.14 &
\phantom{0}1.53 & 3.66 & 2.30 & \cite{Begeman1987} \cite{Begeman1987} \cite{Wevers1986} \cite{Begeman1987}
\\
NGC 6674 & HSB & \phantom{0}42.0 &   \phantom{0}4.935   &\phantom{0}7.1 & 59.1 & 2.18 & \phantom{0}2.00
&2.52 & 3.57 & \cite{Broeils1992} \cite{Broeils1992}  \cite{Broeils1991}  \cite{Broeils1992}
\\
UGC 2259 &LSB& \phantom{0}10.0 &   \phantom{0}0.110   & \phantom{0}1.4 & \phantom{0}7.8 &0.04 &
\phantom{0}0.47 & 4.23 & 3.76 & \cite{Carignan1988} \cite{Carignan1988} \cite{Kent1987a} \cite{Carignan1988}
\\
UGC 2885 &HSB& \phantom{0}80.4 &   23.955   & 13.3 & 74.1 &3.98 &\phantom{0}8.47 & 0.72 & 4.31& \cite{Roelfsema1985} \cite{Kent1986} \cite{Kent1986} \cite{Sanders1996}
\\
Malin~1 &LSB& 338.5 &   \phantom{0}7.912   & 84.2 & 98.0 &5.40 &\phantom{0}1.00 & 1.32 & 1.77& \cite{Lelli2010} \cite{Lelli2010} \cite{Pickering1997} \cite{Lelli2010}
\\
\hline 
\end{tabular}
\label{table:Misc} 
\end{table}

\begin{figure}[t]
\epsfig{file=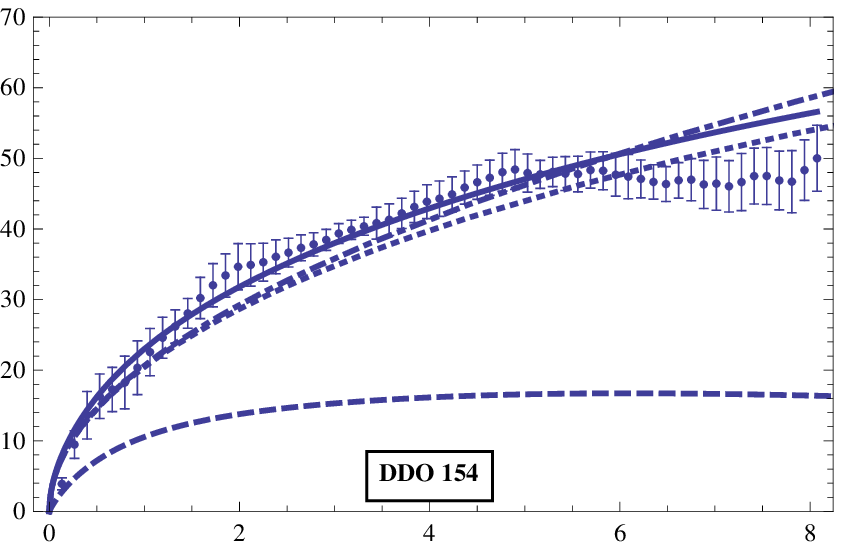,width=2.11in,height=1.2in}~~~
\epsfig{file=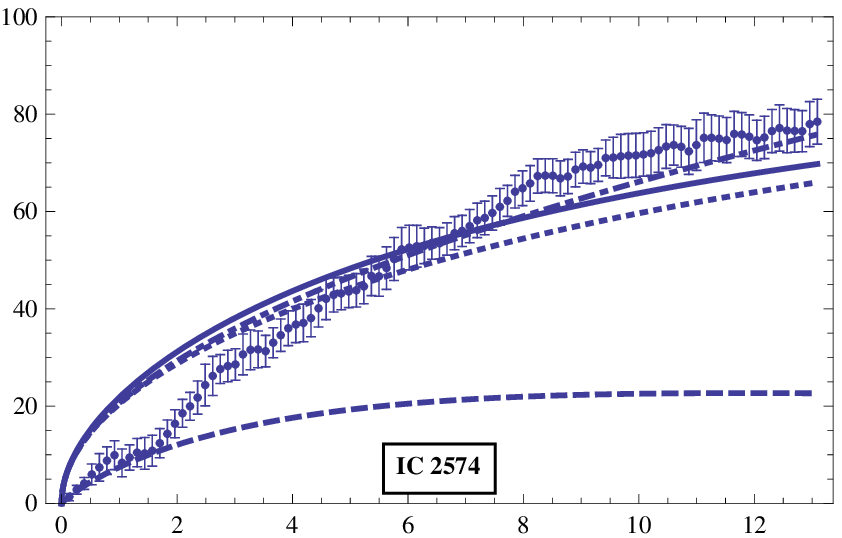,width=2.11in,height=1.2in}~~~
\epsfig{file=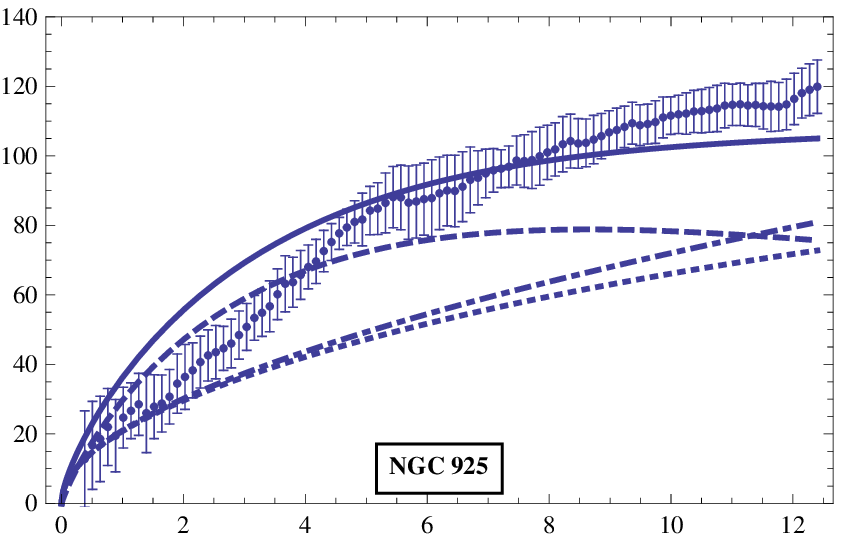,width=2.11in,height=1.2in}\\
\smallskip
\epsfig{file=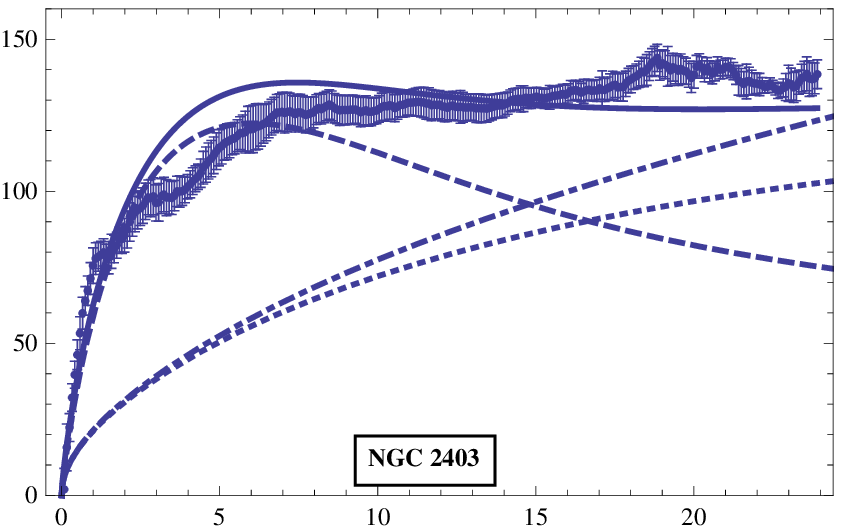,width=2.11in,height=1.2in}~~~
\epsfig{file=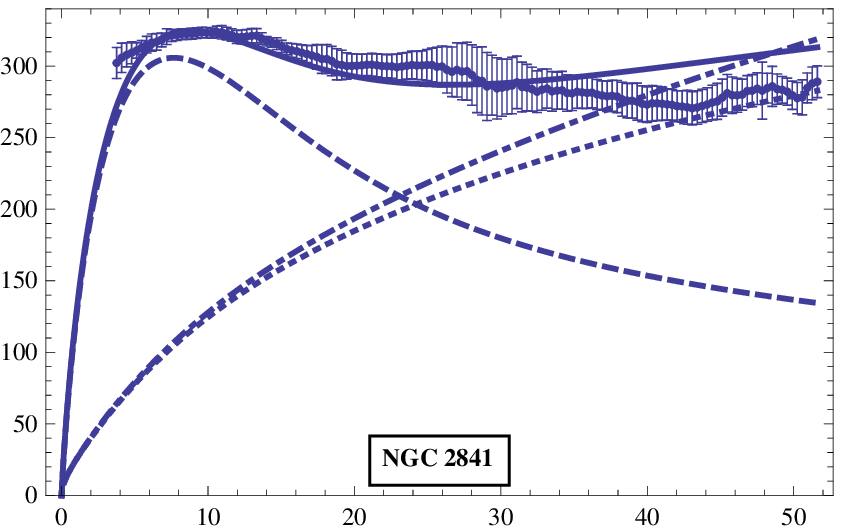,width=2.11in,height=1.2in}~~~
\epsfig{file=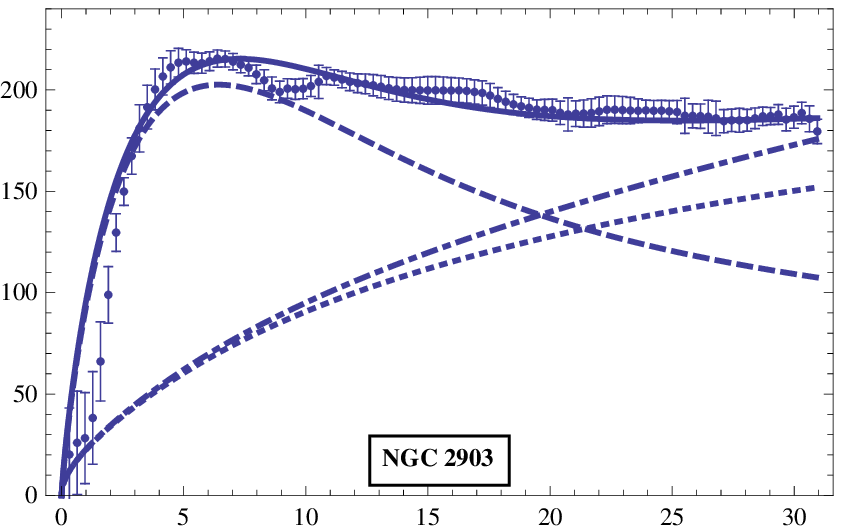, width=2.11in,height=1.2in}\\
\smallskip
\epsfig{file=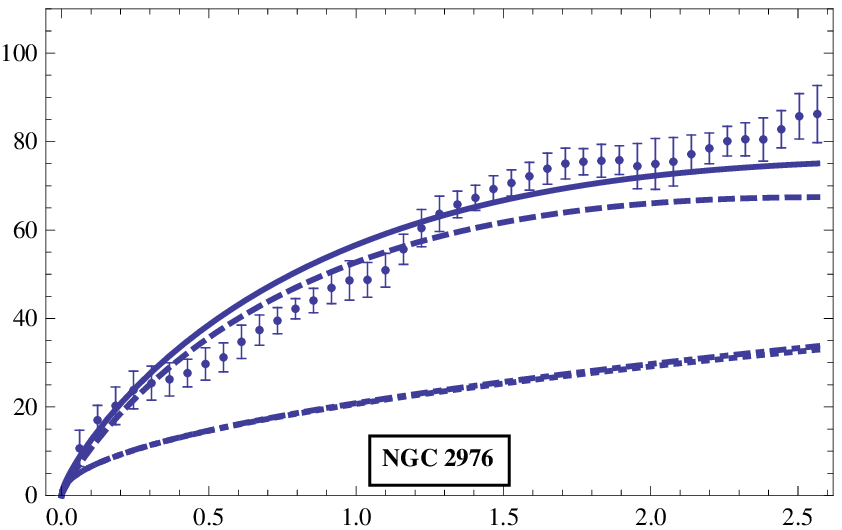,width=2.11in,height=1.2in}~~~
\epsfig{file=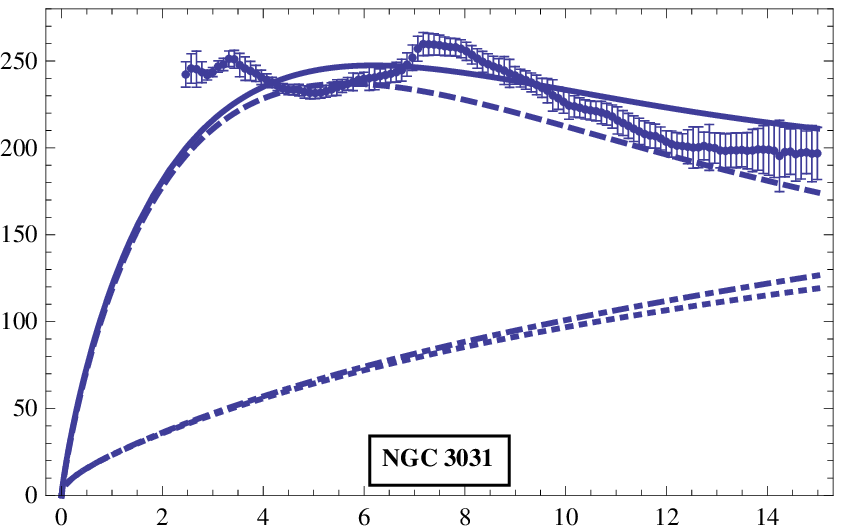,width=2.11in,height=1.2in}~~~
\epsfig{file=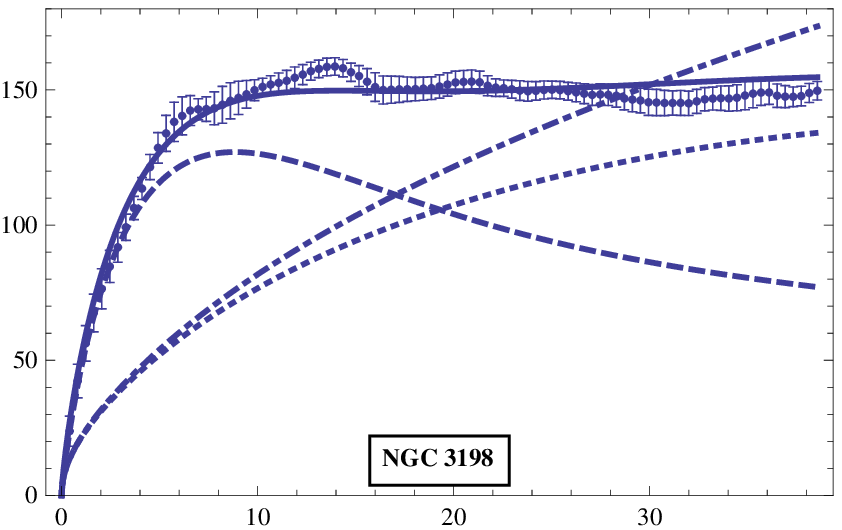,width=2.11in,height=1.2in}\\
\smallskip
\epsfig{file=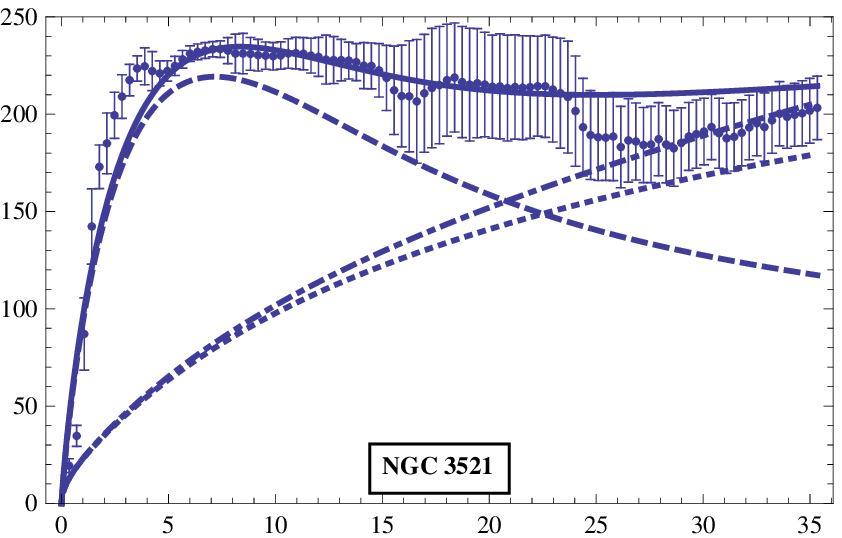,width=2.11in,height=1.2in}~~~
\epsfig{file=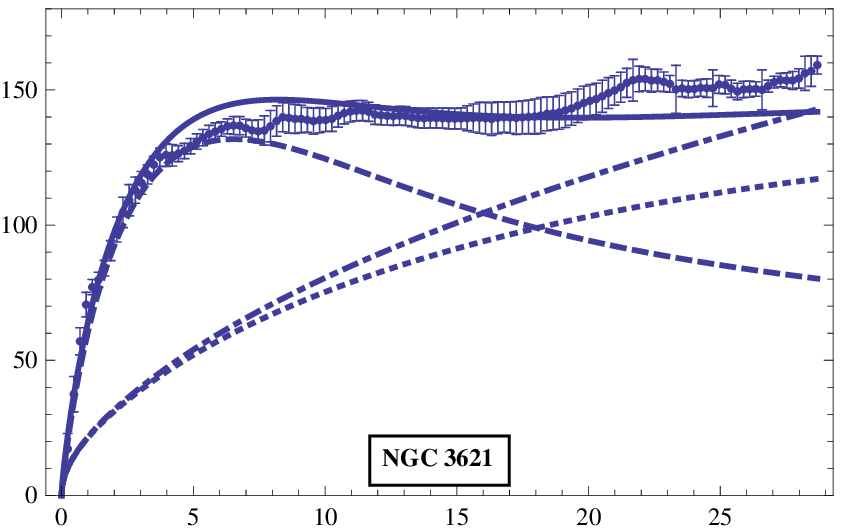,width=2.11in,height=1.2in}~~~
\epsfig{file=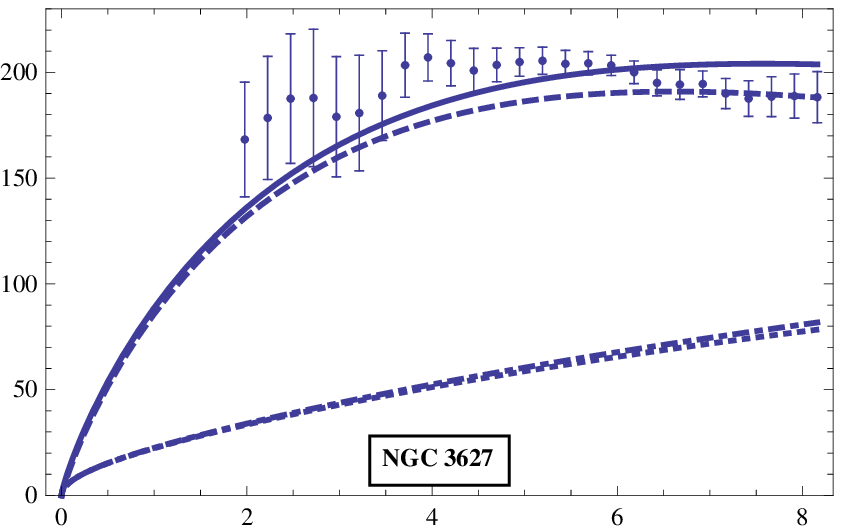,width=2.11in,height=1.2in}\\
\smallskip
\epsfig{file=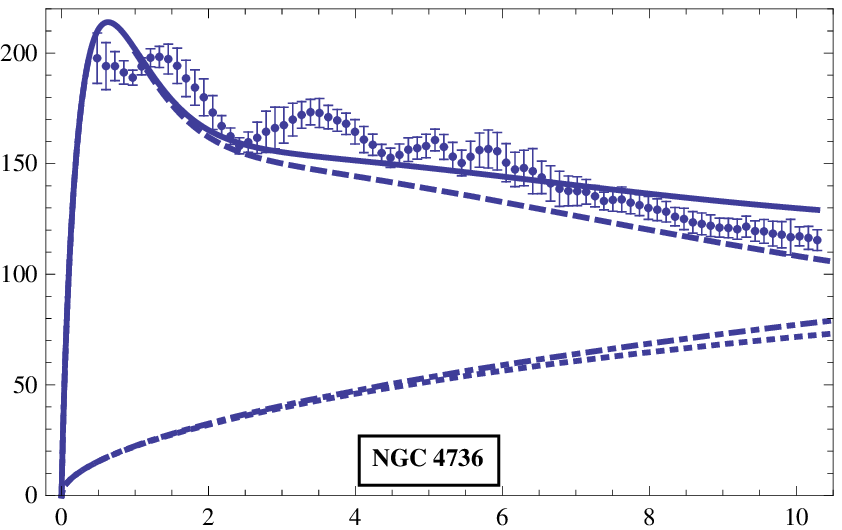,width=2.11in,height=1.2in}~~~
\epsfig{file=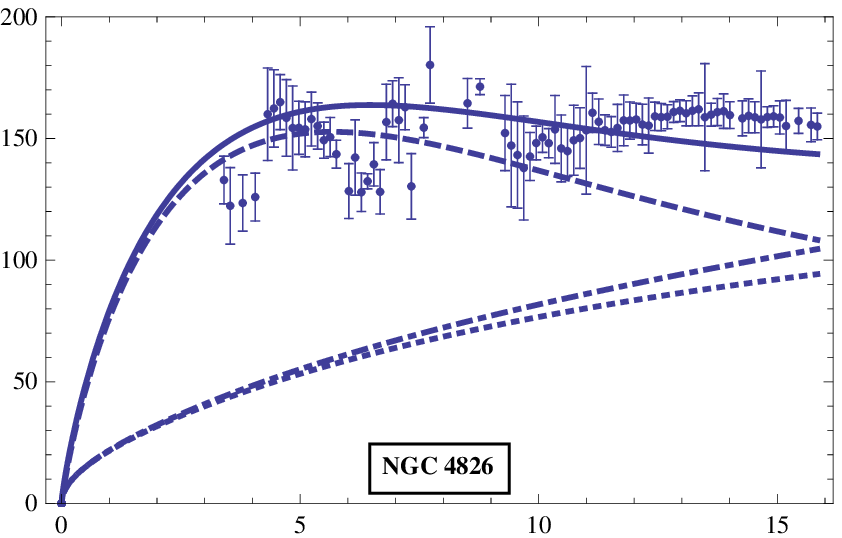,width=2.11in,height=1.2in}~~~
\epsfig{file=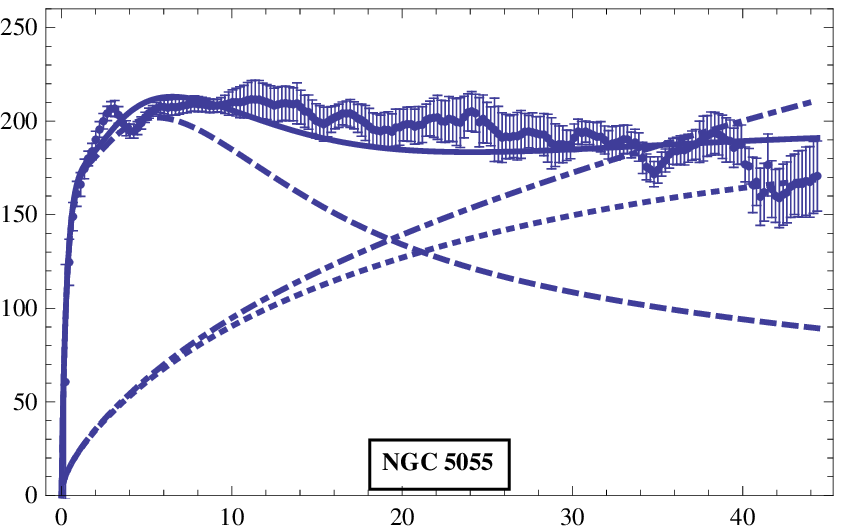,width=2.11in,height=1.2in}\\
\smallskip
\epsfig{file=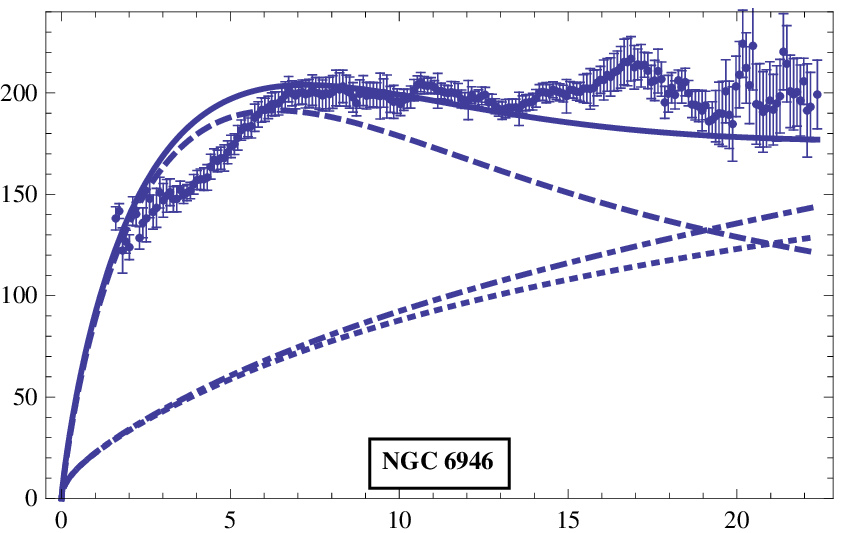,width=2.11in,height=1.2in}~~~
\epsfig{file=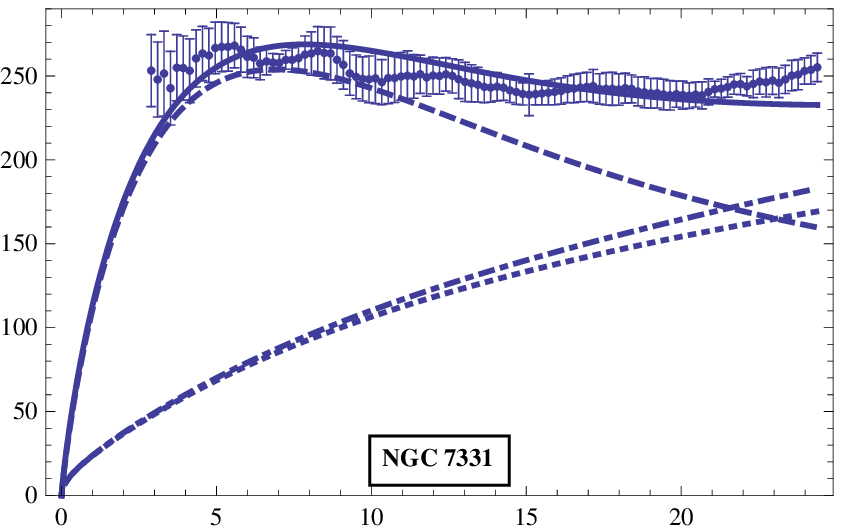,width=2.11in,height=1.2in}~~~
\epsfig{file=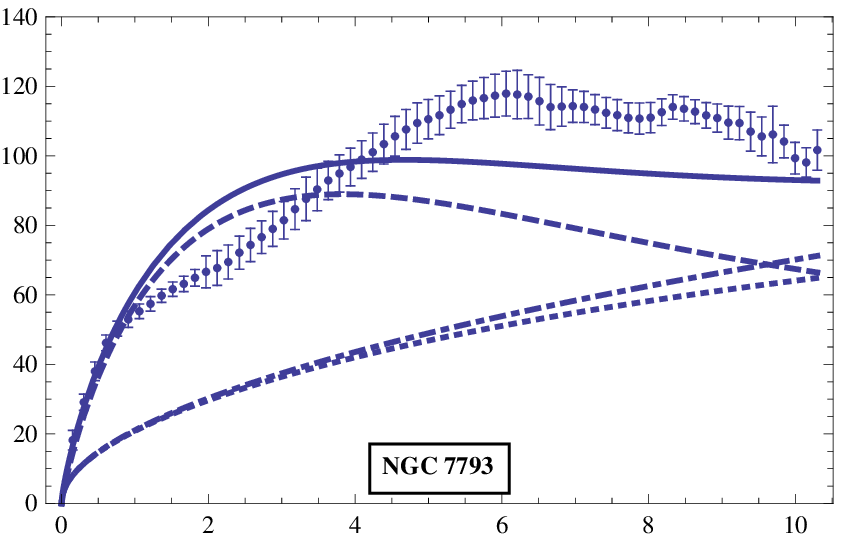,width=2.11in,height=1.2in}\\
\medskip
FIG.~1:~Fitting to the rotational velocities (in ${\rm km}~{\rm sec}^{-1}$) of  the THINGS 18 galaxy sample with their quoted errors as plotted as a function of radial distance (in ${\rm kpc}$). For each galaxy we have exhibited the contribution due to the luminous Newtonian term alone (dashed curve), the contribution from the two linear terms alone (dot dashed curve), the contribution from the two linear terms and the quadratic terms combined (dotted curve), with the full curve showing the total contribution. No dark matter is assumed.
\label{Fig. 1}
\end{figure}

\begin{figure}[t]
\epsfig{file=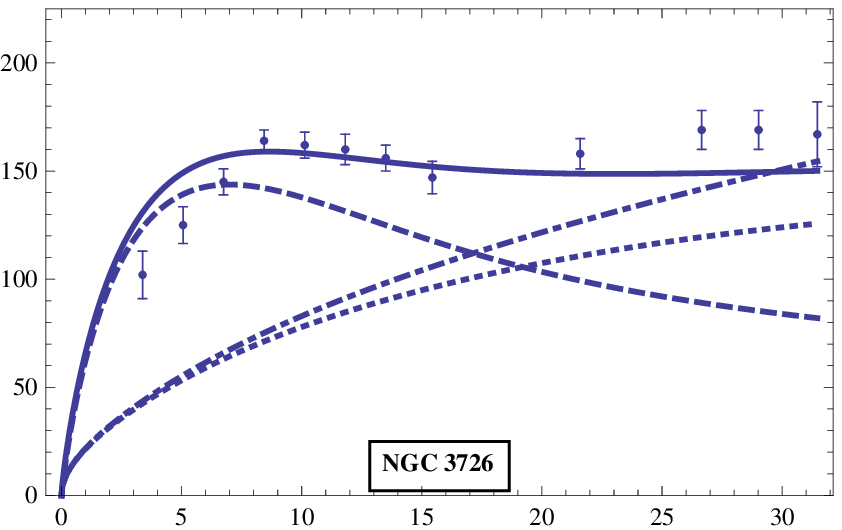,width=2.11in,height=1.2in}~~~
\epsfig{file=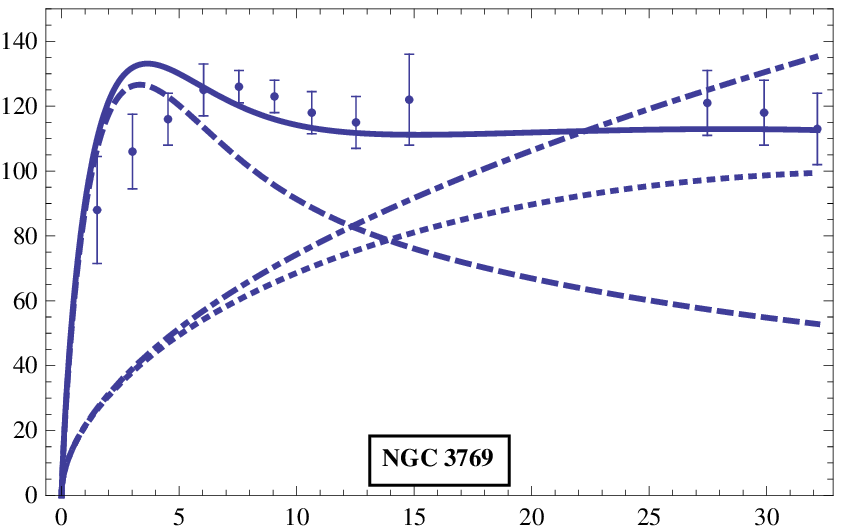,width=2.11in,height=1.2in}~~~
\epsfig{file=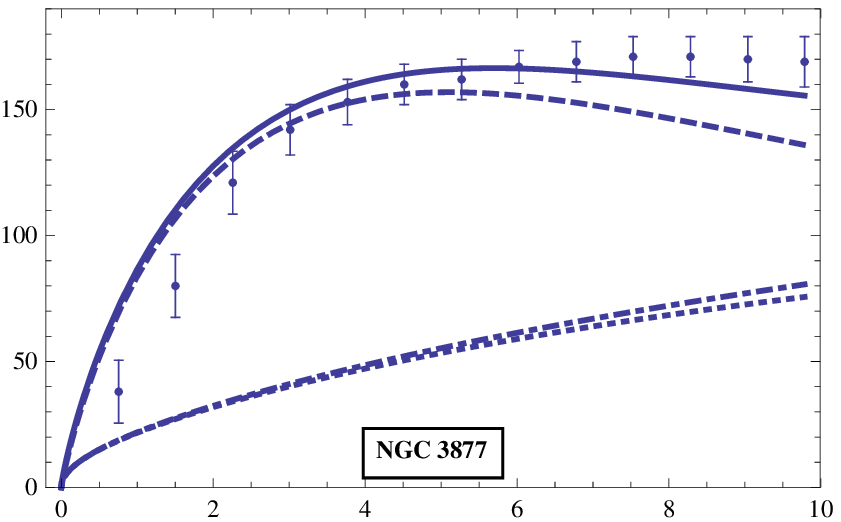,  width=2.11in,height=1.2in}\\
\smallskip
\epsfig{file=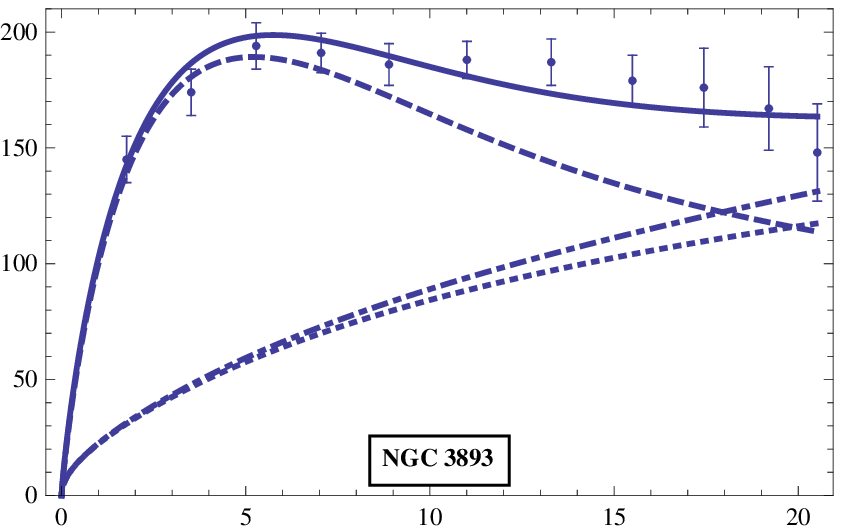,width=2.11in,height=1.2in}~~~
\epsfig{file=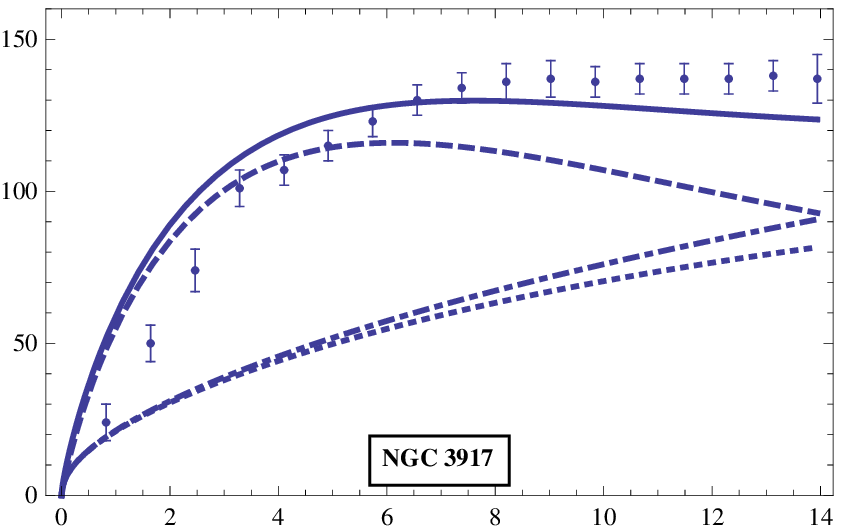,width=2.11in,height=1.2in}~~~
\epsfig{file=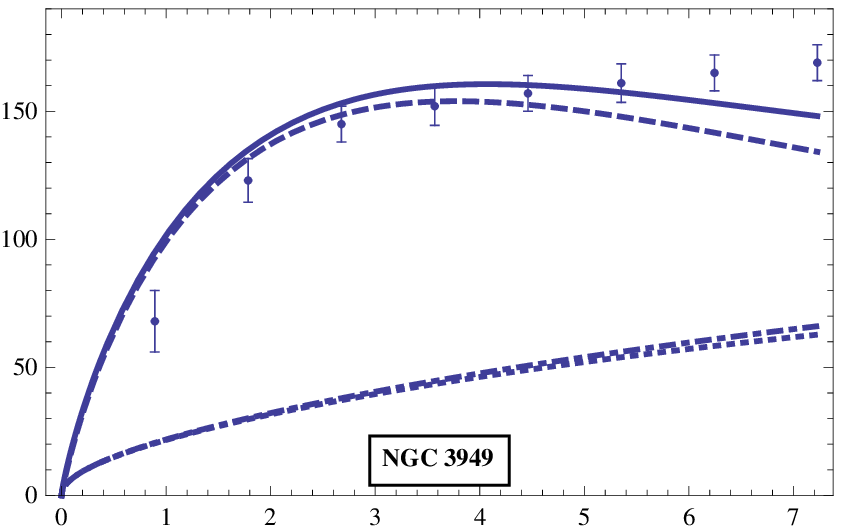,width=2.11in,height=1.2in}\\
\smallskip
\epsfig{file=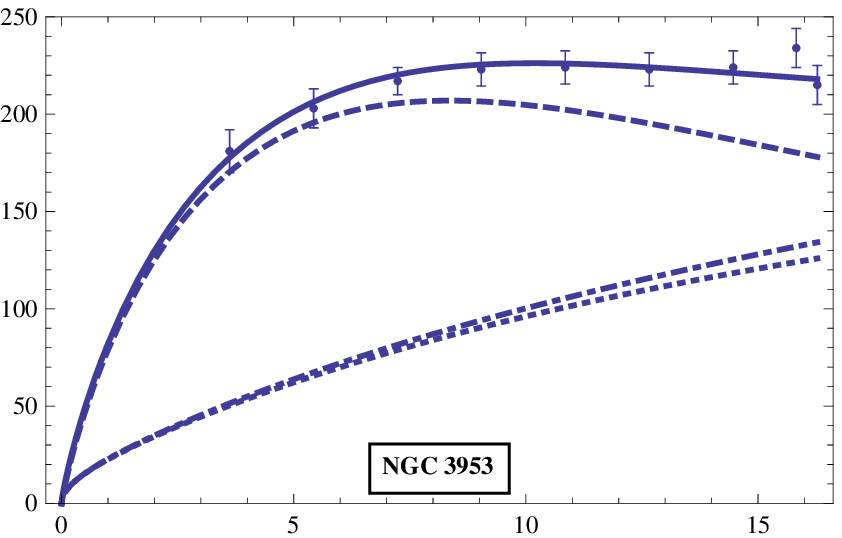,width=2.11in,height=1.2in}~~~
\epsfig{file=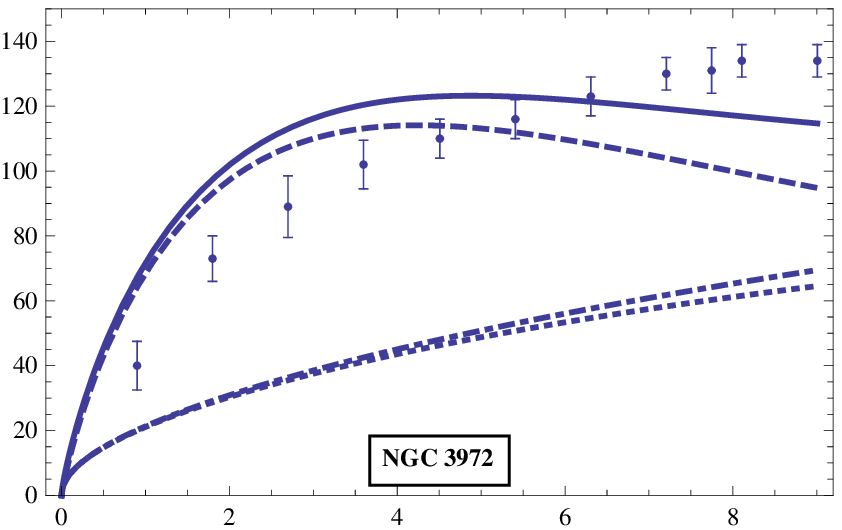,width=2.11in,height=1.2in}~~~
\epsfig{file=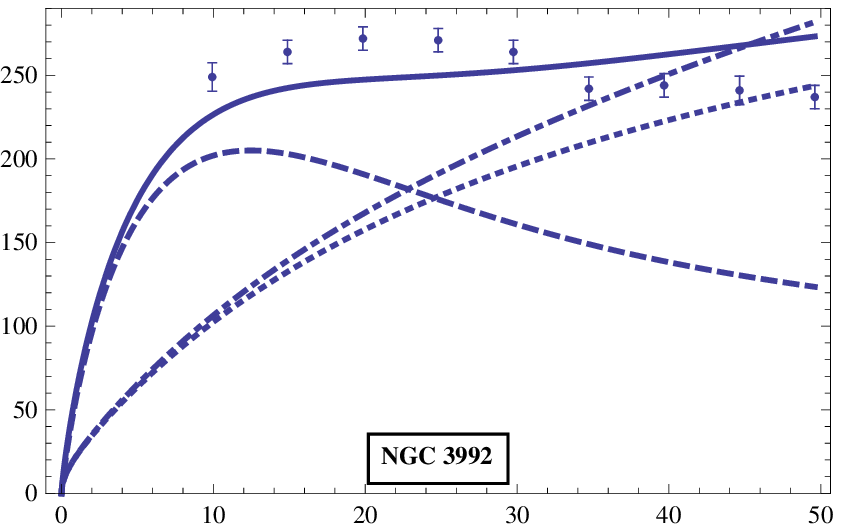,width=2.11in,height=1.2in}\\
\smallskip
\epsfig{file=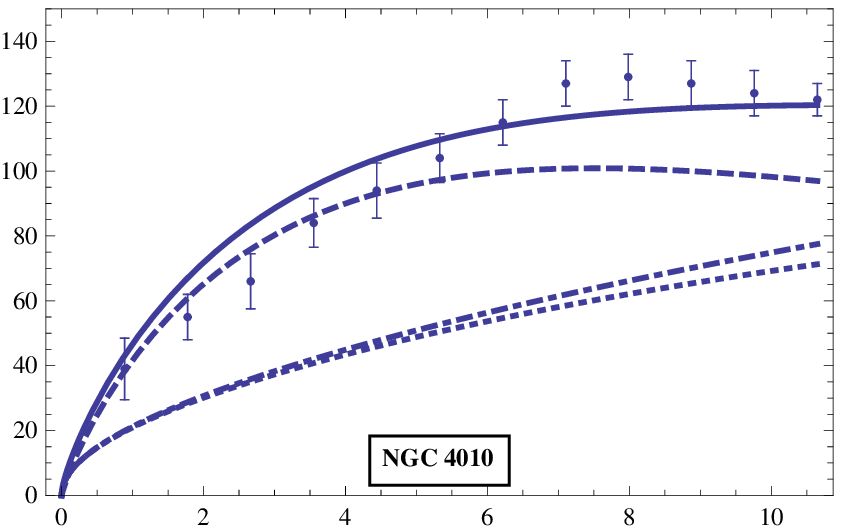,width=2.11in,height=1.2in}~~~
\epsfig{file=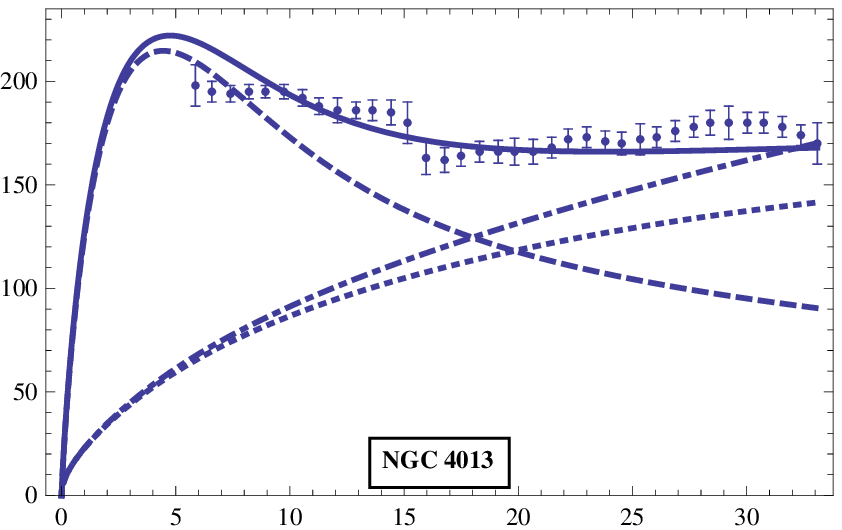,width=2.11in,height=1.2in}~~~
\epsfig{file=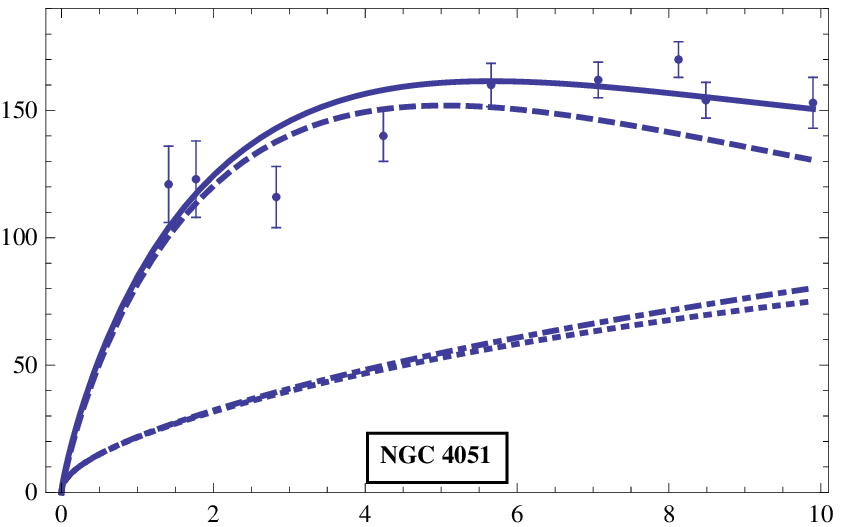,width=2.11in,height=1.2in}\\
\smallskip
\epsfig{file=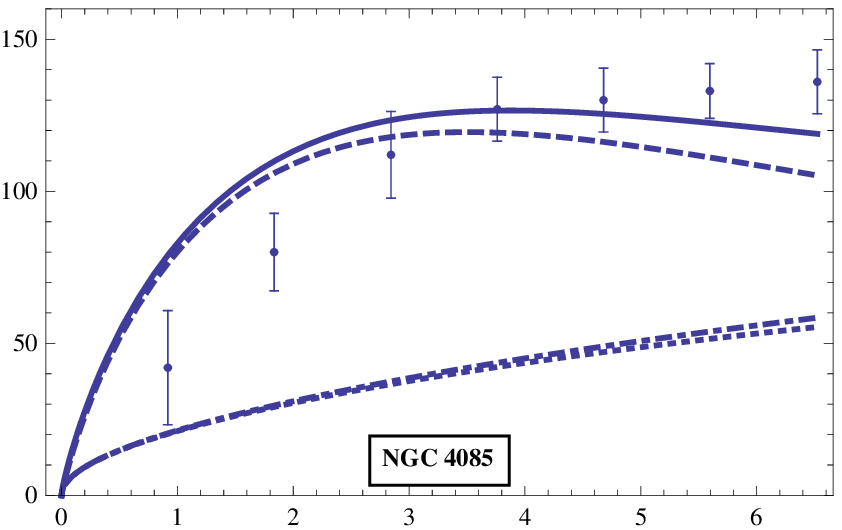,width=2.11in,height=1.2in}~~~
\epsfig{file=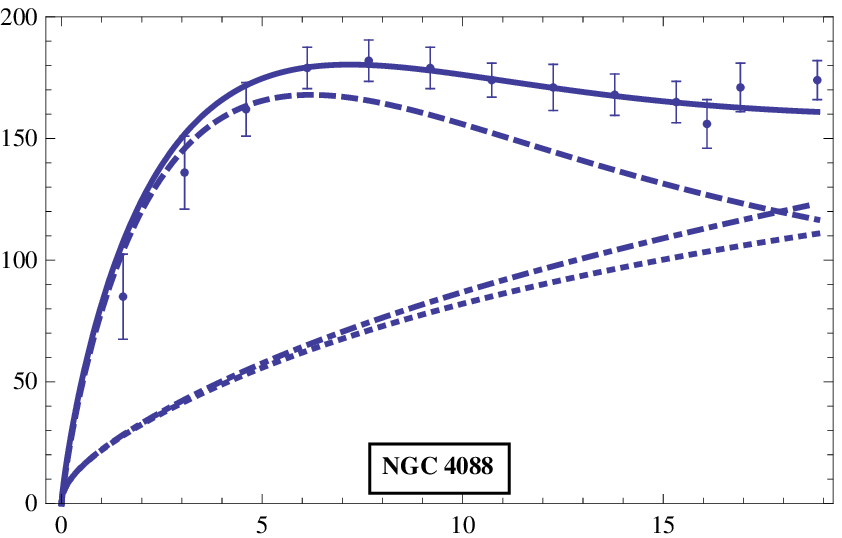,width=2.11in,height=1.2in}~~~
\epsfig{file=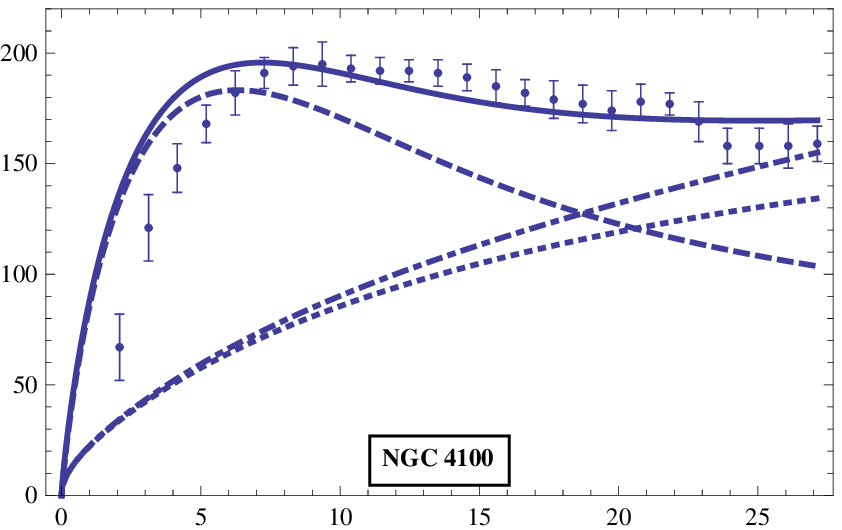,width=2.11in,height=1.2in}\\
\medskip
FIG.~2:~Fitting to the rotational velocities of the Ursa Major 30 galaxy sample -- Part 1
\label{Fig. 2.1}
\end{figure}

\begin{figure}[t]
\epsfig{file=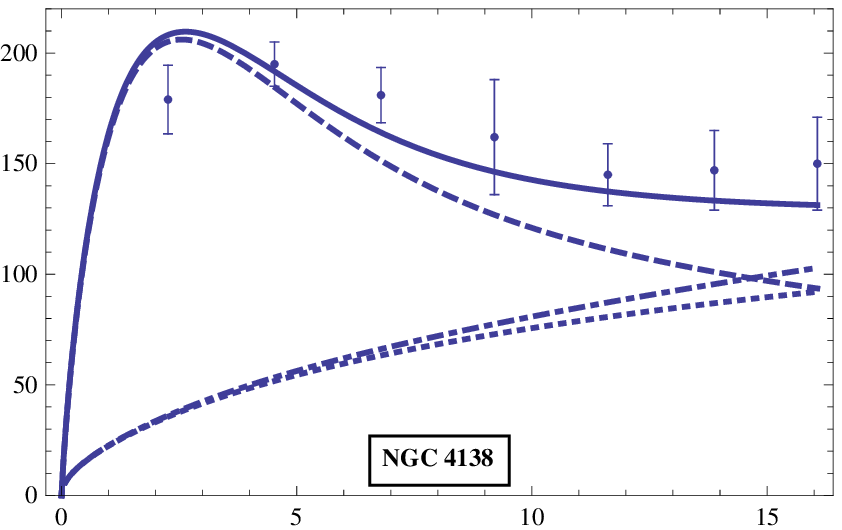,width=2.11in,height=1.2in}~~~
\epsfig{file=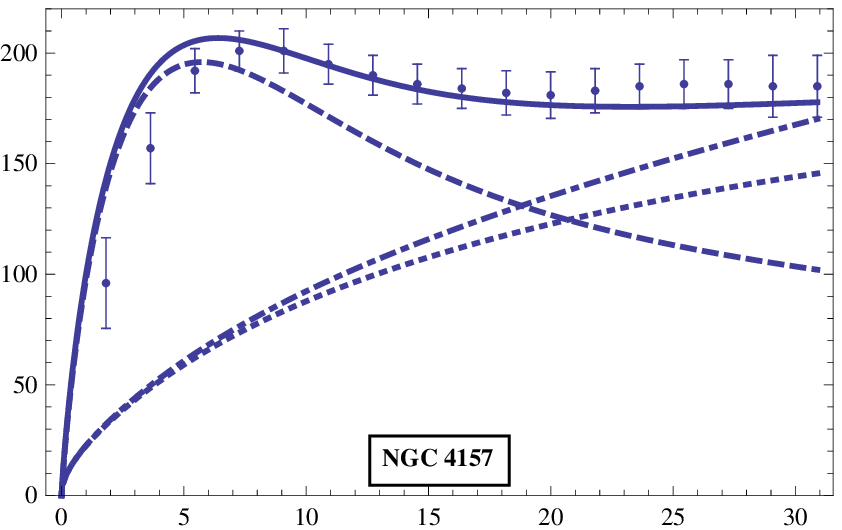,width=2.11in,height=1.2in}~~~
\epsfig{file=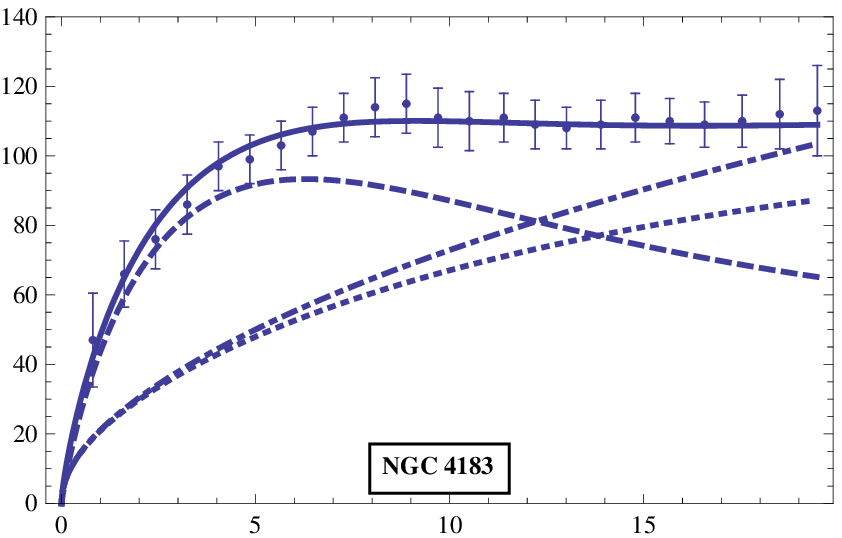,  width=2.11in,height=1.2in}\\
\smallskip
\epsfig{file=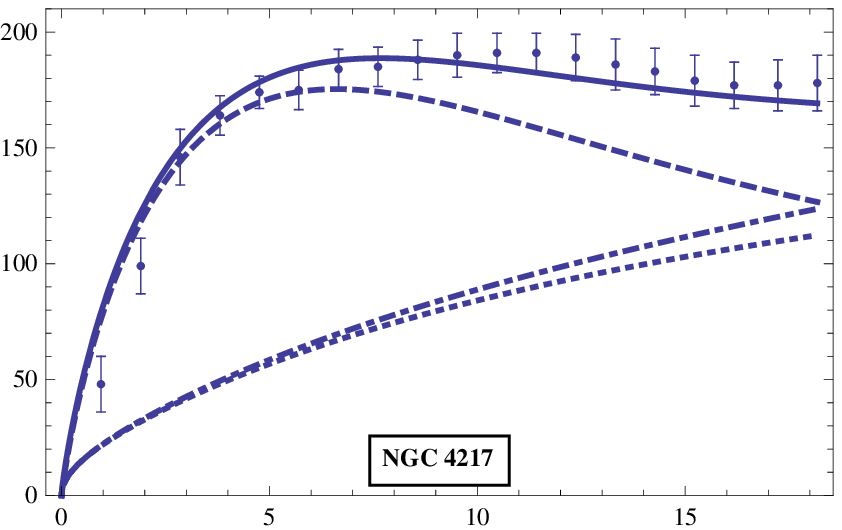,width=2.11in,height=1.2in}~~~
\epsfig{file=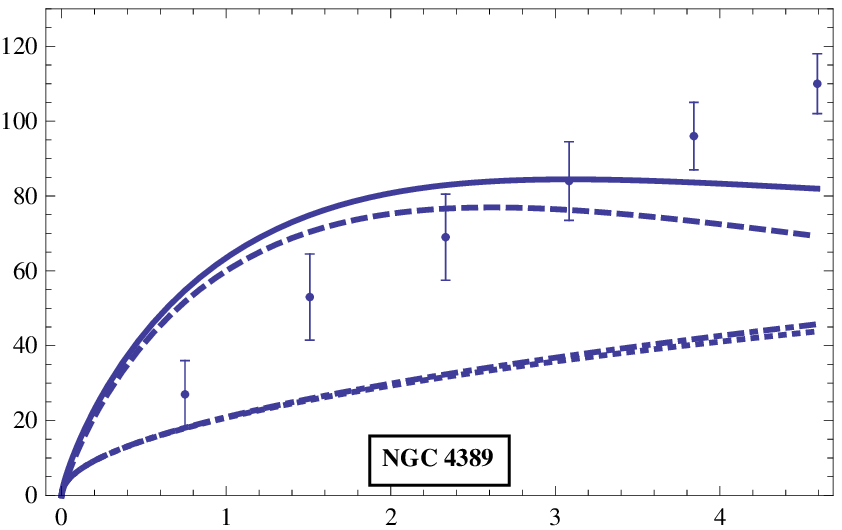,width=2.11in,height=1.2in}~~~
\epsfig{file=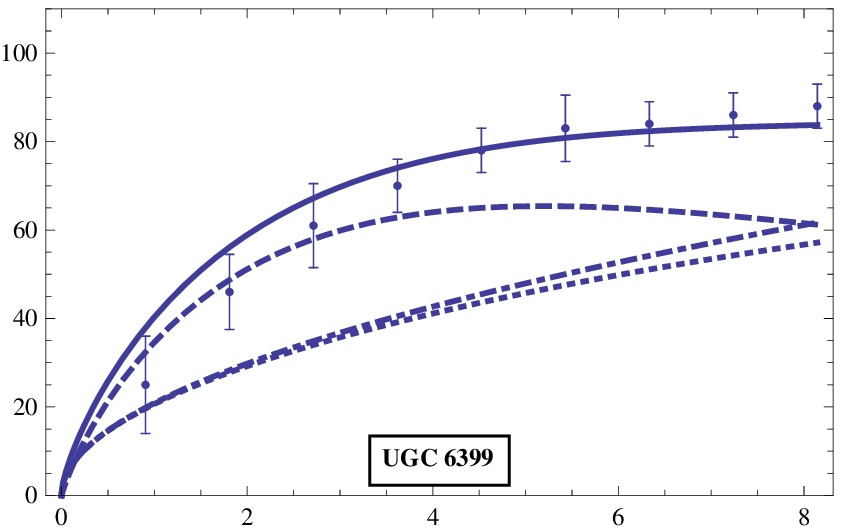,width=2.11in,height=1.2in}\\
\smallskip
\epsfig{file=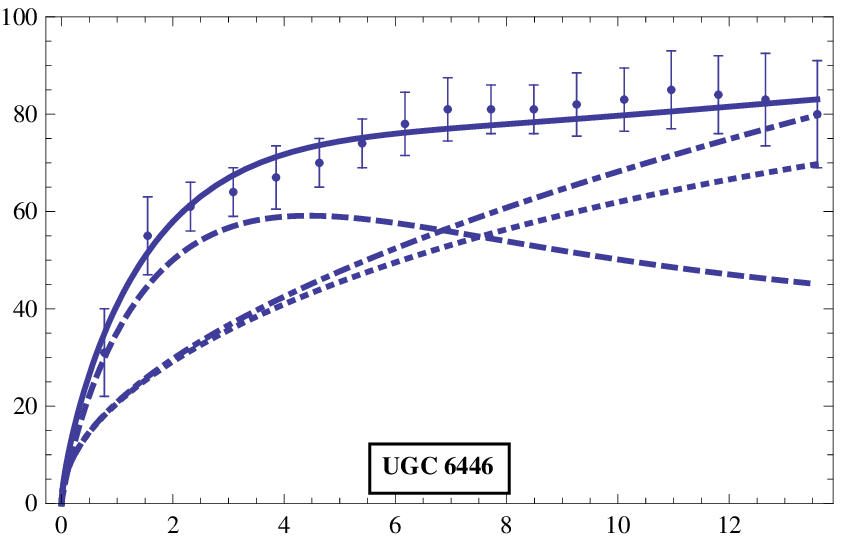,width=2.11in,height=1.2in}~~~
\epsfig{file=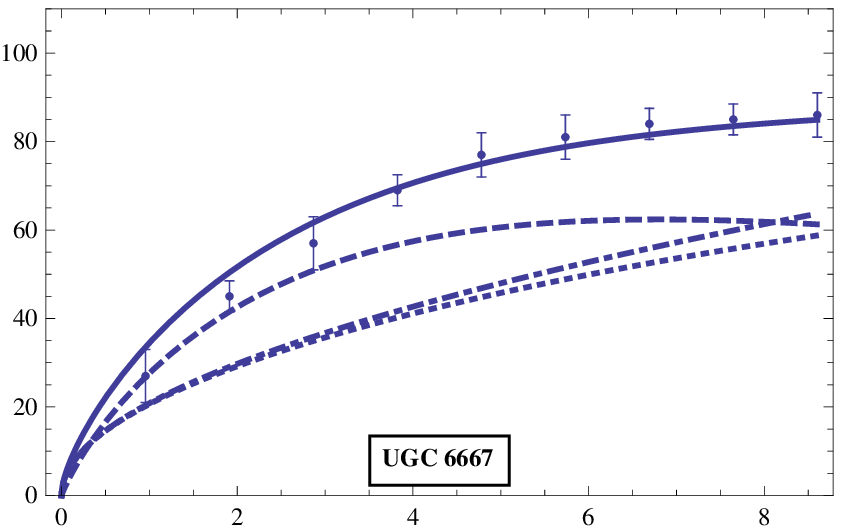,width=2.11in,height=1.2in}~~~
\epsfig{file=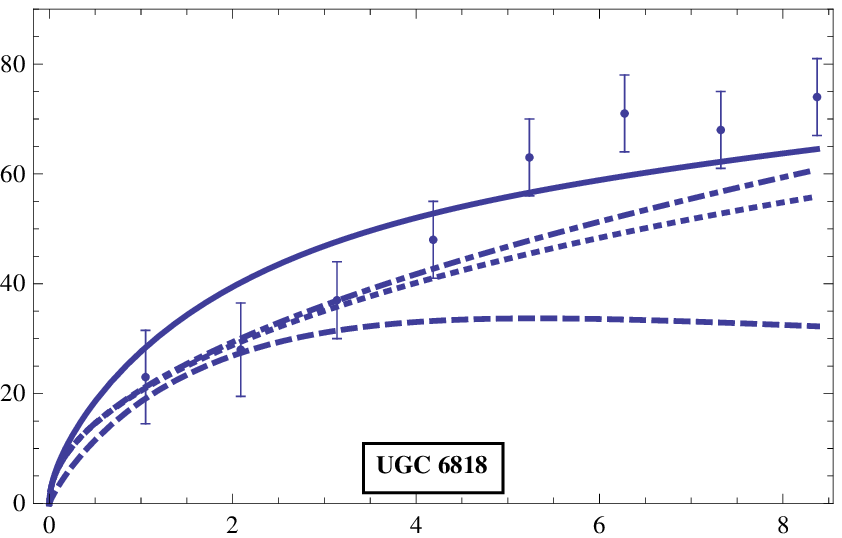,width=2.11in,height=1.2in}\\
\smallskip
\epsfig{file=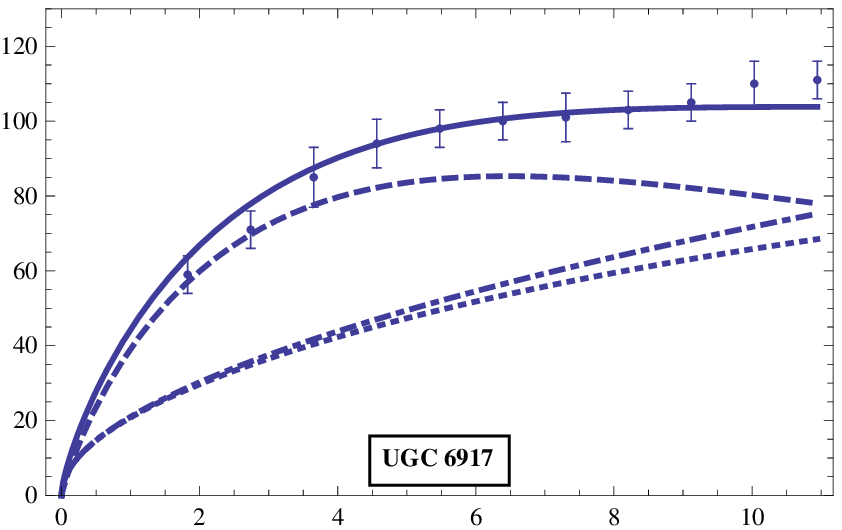,width=2.11in,height=1.2in}~~~
\epsfig{file=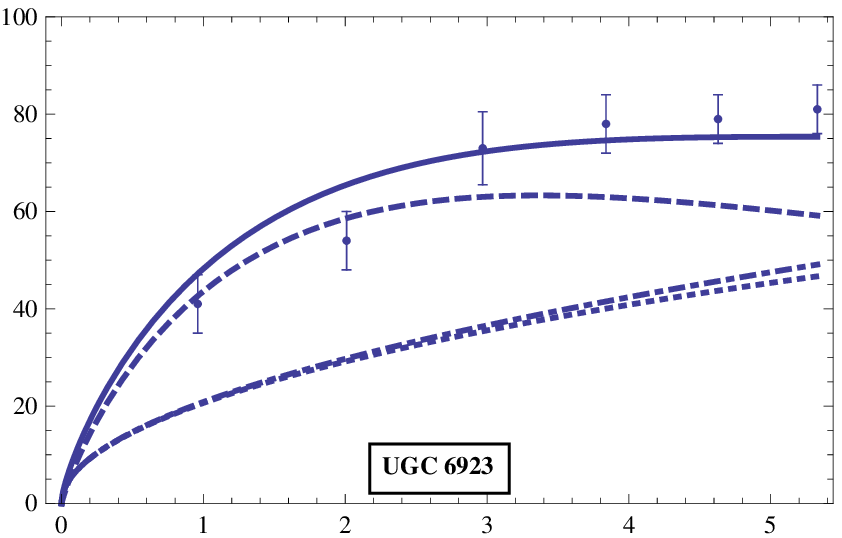,width=2.11in,height=1.2in}~~~
\epsfig{file=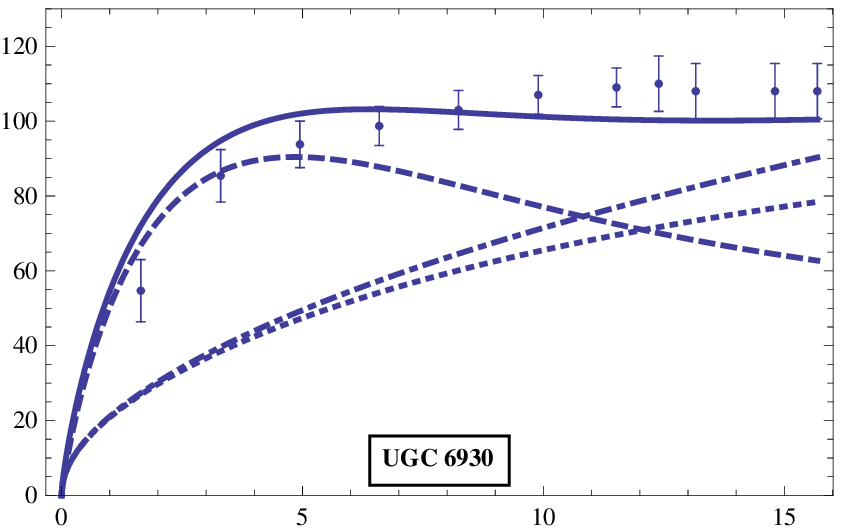,width=2.11in,height=1.2in}\\
\smallskip
\epsfig{file=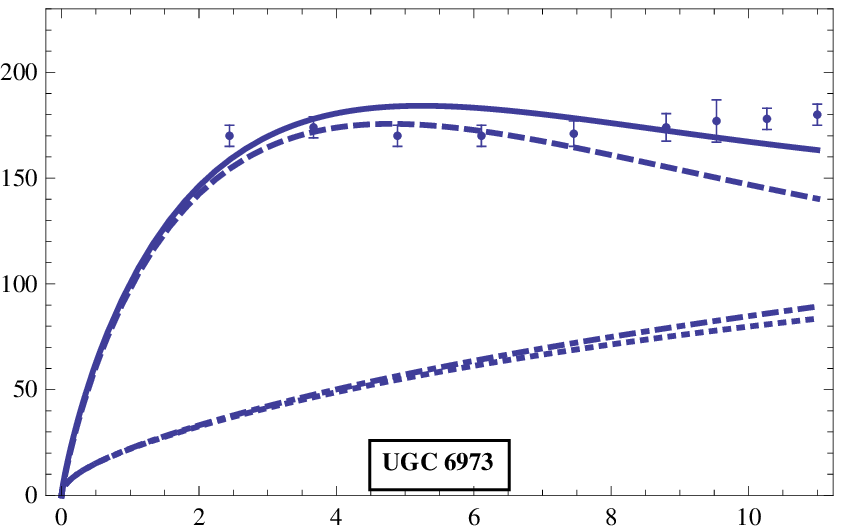,width=2.11in,height=1.2in}~~~
\epsfig{file=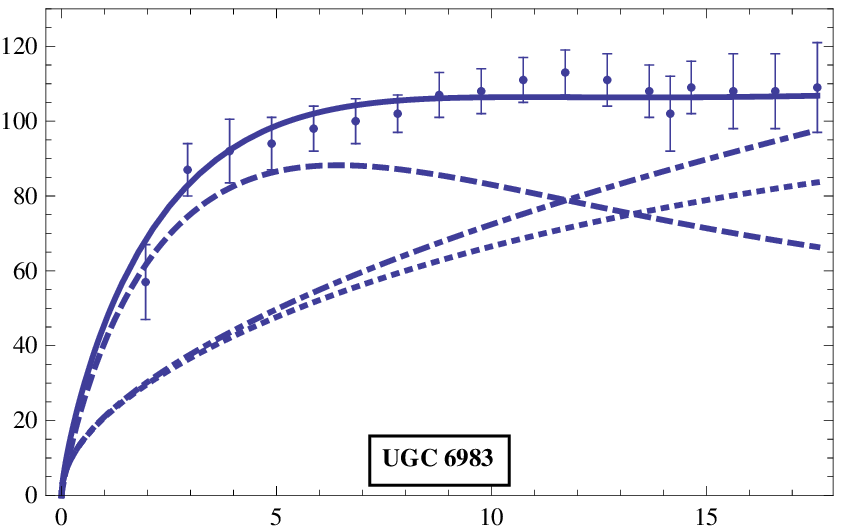,width=2.11in,height=1.2in}~~~
\epsfig{file=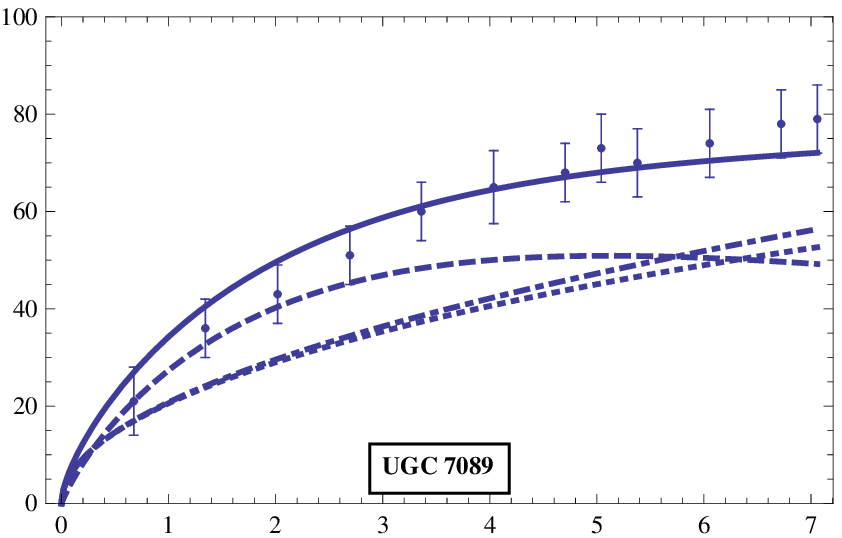,width=2.11in,height=1.2in}\\
\medskip
FIG.~2:~Fitting to the rotational velocities of the Ursa Major 30 galaxy sample -- Part 2
\label{Fig. 2.2}
\end{figure}

\begin{figure}[t]
\epsfig{file=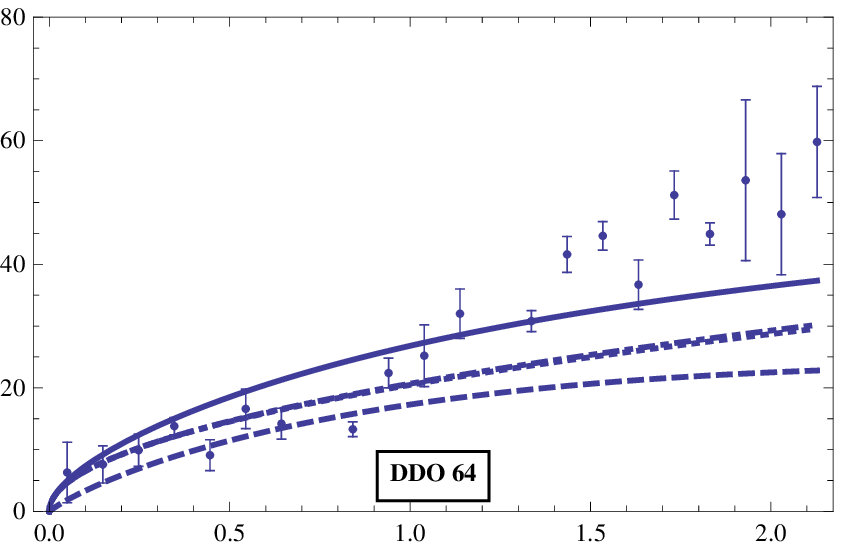,width=2.11in,height=1.2in}~~~
\epsfig{file=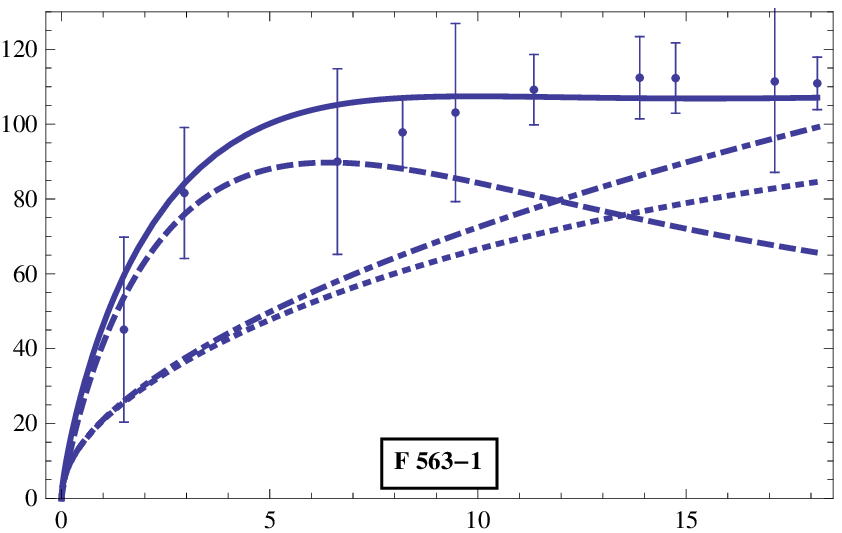,width=2.11in,height=1.2in}~~~
\epsfig{file=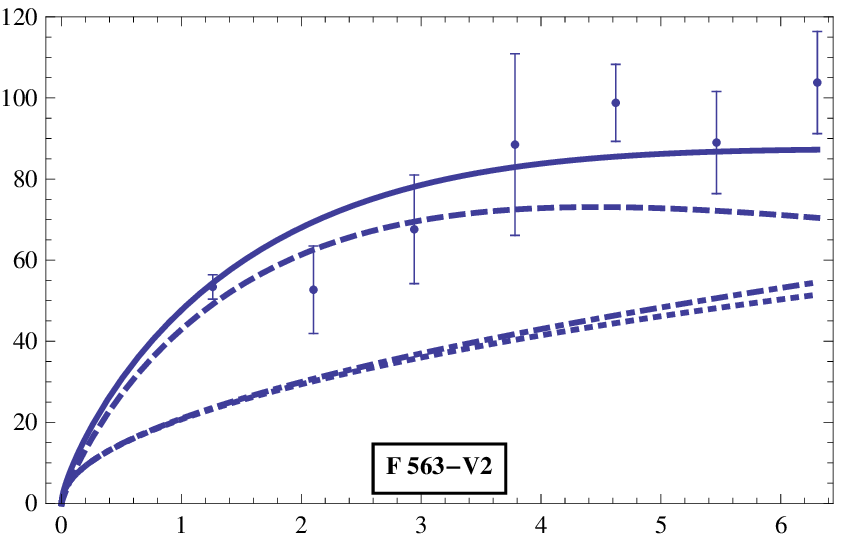,width=2.11in,height=1.2in}\\
\smallskip
\epsfig{file=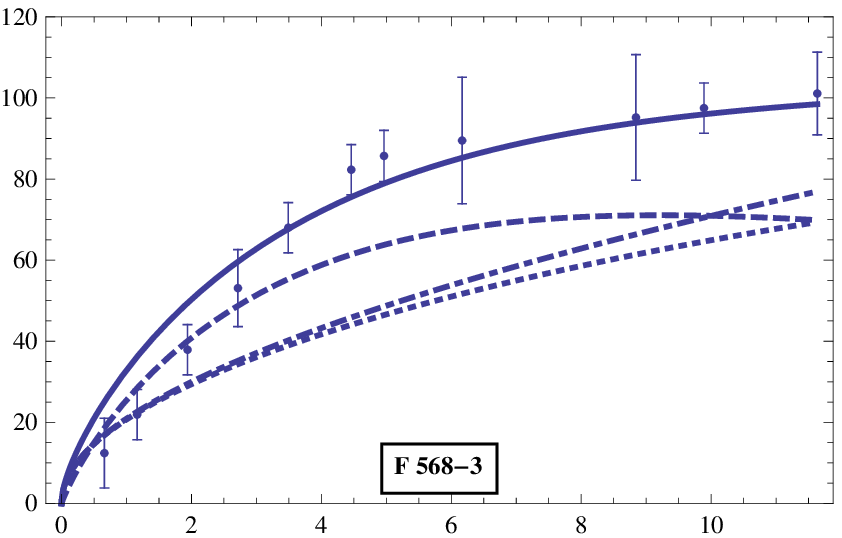,  width=2.11in,height=1.2in}~~~
\epsfig{file=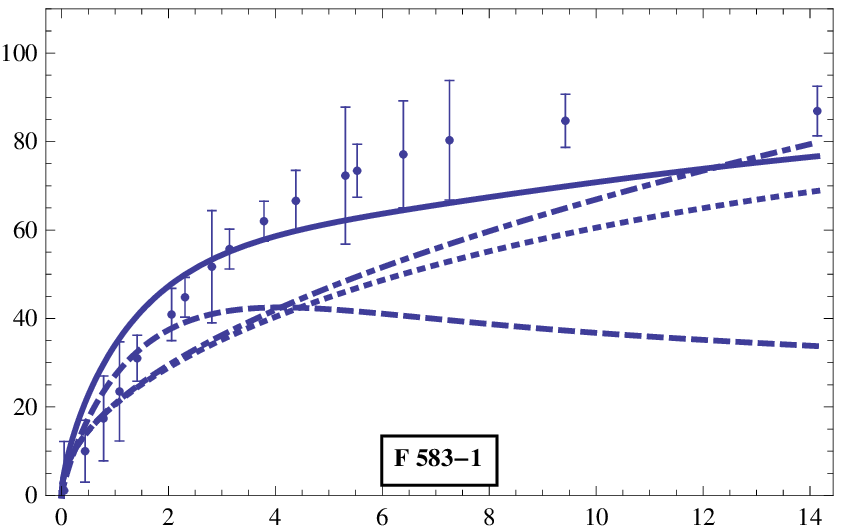,width=2.11in,height=1.2in}~~~
\epsfig{file=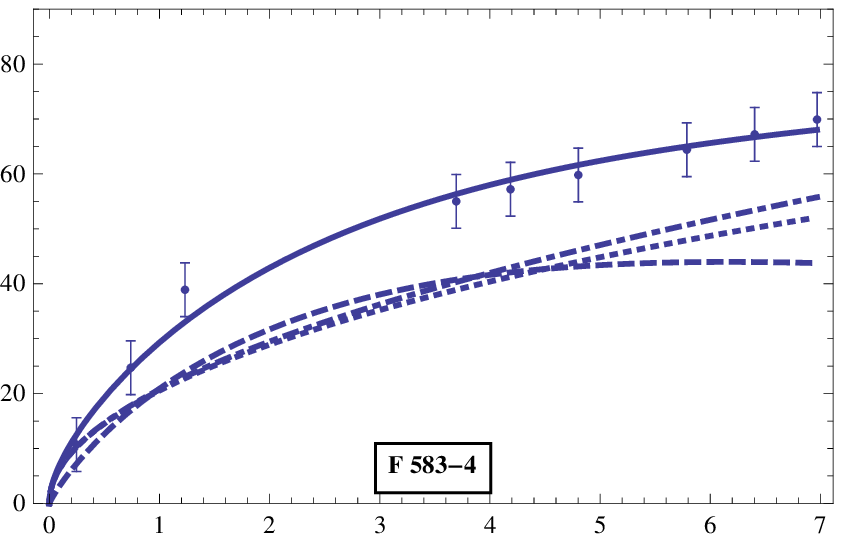,width=2.11in,height=1.2in}\\
\smallskip
\epsfig{file=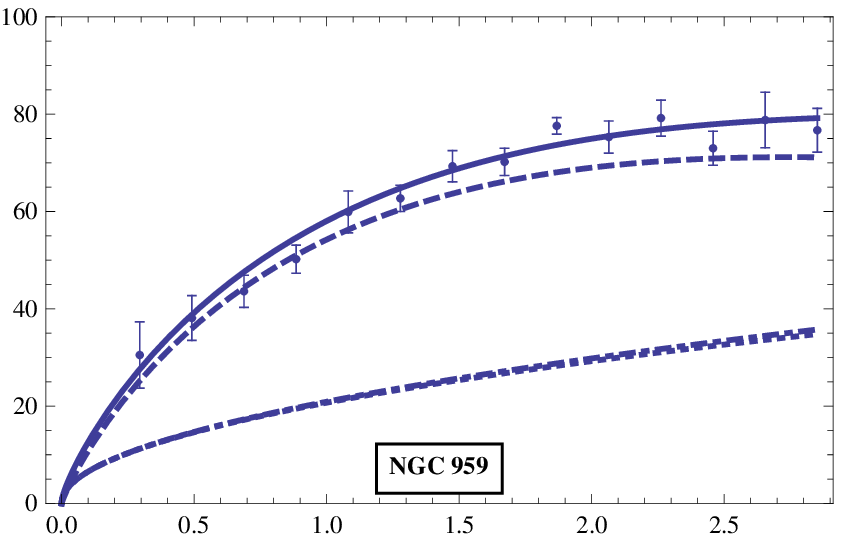,width=2.11in,height=1.2in}~~~
\epsfig{file=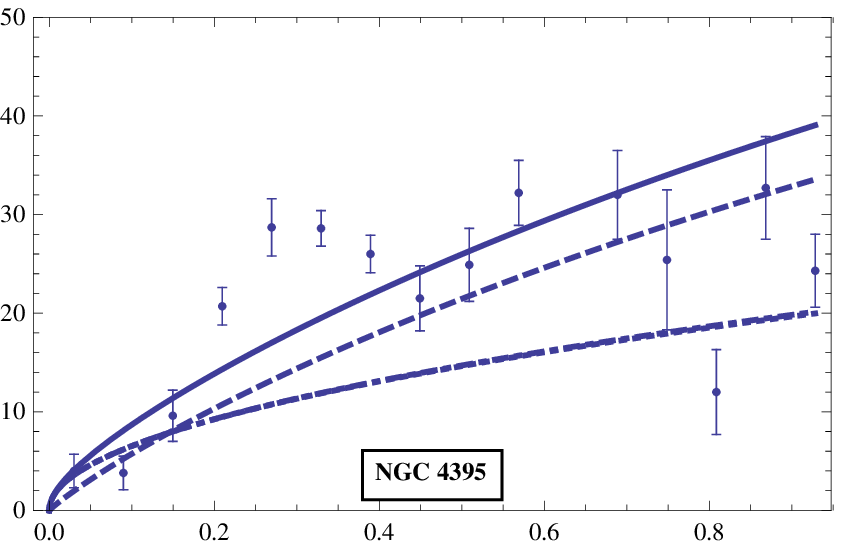,width=2.11in,height=1.2in}~~~
\epsfig{file=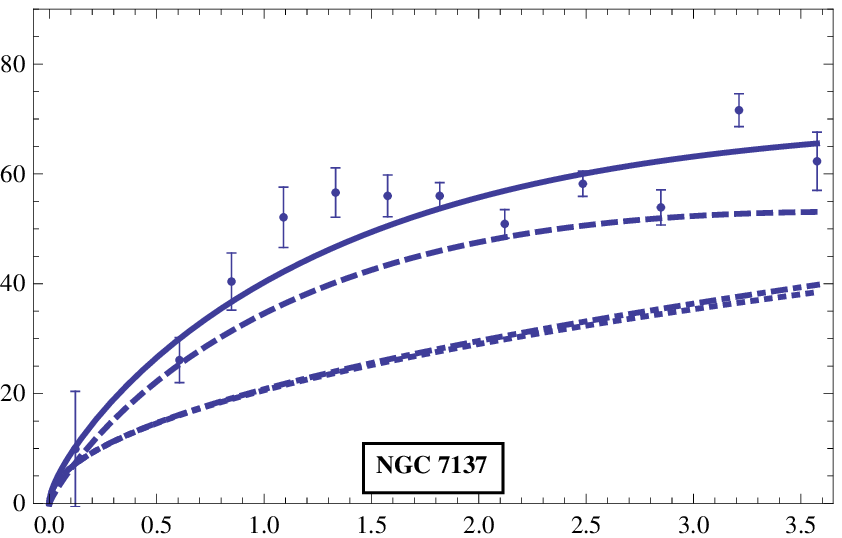,width=2.11in,height=1.2in}\\
\smallskip
\epsfig{file=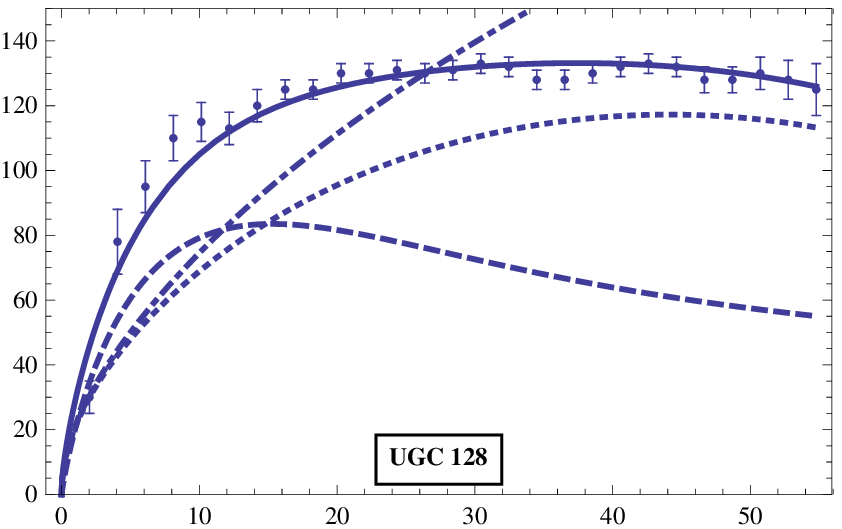,width=2.11in,height=1.2in}~~~
\epsfig{file=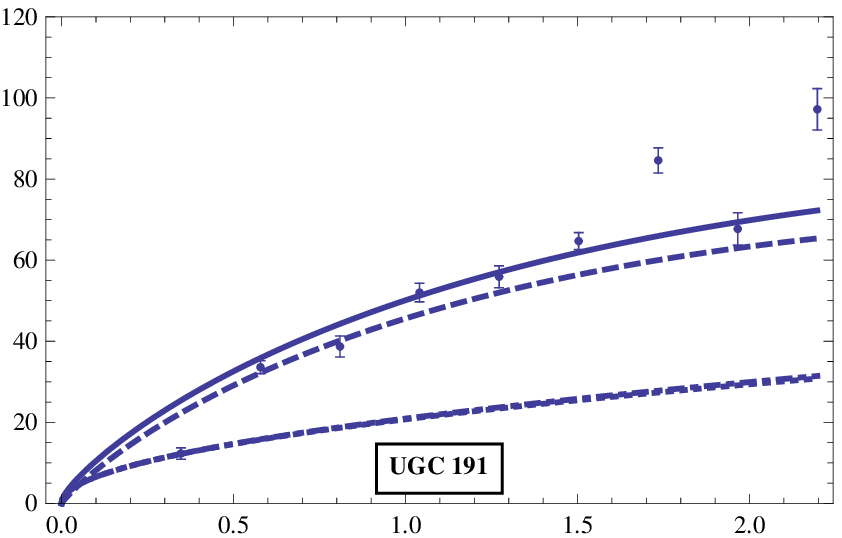,width=2.11in,height=1.2in}~~~
\epsfig{file=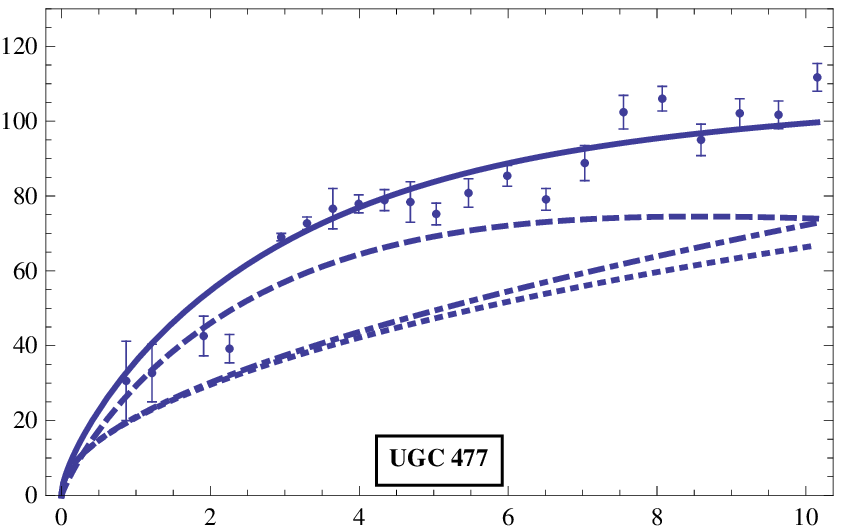,width=2.11in,height=1.2in}\\
\smallskip
\epsfig{file=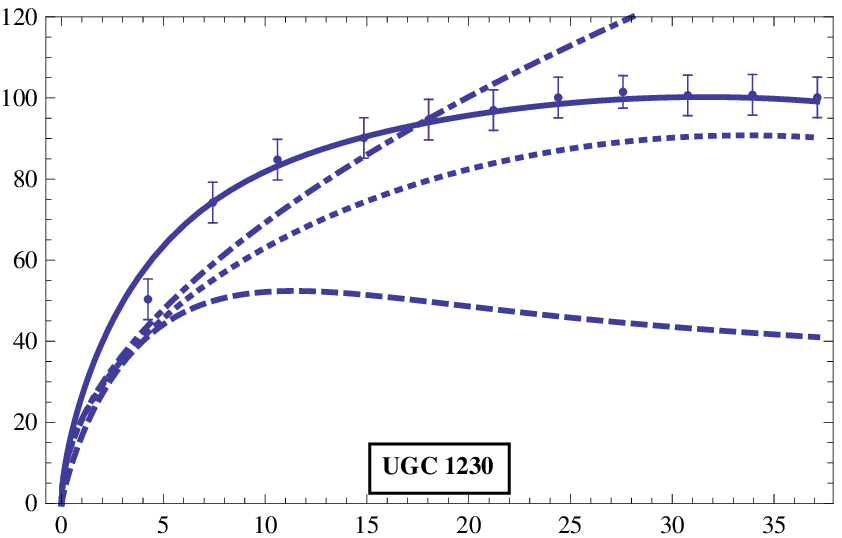,width=2.11in,height=1.2in}~~~
\epsfig{file=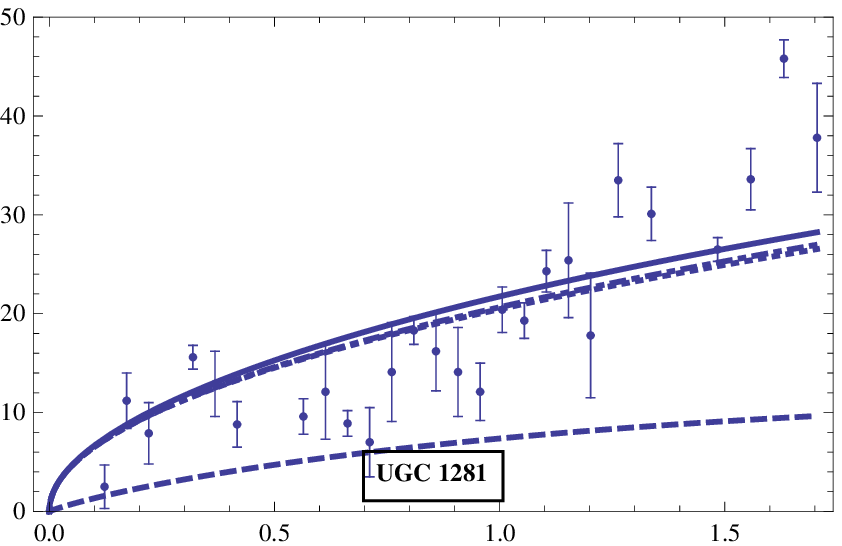,width=2.11in,height=1.2in}~~~
\epsfig{file=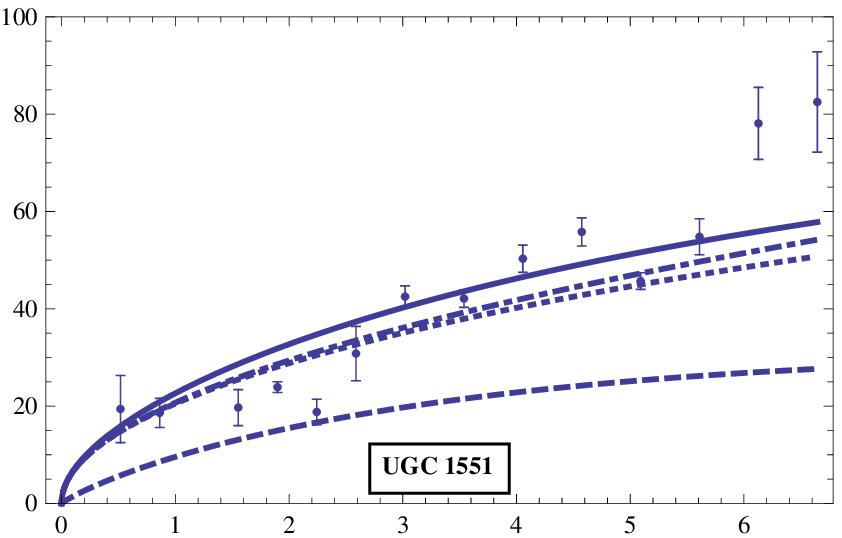,width=2.11in,height=1.2in}\\
\smallskip
\epsfig{file=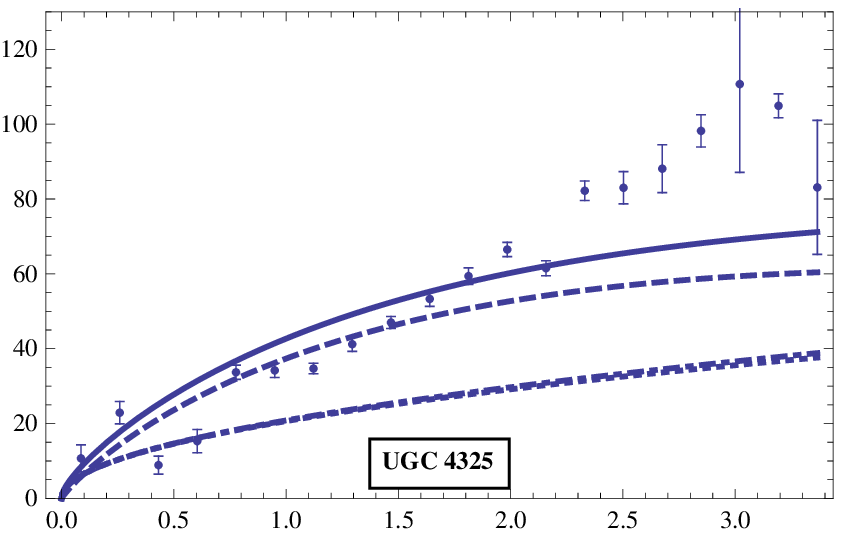,width=2.11in,height=1.2in}~~~
\epsfig{file=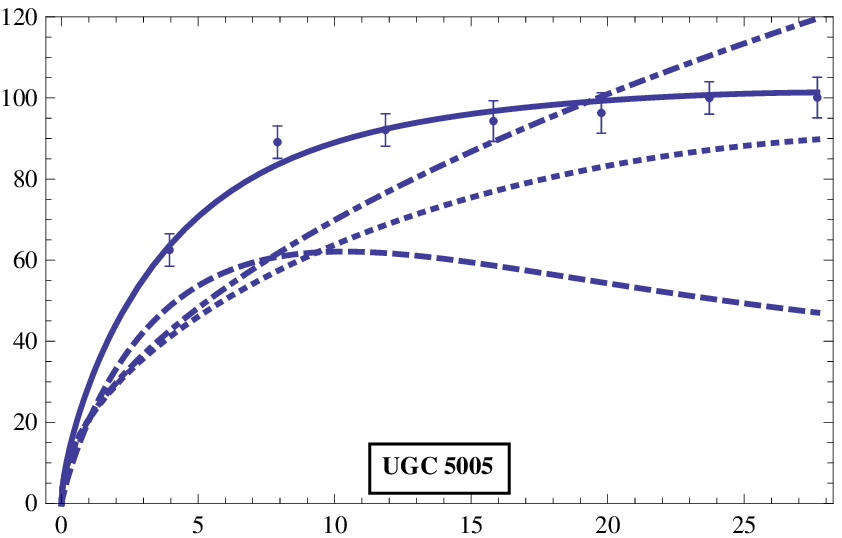,width=2.11in,height=1.2in}~~~
\epsfig{file=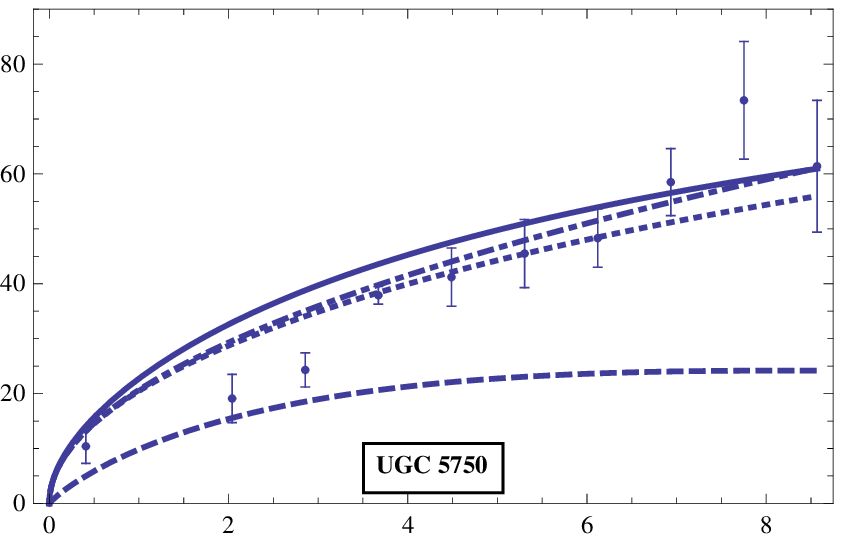,width=2.11in,height=1.2in}\\
\smallskip
\epsfig{file=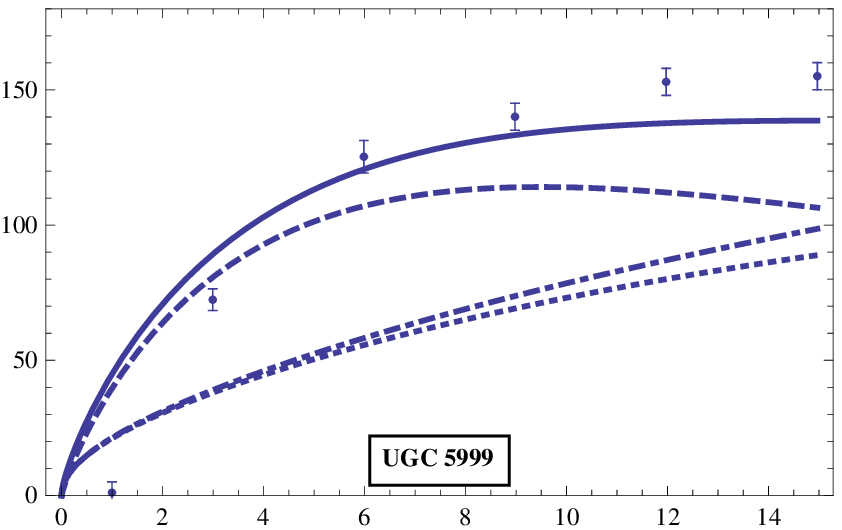,width=2.11in,height=1.2in}~~~
\epsfig{file=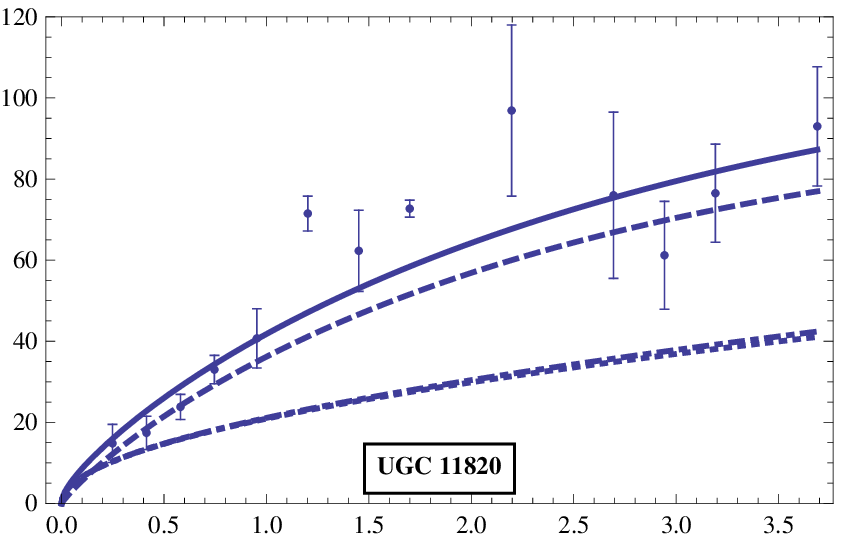,width=2.11in,height=1.2in}\\
\medskip
FIG.~3:~Fitting to the rotational velocities of the LSB 20 galaxy sample 
\label{Fig. 3}
\end{figure}

\begin{figure}[t]
\epsfig{file=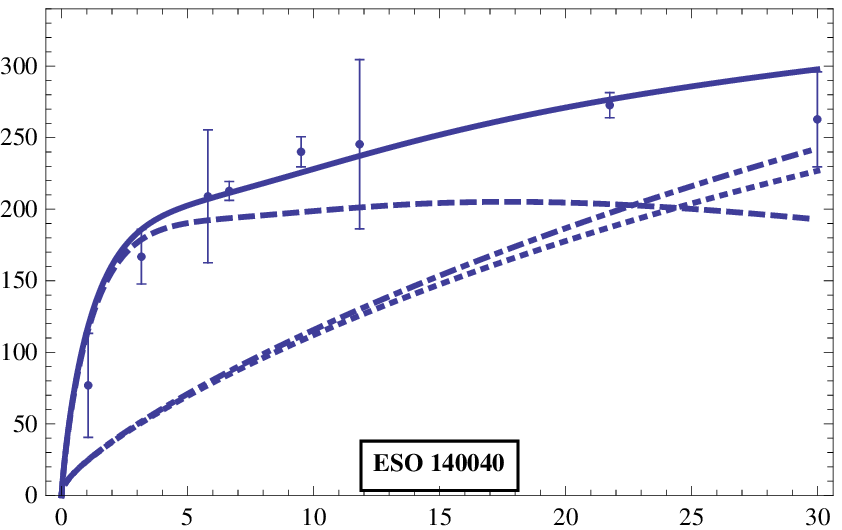,width=2.11in,height=1.2in}~~~
\epsfig{file=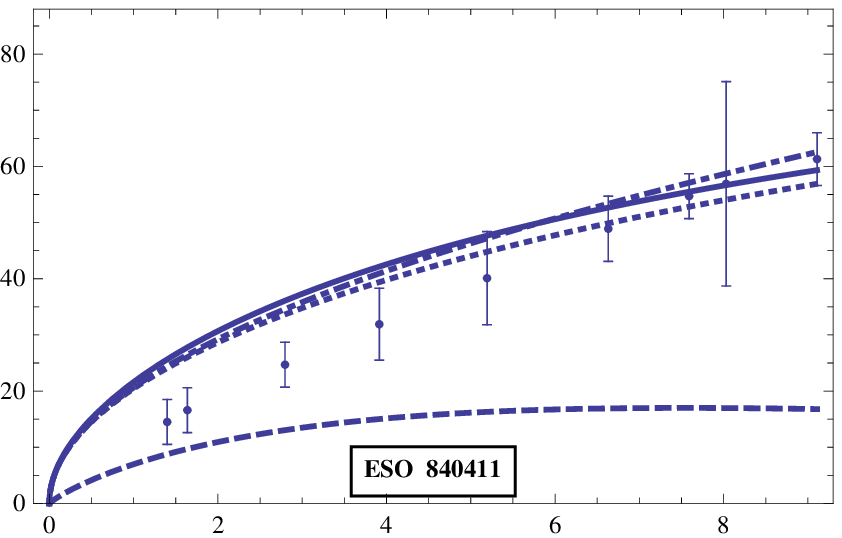,width=2.11in,height=1.2in}~~~
\epsfig{file=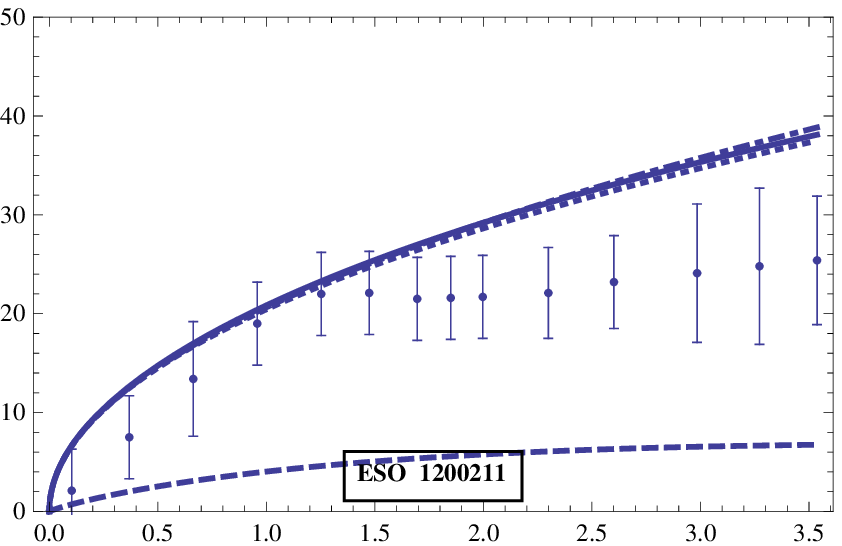,  width=2.11in,height=1.2in}\\
\smallskip
\epsfig{file=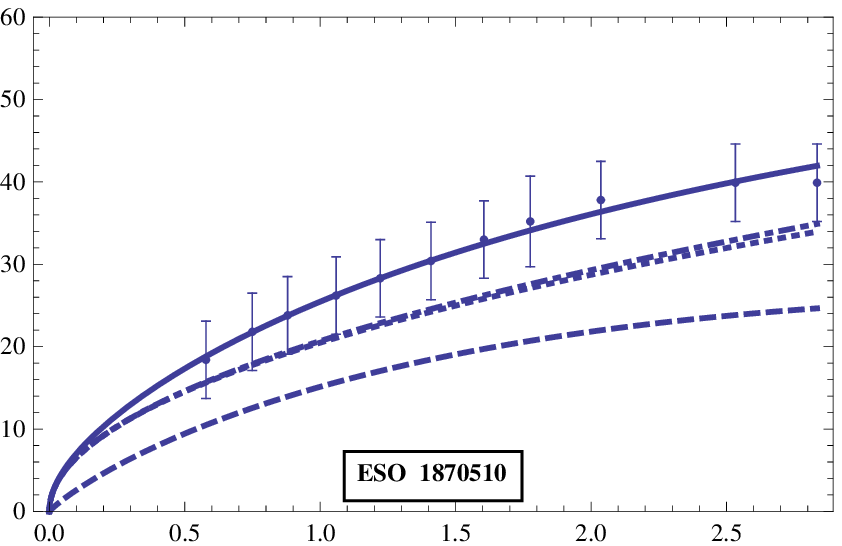,width=2.11in,height=1.2in}~~~
\epsfig{file=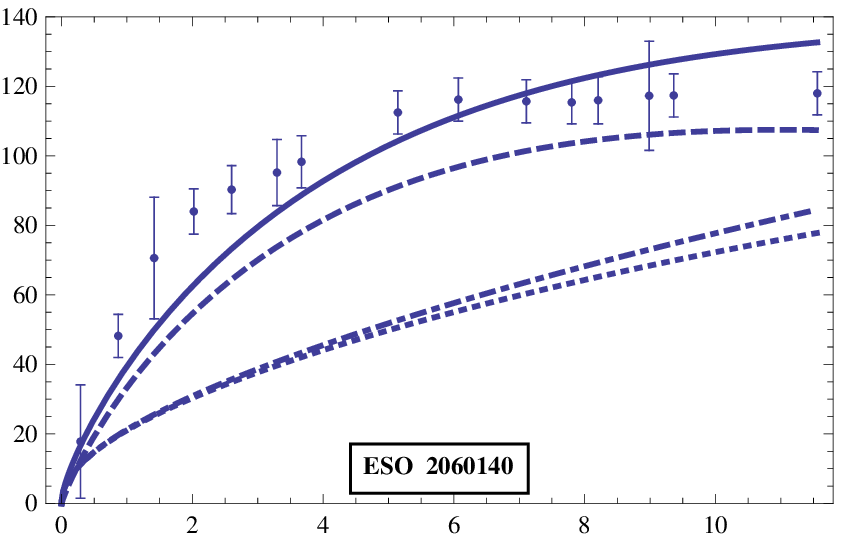,width=2.11in,height=1.2in}~~~
\epsfig{file=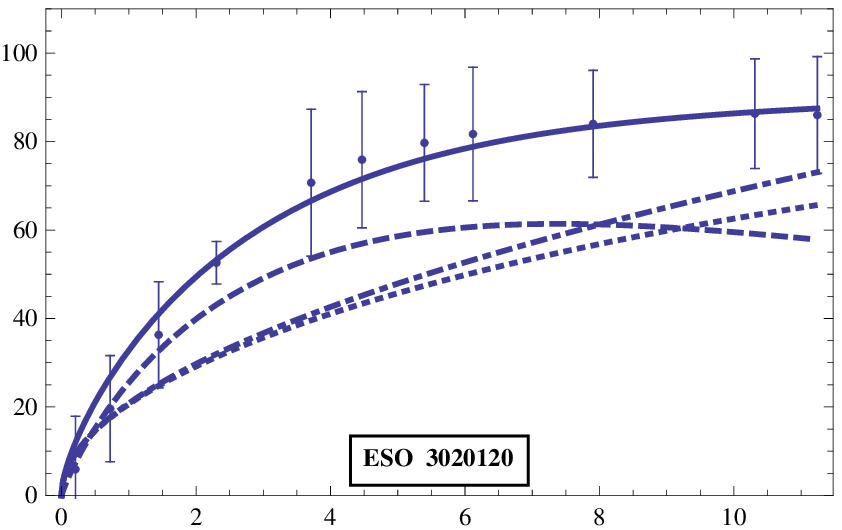,width=2.11in,height=1.2in}\\
\smallskip
\epsfig{file=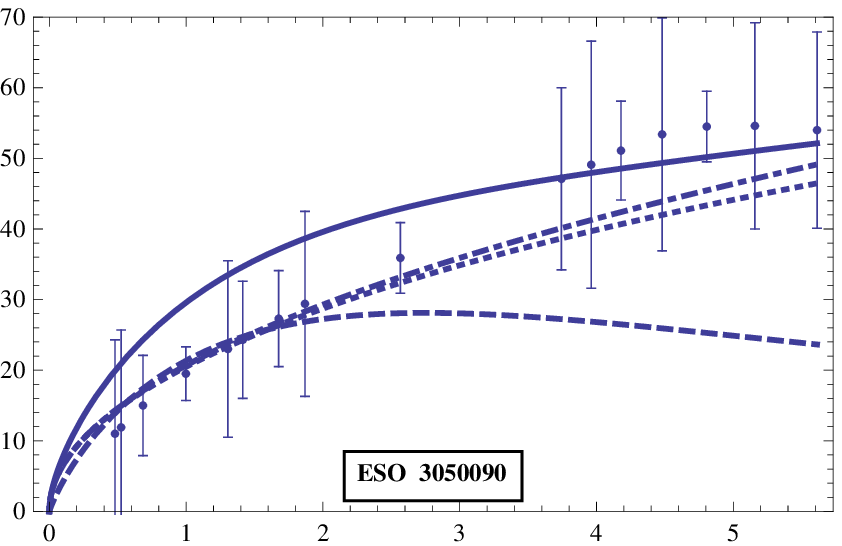,width=2.11in,height=1.2in}~~~
\epsfig{file=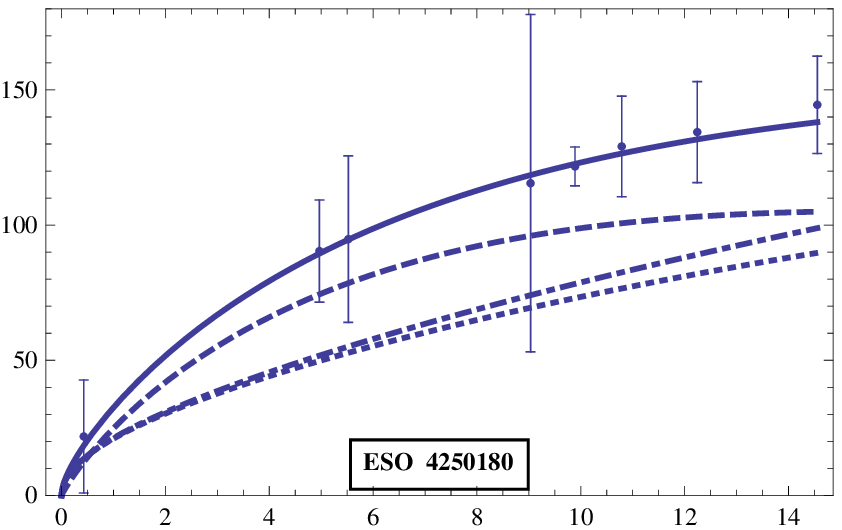,width=2.11in,height=1.2in}~~~
\epsfig{file=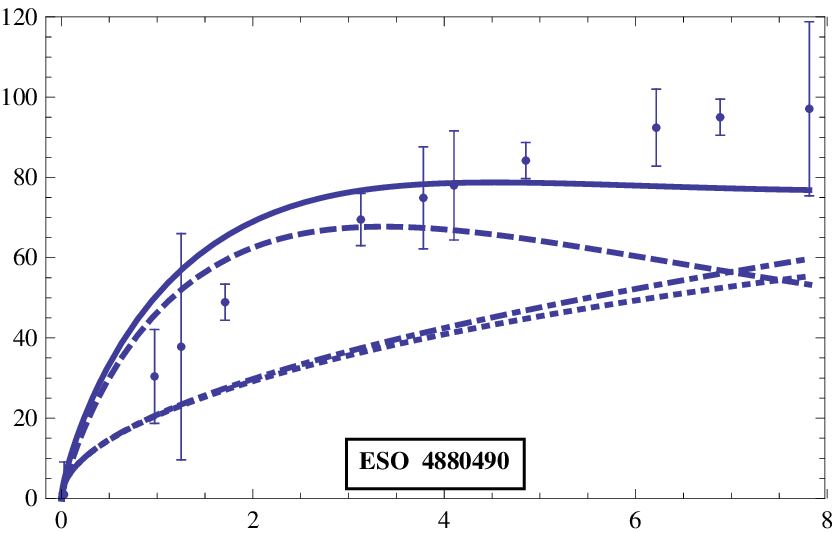,width=2.11in,height=1.2in}\\
\smallskip
\epsfig{file=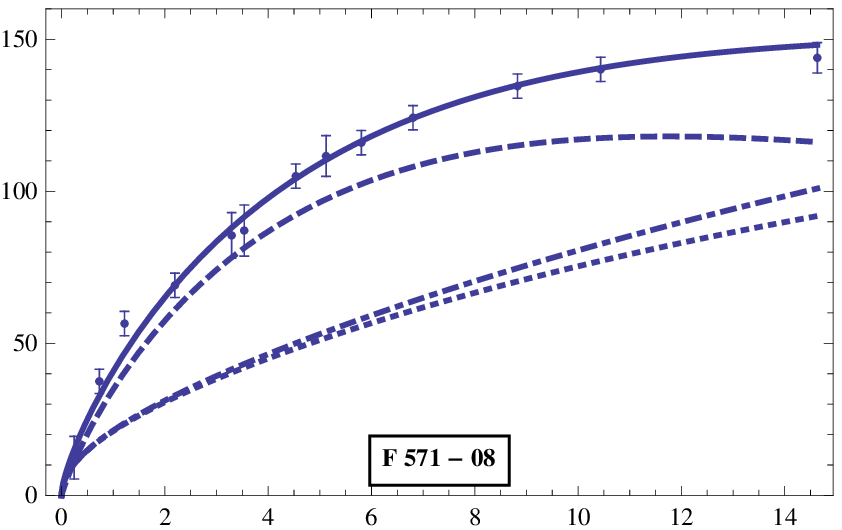,width=2.11in,height=1.2in}~~~
\epsfig{file=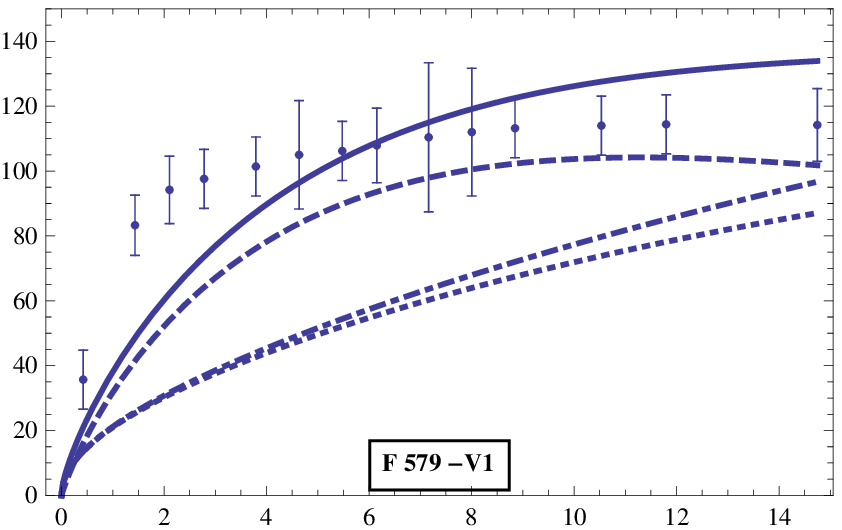,width=2.11in,height=1.2in}~~~
\epsfig{file=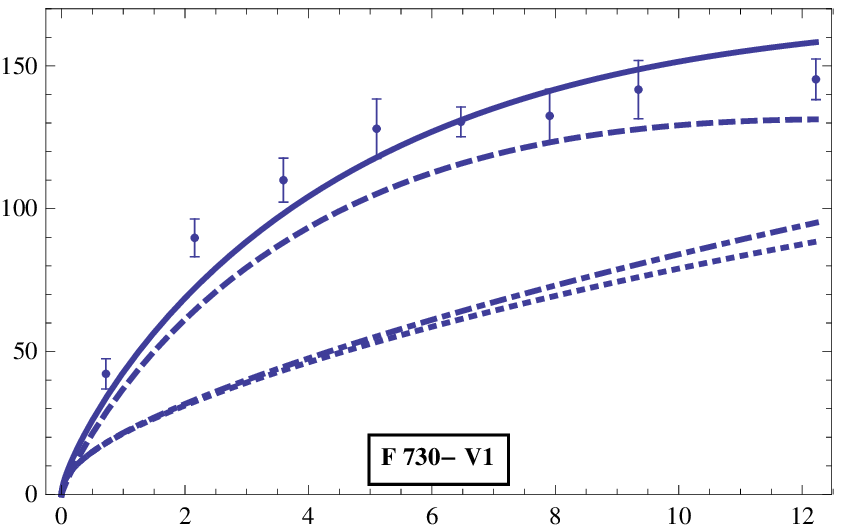,width=2.11in,height=1.2in}\\
\smallskip
\epsfig{file=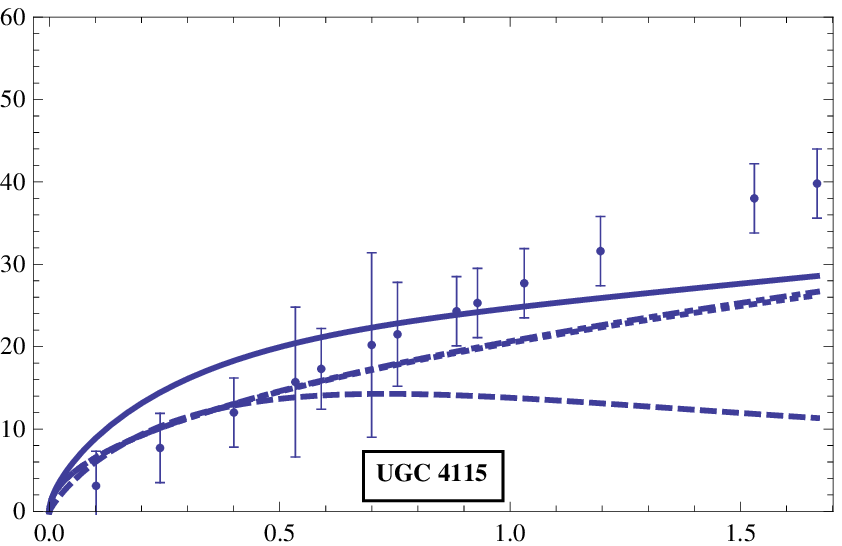,width=2.11in,height=1.2in}~~~
\epsfig{file=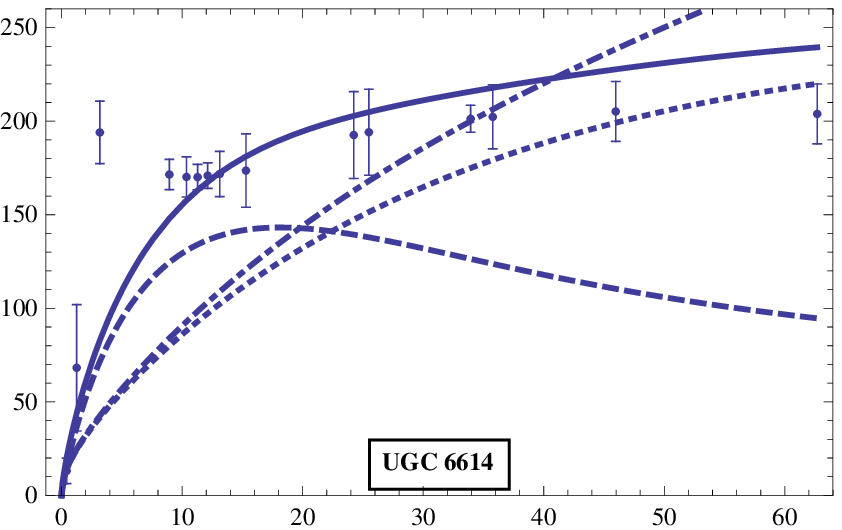,width=2.11in,height=1.2in}~~~
\epsfig{file=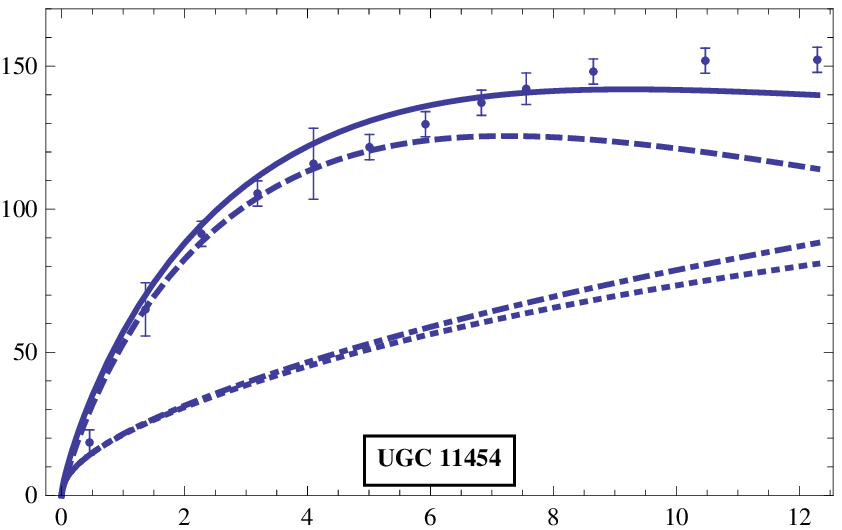,width=2.11in,height=1.2in}\\
\smallskip
\epsfig{file=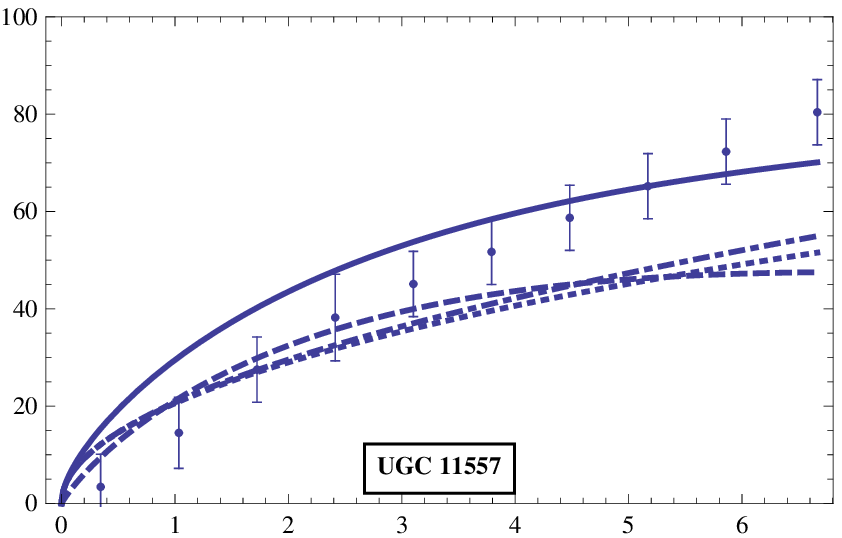,width=2.11in,height=1.2in}~~~
\epsfig{file=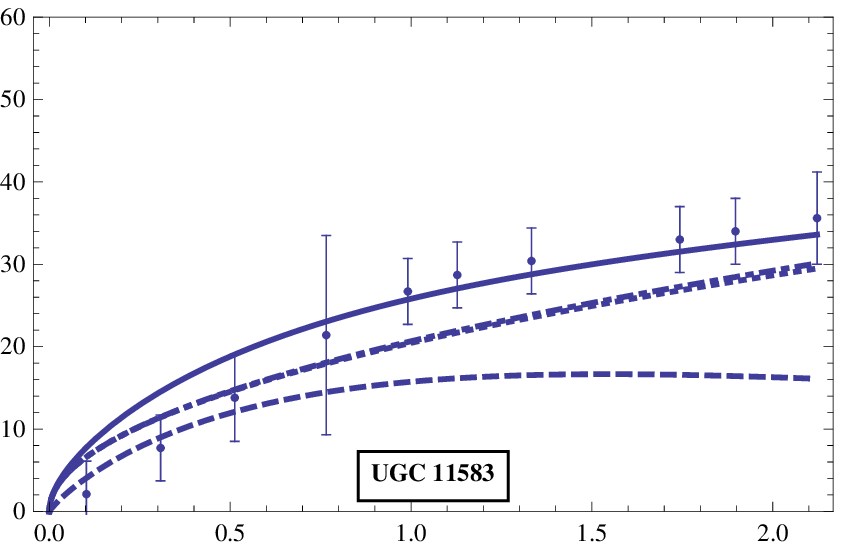,  width=2.11in,height=1.2in}~~~
\epsfig{file=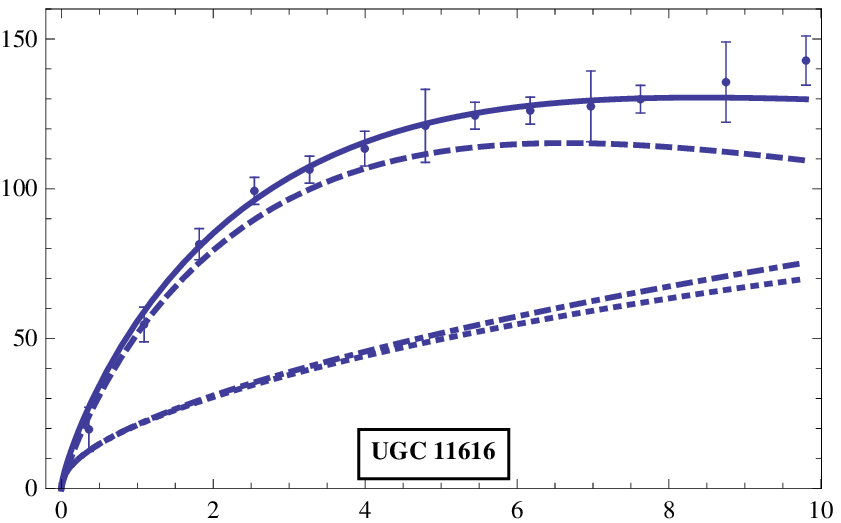,width=2.11in,height=1.2in}\\
\smallskip
\epsfig{file=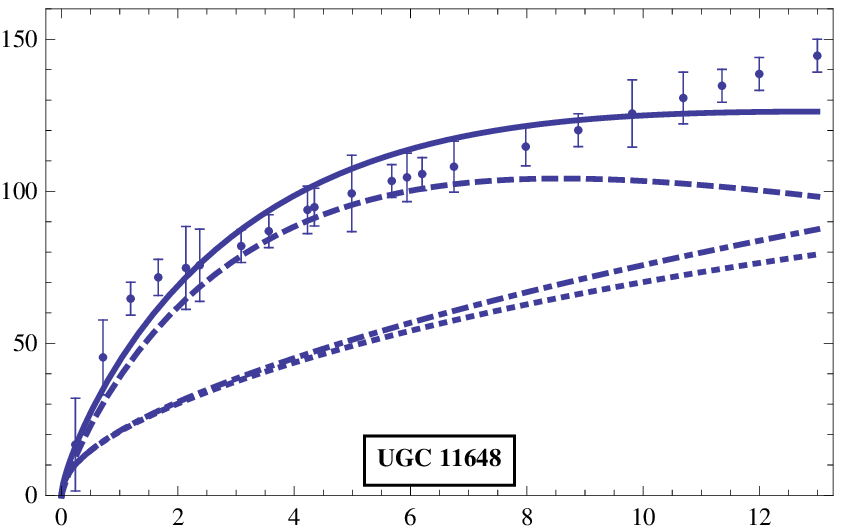,width=2.11in,height=1.2in}~~~
\epsfig{file=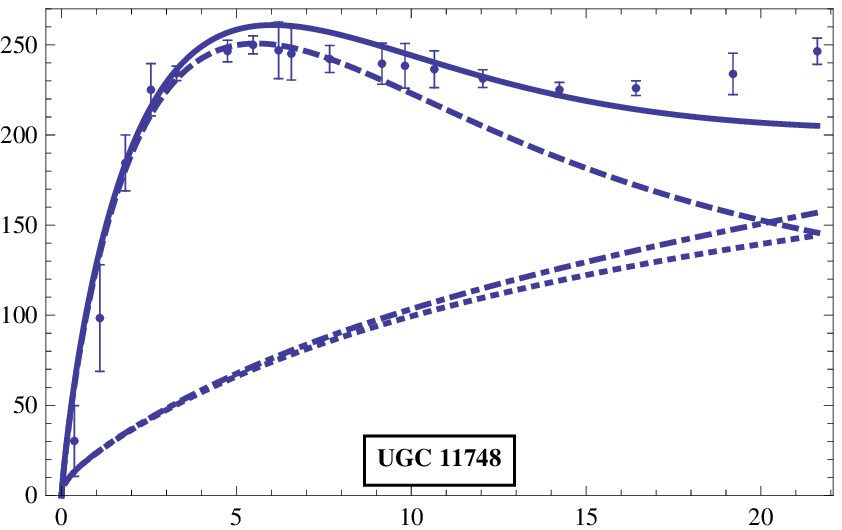,width=2.11in,height=1.2in}~~~
\epsfig{file=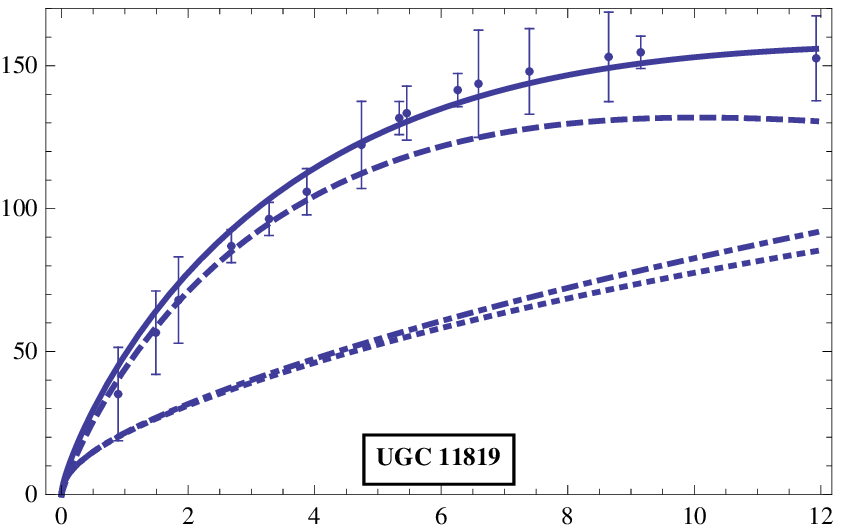,width=2.11in,height=1.2in}\\
\medskip
FIG.~4:~Fitting to the rotational velocities of the LSB 21 galaxy sample 
\label{Fig. 4}
\end{figure}

\begin{figure}[t]
\epsfig{file=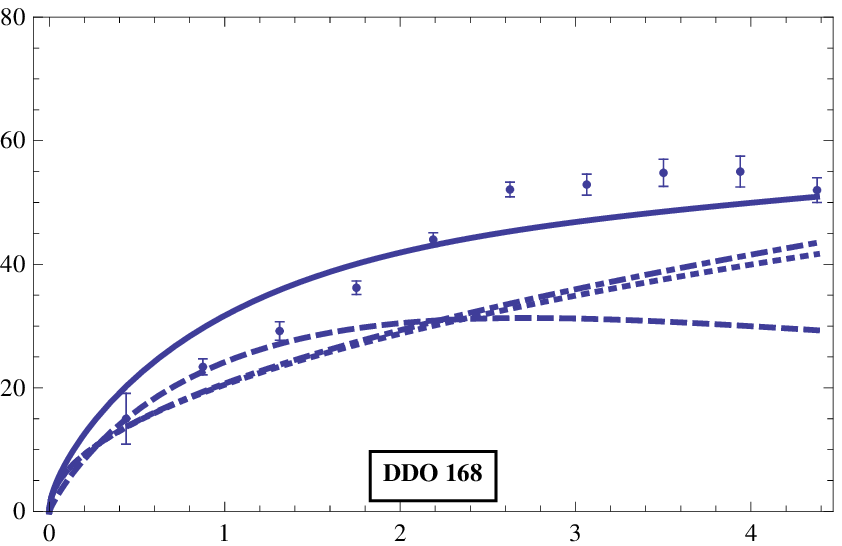,width=2.11in,height=1.2in}~~~
\epsfig{file=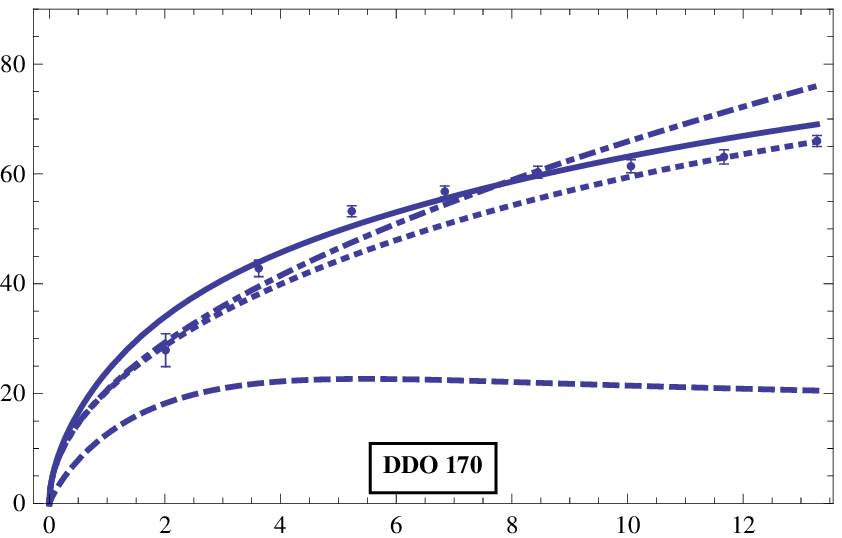,width=2.11in,height=1.2in}~~~
\epsfig{file=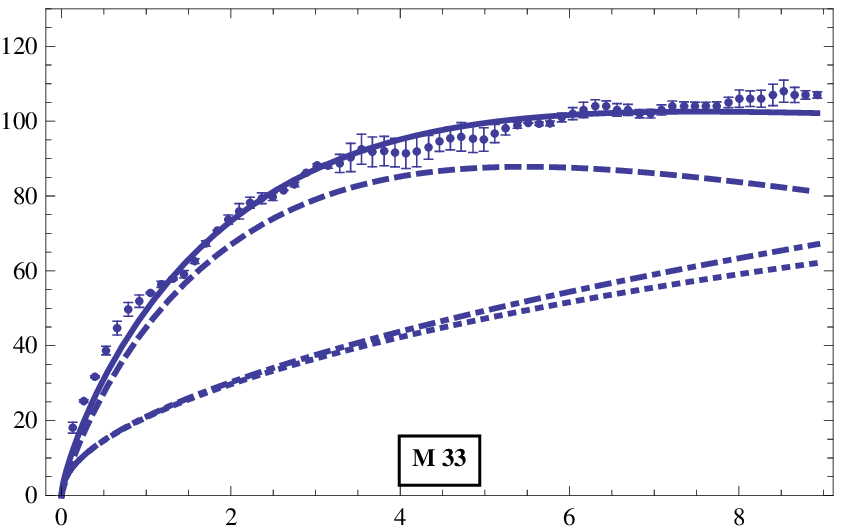,width=2.11in,height=1.2in}\\
\smallskip
\epsfig{file=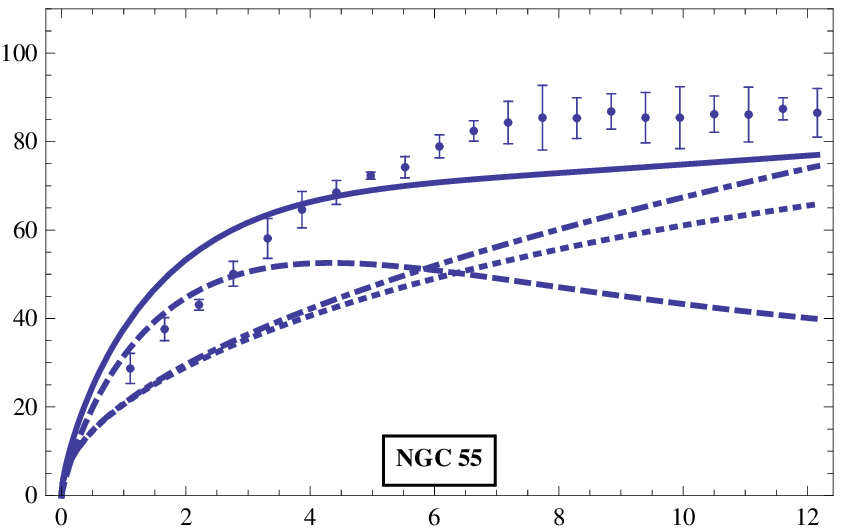,width=2.11in,height=1.2in}~~~
\epsfig{file=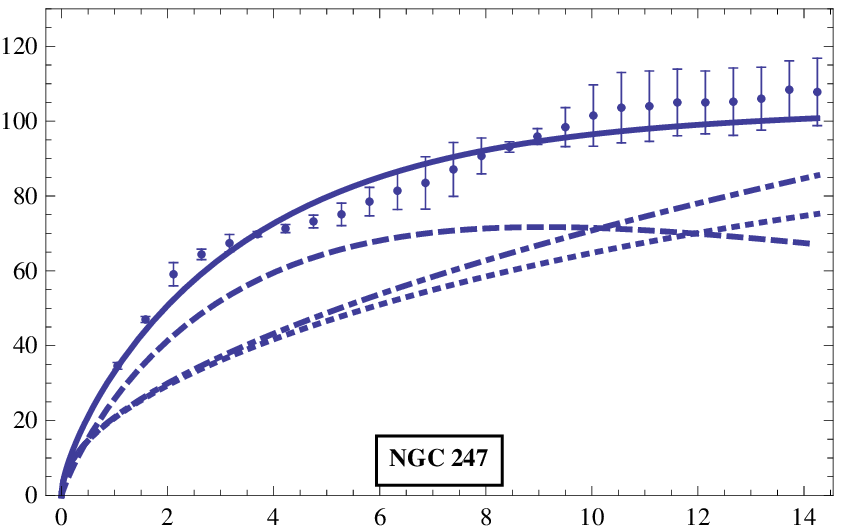,width=2.11in,height=1.2in}~~~
\epsfig{file=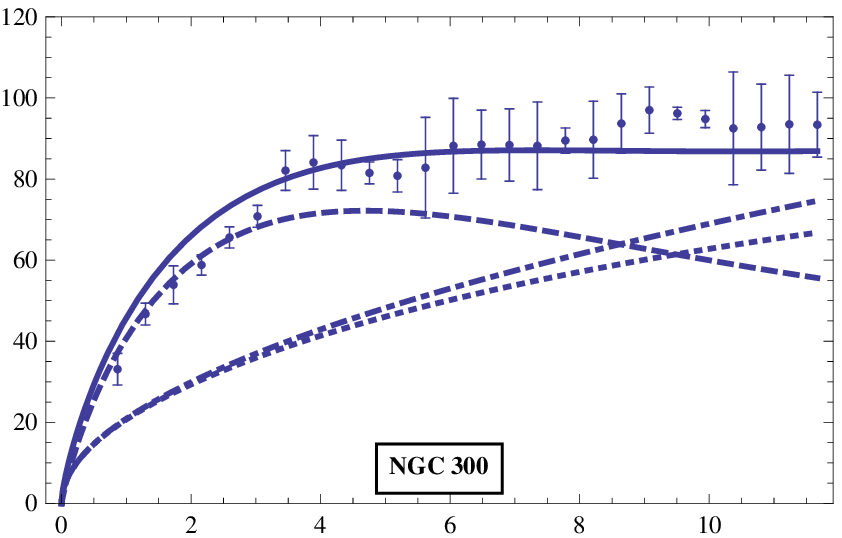,width=2.11in,height=1.2in}\\
\smallskip
\epsfig{file=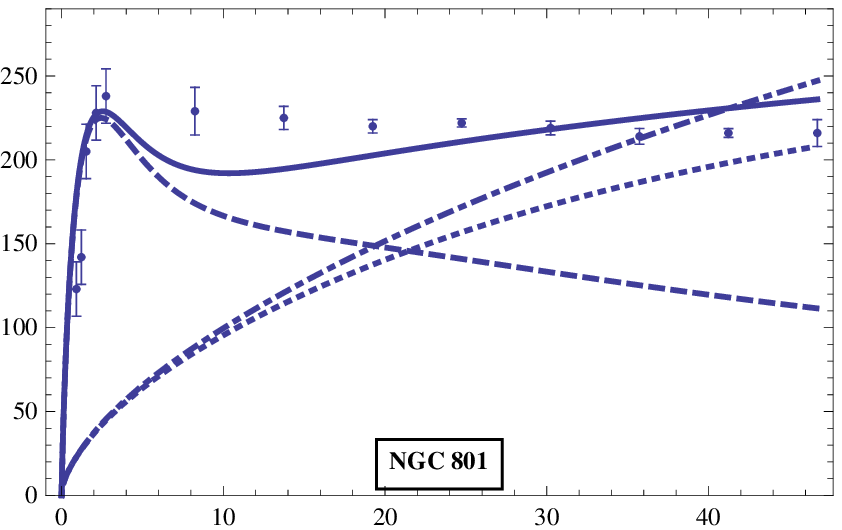,width=2.11in,height=1.2in}~~~
\epsfig{file=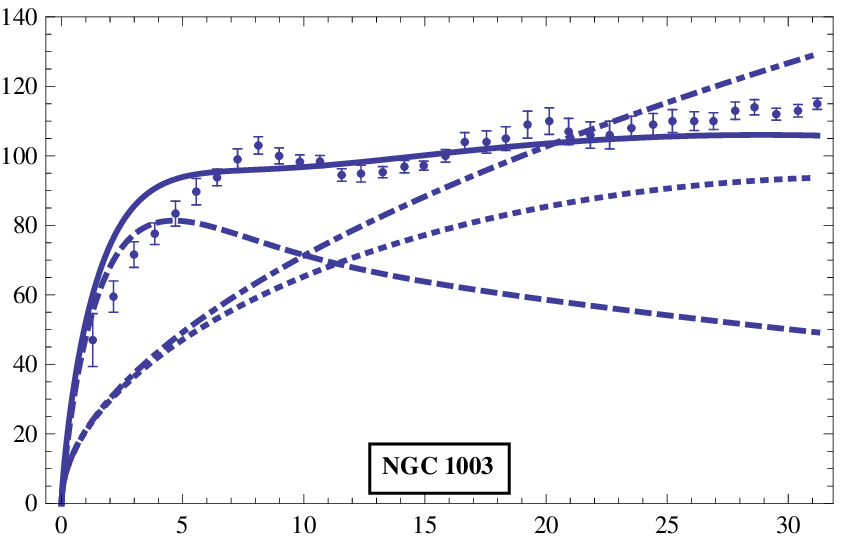,width=2.11in,height=1.2in}~~
\epsfig{file=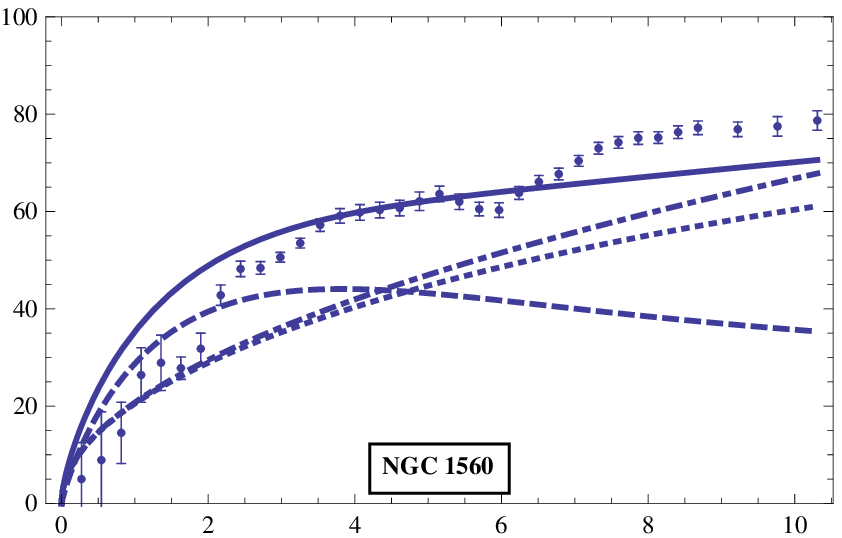,width=2.11in,height=1.2in}\\
\smallskip
\epsfig{file=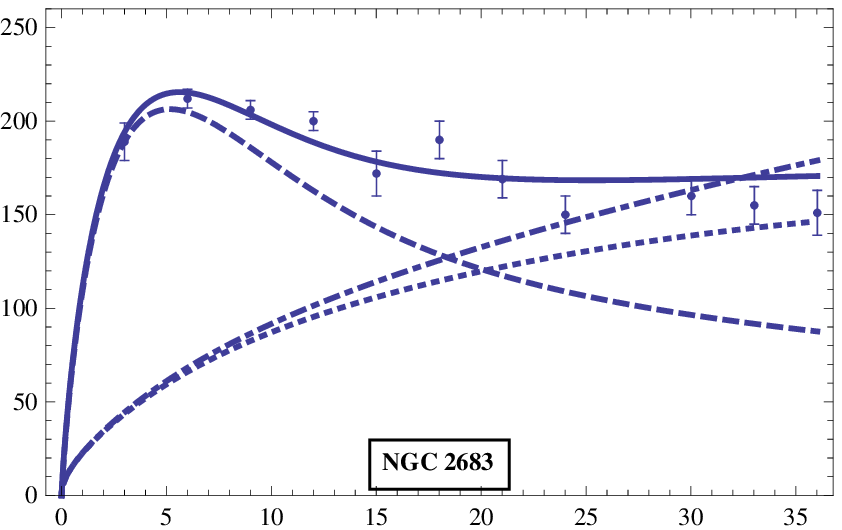,width=2.11in,height=1.2in}~~~
\epsfig{file=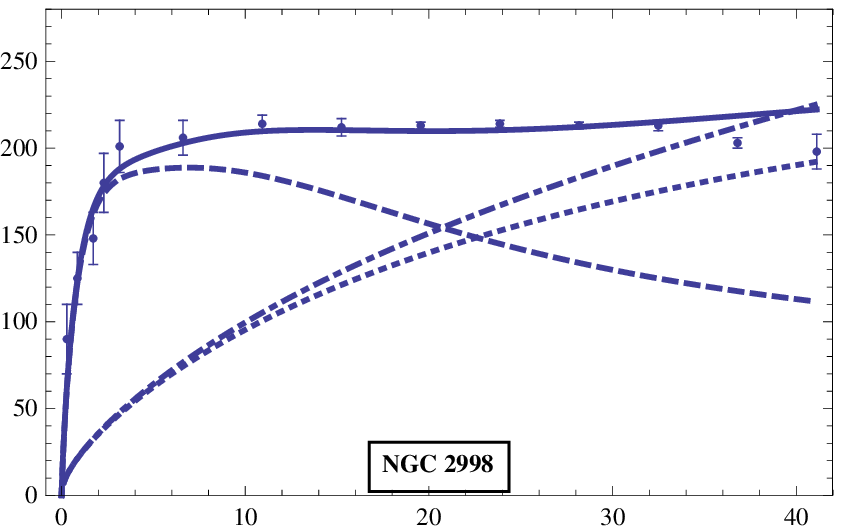,  width=2.11in,height=1.2in}~~~
\epsfig{file=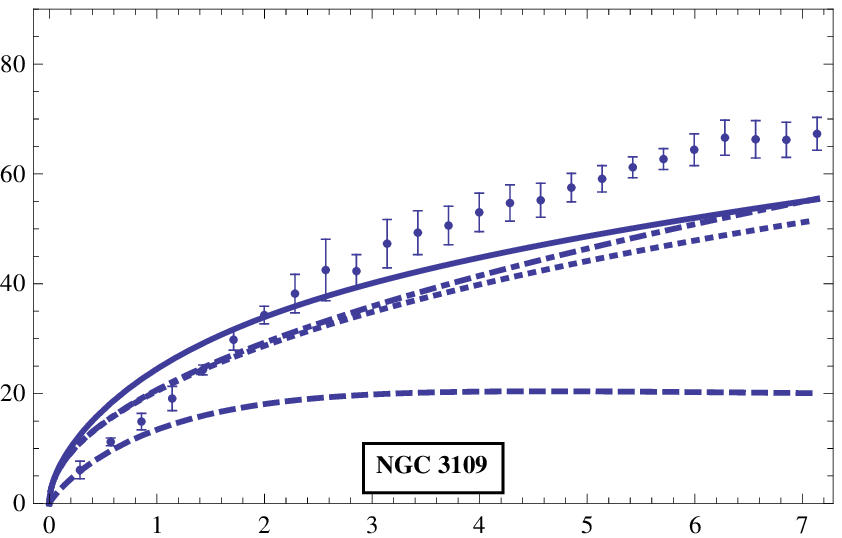,width=2.11in,height=1.2in}\\
\smallskip
\epsfig{file=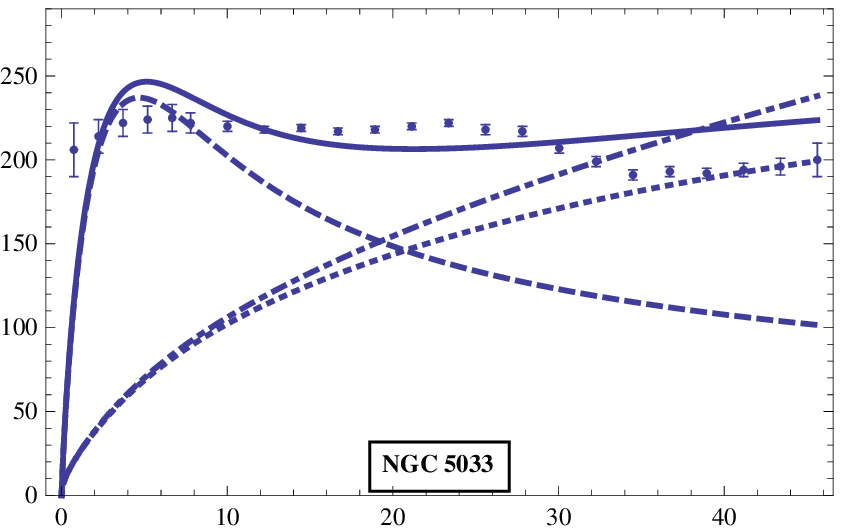,width=2.11in,height=1.2in}~~~
\epsfig{file=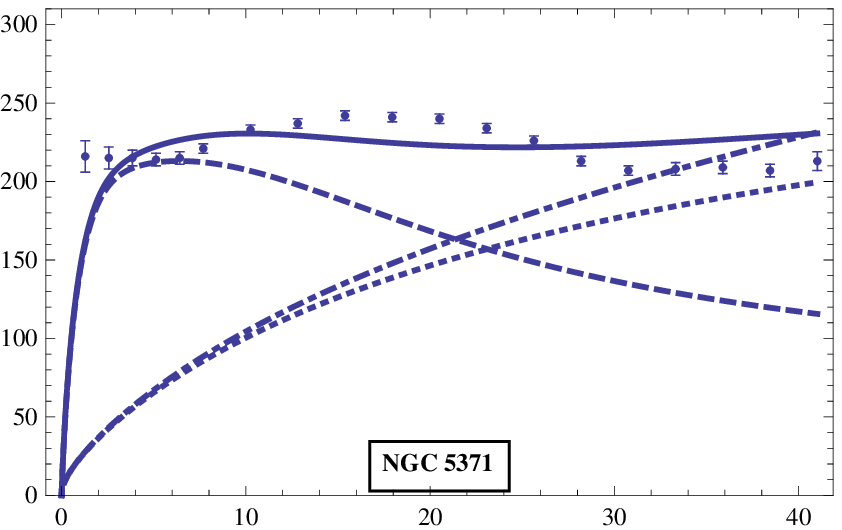,width=2.11in,height=1.2in}~~~
\epsfig{file=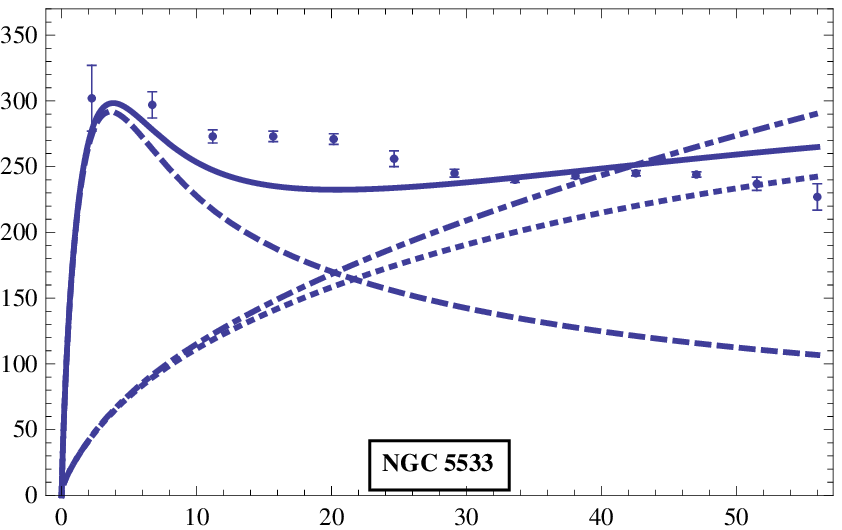,width=2.11in,height=1.2in}\\
\smallskip
\epsfig{file=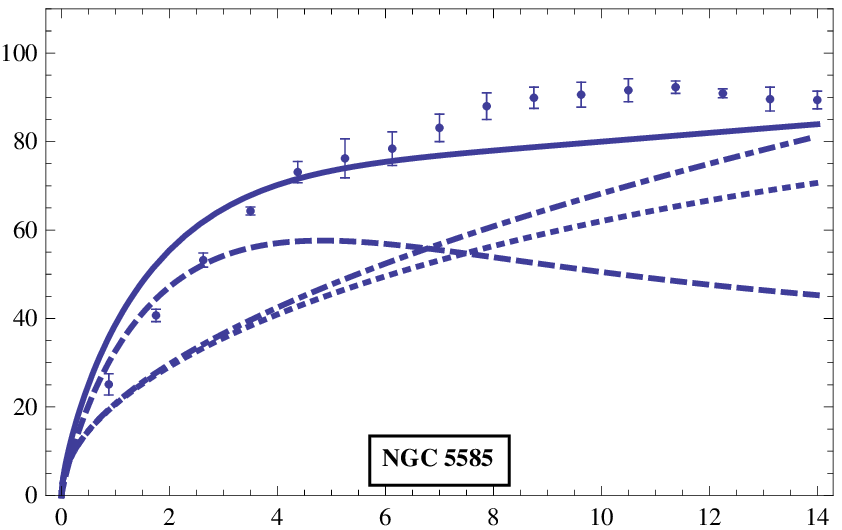,width=2.11in,height=1.2in}~~~
\epsfig{file=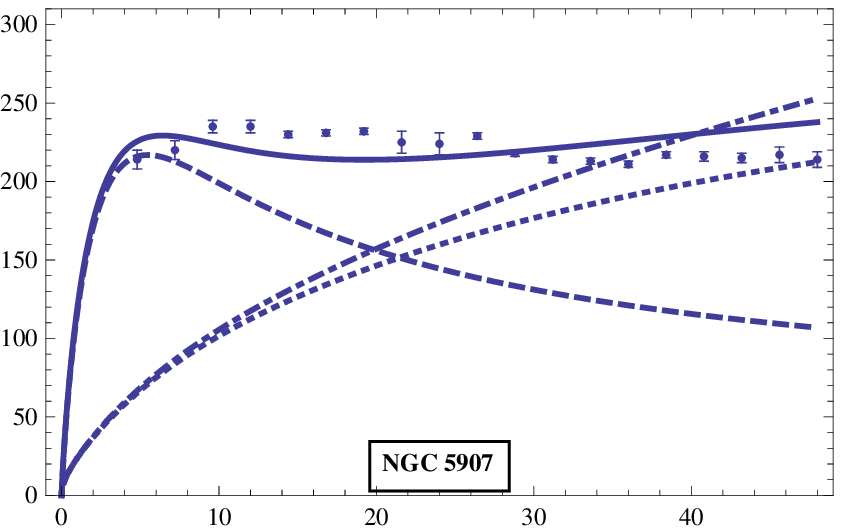,width=2.11in,height=1.2in}~~~
\epsfig{file=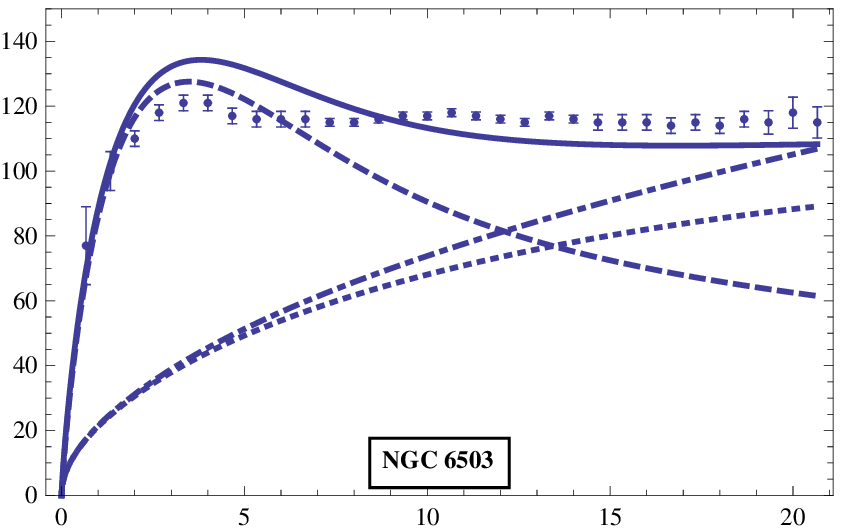,width=2.11in,height=1.2in}\\
\smallskip
\epsfig{file=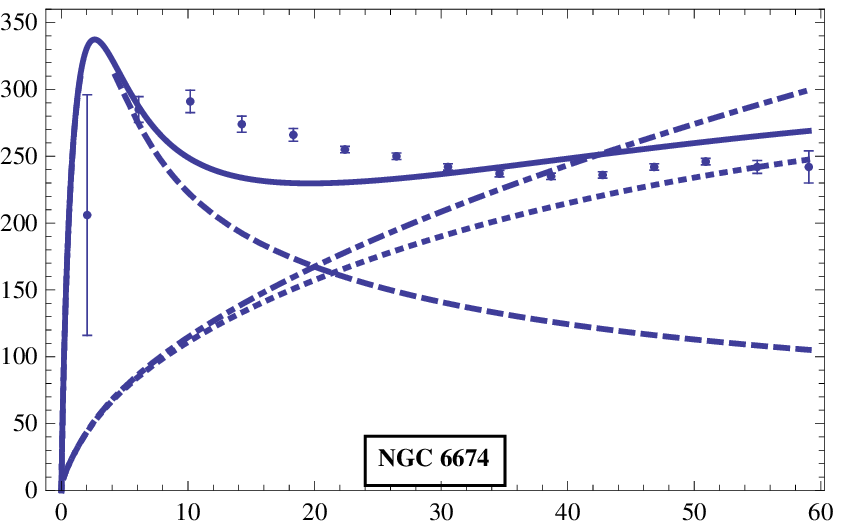,width=1.55in,height=1.2in}~~~
\epsfig{file=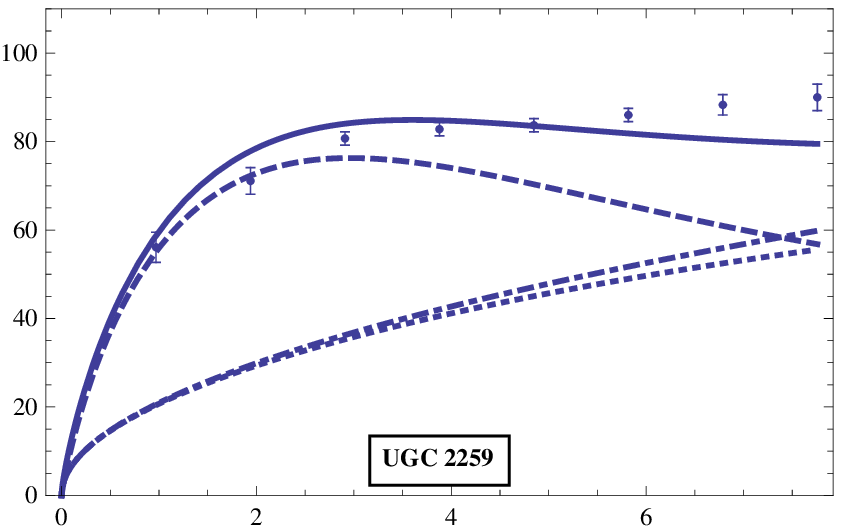,width=1.55in,height=1.2in}~~~
\epsfig{file=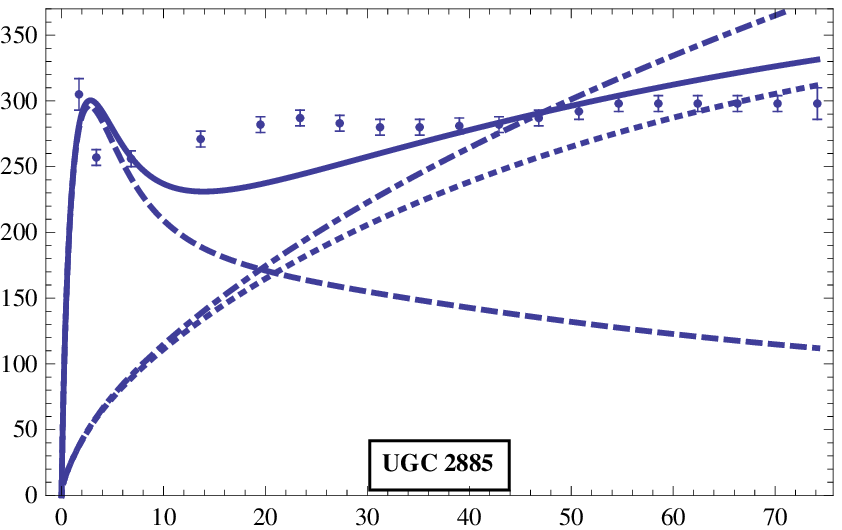,width=1.55in,height=1.2in}~~~
\epsfig{file=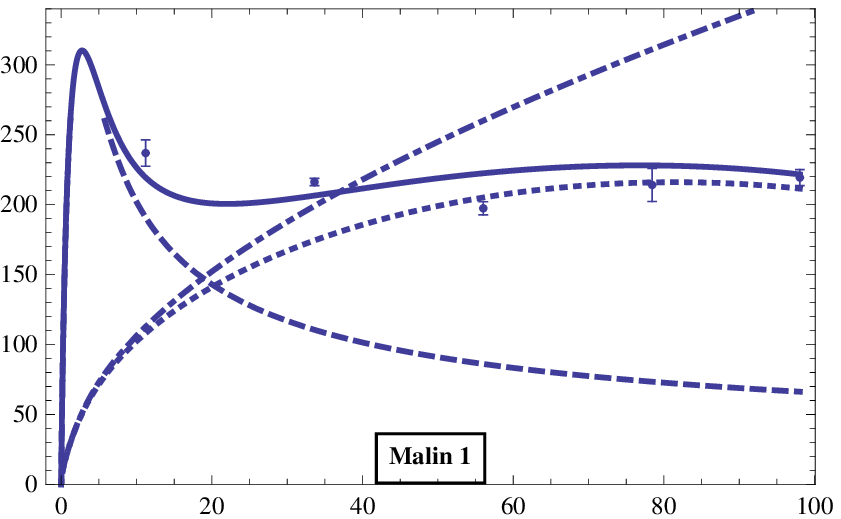,width=1.55in,height=1.2in}\\
\medskip
FIG.~5:~Fitting to the rotational velocities of the miscellaneous 22 galaxy sample 
\label{Fig. 5}
\end{figure}

\begin{figure}[t]
\epsfig{file=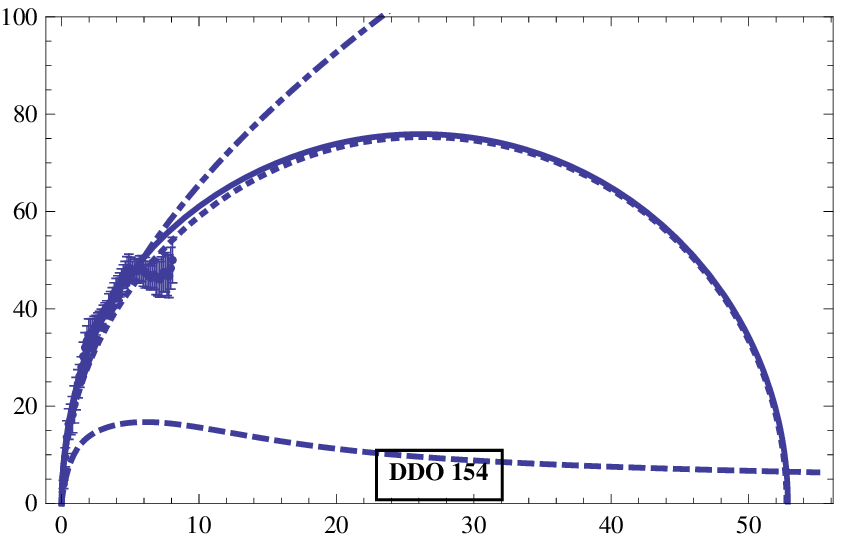, width=2.23in,height=1.2in}~~~
\epsfig{file=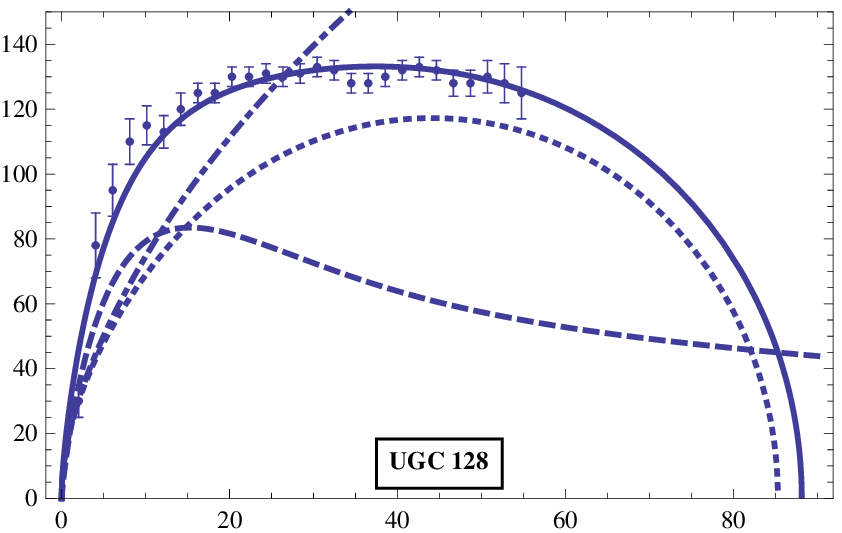, width=2.23in,height=1.2in}~~~
\epsfig{file=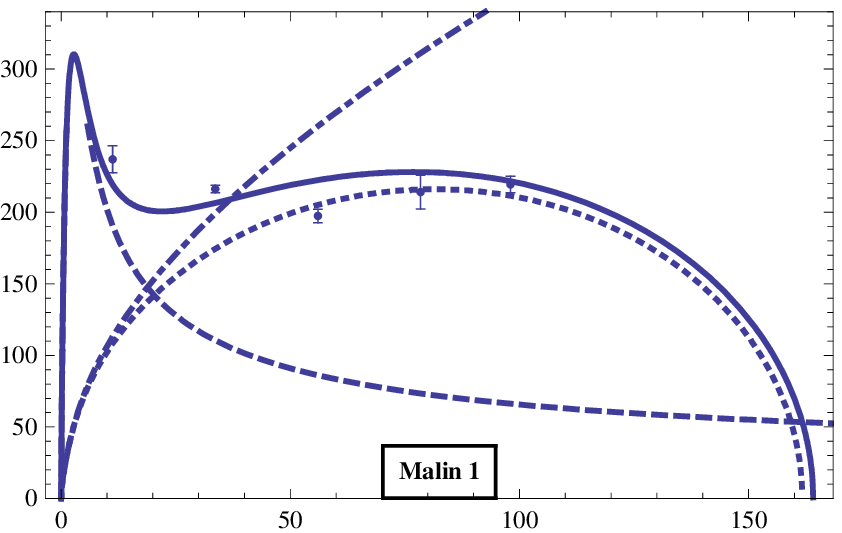, width=2.23in,height=1.2in}\\
\medskip
FIG.~6:~Extended distance predictions for DDO 154, UGC 128, and Malin 1.
\label{Fig. 6}
\end{figure}

\end{document}